\documentclass[11pt]{article}
\usepackage{etex}
\usepackage{youngtab}
\usepackage{amscd,amsmath,amssymb,amsfonts,xspace,mathrsfs,mathtools,amsthm}
\usepackage{xr-hyper}
\usepackage{jheppub}
\usepackage[dvipsnames]{xcolor}
\usepackage[utf8]{inputenc}
\usepackage{bbm}
\usepackage{graphicx}
\usepackage{tabu} 
\usepackage[color,matrix,arrow]{xy}
\usepackage{wasysym} 
\usepackage{stmaryrd}
\usepackage[shortlabels]{enumitem}
\usepackage{array,amsmath}
\usepackage{slashed}
\usepackage{tensor}
\usepackage{caption}
\usepackage{subcaption}
\usepackage{enumitem}
\usepackage[textsize=tiny]{todonotes}
\usepackage{cellspace, hhline}
\usepackage{lineno}
\usepackage{multirow}
\usepackage{svg}
\usepackage{float}
\usepackage{comment}
\usepackage{booktabs, cellspace, hhline}
\usepackage{longtable}
\usepackage{lscape}

\usepackage{tikz}
\usepackage{tikz-cd}
\usepackage{diagbox}
\usetikzlibrary{positioning}
\usetikzlibrary{calc}
\usetikzlibrary{decorations.pathreplacing,calligraphy}

\usepackage{xstring}
\usetikzlibrary{decorations.pathmorphing} 
\usetikzlibrary{decorations.markings} 
\usetikzlibrary{arrows} 
\usetikzlibrary{shapes} 
\usetikzlibrary{matrix} 
\usetikzlibrary{positioning} 
\usepackage[english]{babel} 
\usepackage[autostyle]{csquotes}
\usetikzlibrary{positioning,fit}
\usepackage[dvipsnames]{xcolor}

\DeclarePairedDelimiterX\braket[2]{\langle}{\rangle}{#1\delimsize\vert\mathopen{}#2}%

\newcommand{\rk}{\text{rk}}

\newcommand{\be}{\begin{equation}} 
\newcommand{\ee}{\end{equation}} 
\newcommand{\bea}{\begin{equation} \begin{aligned}} \newcommand{\eea}{\end{aligned} \end{equation}}
\newcommand{\bes}{\begin{equation*}}
\newcommand{\ees}{\end{equation*}}

\newcommand{\ov}{\over}

\newcommand{\kk}{\mathsf{k}}

\newcommand{\CE}{\mathcal{E}}

\newcommand{\CI}{\mathcal{I}}

\newcommand{\CN}{\mathcal{N}}
\newcommand{\CO}{\mathcal{O}} 

\newcommand{\CQ}{\mathcal{Q}}
\newcommand{\CR}{\mathcal{R}}
\newcommand{\CS}{\mathcal{S}}

\newcommand{\Tr}{\text{Tr}}

\newcommand{\SL}{\text{SL}(2,\BZ)}

\renewcommand{\t}{\widetilde }
\newcommand{\Z}{\mathbb{Z}}
\newcommand{\C}{\mathbb{C}}

\newcommand{\R}{\mathbb{R}}

\newcommand{\FR}{\mathfrak{R}}

\renewcommand{\SL}{{\mathscr L}}
\renewcommand{\l}{\mathsf{l}}

\newcommand{\qcoh}{\mathsf{q}}
\newcommand{\qk}{q}

\numberwithin{equation}{section}  
\allowdisplaybreaks  
\numberwithin{table}{section}

\title{
Schubert line defects in 3d GLSMs, part I: 
Complete flag manifolds and quantum Grothendieck polynomials}

\abstract{
We construct new half-BPS line defects in 3d $\mathcal{N}=2$ supersymmetric quiver gauge theories whose Higgs branches are complete flag manifolds $X = {\rm Fl}(n)$. Upon circle compactification, the bulk theory flows to a non-linear sigma model (NLSM) with target space $X$ and the line defects flow to objects supported on Schubert varieties $X_w \subseteq X$.
 These {\it Schubert line defects} form an important basis of the quantum K-theory of $X$. 
    They are realized as $\mathcal{N}=2$ supersymmetric quantum mechanics (SQM) quivers coupled to the 3d gauge theory. We show that the insertion of the Schubert line defect restricts the target space of the 3d gauged linear sigma model (GLSM) to the Schubert variety $X_w$, with the 1d degrees of freedom physically realizing a Bott--Samelson resolution of $X_w$. Moreover, we verify in examples that the 1d flavored Witten index of the quiver SQM reproduces the (equivariant) Chern character of the structure sheaf $\mathcal{O}_{X_w}$ as a (double) quantum Grothendieck polynomial, generalizing previous results for $X$ a Grassmannian manifold. Our construction thus provides a more direct realization of the 3d GLSM/quantum K-theory correspondence for complete flag manifolds. Finally, in the small-circle limit, we obtain a 0d-2d coupled system that realizes the Schubert classes $[X_w]$ in the quantum cohomology ring of $X$. 
}

\author[a]{Cyril Closset,}
\affiliation[a]{ School of Mathematics, University of Birmingham,\\ 
 Watson Building, Edgbaston, Birmingham B15 2TT, UK}
\emailAdd{c.closset@bham.ac.uk}

\author[b]{Wei Gu,}
\affiliation[b]{Zhejiang Institute of Modern Physics, School of Physics, Zhejiang University,\\
Hangzhou, Zhejiang 310058, China}
\emailAdd{guwei2875@zju.edu.cn}

\author[a,c]{Osama Khlaif,}
\emailAdd{osama.khlaif@phys.ens.fr}
\affiliation[c]{Philippe Meyer Institute, Physics Department, École Normale Supérieure (ENS), Université PSL, 24 rue Lhomond, F-75231 Paris, France}

\author[d]{Eric Sharpe,}
\affiliation[d]{Physics Department, 850 West Campus Drive, Virginia Tech,\\
Blacksburg, VA 24061, US}
\emailAdd{ersharpe@vt.edu}

\author[d]{Hao Zhang,}
\emailAdd{hzhang96@vt.edu}

\author[e,f]{Hao Zou}
\affiliation[e]{Center for Mathematics and Interdisciplinary Sciences, Fudan University, Shanghai 200433, China}
\affiliation[f]{Shanghai Institute for Mathematics and Interdisciplinary Sciences, Shanghai 200433, China}
\emailAdd{haozou@fudan.edu.cn}

\begin{document}

\maketitle

\newpage

\section{Introduction}

Two-dimensional non-linear sigma models (NLSM) onto K\"ahler manifolds $X$ often arise as infrared (IR) limits of 2d $\CN=(2,2)$ supersymmetric gauge theories, also known as gauged linear sigma models (GLSM)~\cite{Witten:1993yc}. These models govern the quantum geometry of $X$ as probed by genus-zero string worldsheets $\Sigma$ --- in the NLSM, we have worldsheet instantons corresponding to non-trivial holomorphic maps $\Sigma\rightarrow X$, while in the GLSM we have a sum over gauge vortices~\cite{Morrison:1994fr,Closset:2015rna, Benini:2015noa} that can be formalized as quasimaps~\cite{ciocan2014stable,kim2014stablequasimapsholomorphicsymplectic, Kim:2016jye}. The relevant quantum geometry is encoded mathematically into the quantum cohomology ring ${\rm QH}^\bullet(X)$, which corresponds to the twisted chiral ring of the 2d GLSM.

In this work, we are interested in the 3d uplift, where we consider a 3d $\CN=2$ supersymmetric gauge theory with gauge group $G$ and with a Higgs branch given by the Geometric Invariant Theory (GIT) quotient 
\be
X \,\cong\, V\sslash  G_\C~,
\ee
with the 3d Chern--Simons levels lying within windows such that there are no additional Coulomb or hybrid vacua~\cite{Intriligator:2013lca, Closset:2023jiq, Gu:2023hgm,Gu:2020zpg}. The 3d gauge theory on $\Sigma \times S^1$ gives us an effective 2d $\CN=(2,2)$ supersymmetric description along $\Sigma$, which we call the 3d GLSM description. It naturally flows to an NLSM in 2d, but the presence of the additional circle direction modifies the IR interpretation significantly. 

The twisted chiral operators are half-BPS line operators wrapping the circle direction --- see~{\it e.g.}~\cite{Closset:2019hyt}. 
Instead of the quantum cohomology ring ${\rm QH}^\bullet(X)$, for suitable Chern-Simons levels, the twisted chiral ring one obtains from those line operators in the 3d GLSM can be identified with the quantum K-theory ring of $X$~\cite[section 2.4]{Bullimore:2014awa}, \cite{Jockers:2018sfl, Jockers:2019lwe, Jockers:2019wjh, Jockers:2021omw, Ueda:2019qhg, Koroteev:2017nab, 
Gu:2020zpg, Gu:2022yvj, Gu:2023tcv, Gu:2023fpw, Gu:2025abc, Gu:2021yek, Gu:2021beo, Sharpe:2024ujm, Dedushenko:2023qjq, Huq-Kuruvilla:2024tsg, Huq-Kuruvilla:2025nlf}.
One simple basis of half-BPS line operators are the Wilson lines for the 3d gauge group $G$, which flow to locally free sheaves over $X$. In a previous work, two of the authors considered the case of the Grassmannian manifold $X= {\rm Gr}(k,n) \cong \C^{kn}\sslash { GL}(k)$ and constructed half-BPS line defects that correspond to more general coherent sheaves~\cite{Closset:2023bdr}, namely the Schubert classes with compact support on Schubert varieties inside $X$. It was also shown that these line defects provide a physical construction of the {\it Grothendieck polynomials} that can be used to represent the Schubert varieties in the quantum K-theory ring of $X$. This construction was later generalized to Lagrangian Grassmannians by a complementary subset of the authors~\cite{Gu:2025tda}.

In this work and its sequel (part II)~\cite{Closset:2026bnk}, we generalize the line defect construction of Schubert classes~\cite{Closset:2023bdr,Gu:2025tda} to flag manifolds. Recall that a flag 
\be
V_\bullet \,= \,(V_{1}\,\subset \,V_2\, \subset\, \cdots\,\subset\, V_s\, \subset\, \C^n)
\ee
consists of nested linear subsets $V_i\, \cong\, \C^{{\rm dim}(V_i)}$ of some fixed ambient linear space $\C^n$. The complete flag manifold is defined as the set of all complete flags:
\be
 {\rm Fl}(n)\,=\,\{V_\bullet = (0\,\subset \,V_1\,\subset\, V_2\,\subset\, \cdots\,\subset\, V_{n-1}\,\subset\,\C^n)~|~\dim(V_i) \,= \,i\}~, \ee
and the partial flag manifold consists of all partial flags with fixed dimension vector ${\bf k}=(k_\ell)$:
\be
 {\rm Fl}(k_1, \cdots, k_s; n)\,=\,\{V_\bullet \,=\, (0\,\subset\, V_1\,\subset\, V_2\,\subset\, \cdots\,\subset \,V_{s}\,\subset\,\C^n)~|~\dim(V_\ell) \,=\, k_\ell\}~.
\ee
Obviously, this generalizes the Grassmannian ${\rm Gr}(k,n)={\rm Fl}(k; n)$. A famous property of flag manifolds is that they admit a decomposition as a disjoint union of Schubert cells:
\be
 {\rm Fl}({\bf k}; n) \,=\, \bigsqcup_{w\in {\rm W}^{({\bf k}; n)}} \,X_w^\circ~.
\ee
While the indexing set~${\rm W}^{({\bf k}; n)}$ appearing here will be defined in part II~\cite{Closset:2026bnk}, the Schubert cells of the complete flag are simply indexed by permutations $w\in S_n$:
\be
 {\rm Fl}(n) \,=\, \bigsqcup_{w\in S_n}\, X_w^\circ~.
\ee
The Schubert variety $X_w$ is the closure of the Schubert cell,  $X_w=  \overline{X_w^\circ}$. Note that there are $n!$ distinct Schubert varieties inside $X={\rm Fl}(n)$, including $X$ itself.

\medskip
\noindent
\textbf{Schubert line defects for complete flag manifolds.} Let us focus on the complete flag manifold, $X={\rm Fl}(n)$, which is the subject of this paper. (More general partial flag manifolds will be discussed in the sequel~\cite{Closset:2026bnk}.) The key result of this paper is an explicit construction of a 1d-3d coupled system which defines a half-BPS defect line $\SL_w$ that will flow to 
an object supported on 
the Schubert variety $X_w$.
The defect line $\SL_w$ is defined as an $\CN=2$ supersymmetric quantum mechanics (SQM)~\cite{Witten:1982im} --- more specifically, it is a 1d quiver gauge theory, with 1d unitary gauge group $G_{\rm 1d}= \prod_\ell U(r_\ell)$ and a specific matter content determined by $w\in S_n$, coupled to the 3d bulk fields. Our key result is that, when specialized to the location of the line insertion at a point on $\Sigma$, the 1d-3d coupled system behaves as a 1d GLSM for the Schubert variety $X_w$ itself. Indeed, our construction is closely related to a recent mathematical description of the  Bott--Samelson resolution~\cite{BottSamelsonResolution,DemazureOnBottSamelson} of Schubert varieties using quiver subrepresentations~\cite{iezzi2025quivergrassmanniansbottsamelsonresolution}, or equivalently in terms of bioriented flags~\cite{cibotaru2020bioriented}. Physically, the additional one-dimensional degrees of freedom on the line provide us with the resolution of the Schubert variety.

The 1d $\CN=2$ supersymmetric gauge theory couples to the 3d gauge fields and superpartners as well as to $GL(n)$ fugacities%
\footnote{
We keep the $GL(n)$ manifest in our notation, but of course the global symmetry is only $SL(n)$.
}, realizing a 3d symmetry-breaking pattern specified by $w\in S_n$. Taking the flavored Witten index at fixed 3d gauge fugacities $x^{(i)}$, we obtain a polynomial in the 3d gauge fugacities $x^{(i)}$ as well as in the $T\equiv (\mathbb{C}^\times)^n \subset{GL}(n)$ fugacities $y_i$. Our second key result is that the Witten index of the 1d lines precisely reproduces the {\it (double) quantum Grothendieck polynomials} $\mathfrak{G}_w^{(\qk)}(x,y)$ that are known to represent the (equivariant) Schubert classes in the quantum K-theory of the complete flag manifold~\cite{Buch02,FL94,lascoux1982structure,Las90,lenart2006quantum,MNS23}. We find:
\be\label{WI result intro}
{\bf I}_{\rm W}(\SL_w)\, \cong\,  {\rm ch}_T([\CO_w]) \,=\, \mathfrak{G}_w^{(\qk)}(x,y)~,
\ee
where the second equality holds in the twisted chiral ring of the 3d GLSM. Indeed, the Witten index is naturally computed in terms of $n(n-1)/2$ variables $x^{(i)}_{a_i}$ ($a_i=1,\cdots, i$ for $i=1, \cdots, n-1$), which correspond mathematically to exponentiated (K-theoretic) Chern roots for the tautological vector bundles $\mathcal{S}_i$ over ${\rm Fl}(n)$ with fiber $V_i$. Instead, the quantum K-theory ring relations and the quantum Grothendieck polynomials themselves are usually expressed in terms of $n$ variables $x_i$ corresponding to the quotient line bundles  $\CQ_i$ with fiber $V_i/V_{i-1}$ \cite{lenart2006quantum}. One can systematically transform any polynomial in the $x^{(i)}$ variables into one in the $x_i$ variables using the quantum K-theory relations, which leads us to the advertised result. More precisely, in this paper, we check~\eqref{WI result intro} explicitly for $n\leq 4$; a general proof of the identity is left for future work, though it is clear from the physics that it should hold in general. 
 For the special case of the longest permutation, for general $n$, our proposal matches the result in \cite[Theorem 1.2]{ahkmox}. Indeed, we will explain in Part~II~\cite{Closset:2026bnk} that the Witten indices~${\bf I}_{\rm W}(\SL_w)$, when  expressed in the original 3d gauge variables, match precisely with the Whitney-presentation polynomials introduced in~\cite{ahkmox}; by contrast, the quantum Grothendieck polynomials represent the same quantum K-theory classes in the Toda presentation.

\medskip
\noindent
\textbf{Plan of the paper.} The paper is structured as follows. In section~\ref{sec:defects and dualities}, we provide general comments on coupling 3d $\CN=2$ gauge theories to 1d $\CN=2$ gauge theories, and emphasize the role of 1d dualities in the description of the line defects. In section \ref{sec:GLSM for flag}, we give a brief review of the GLSM construction of complete flag manifolds. We state the quantum cohomology ring relations as well as the quantum K-theory ring relations.  
Section~\ref{sec:schubert line defects} discusses Schubert varieties and presents our construction of the Schubert line defects in the 3d GLSM. We provide a detailed discussion on the way these defects restrict the target space to the corresponding Schubert variety and how the 1d degrees of freedom realize a resolution thereof. We also compute the flavored Witten indices for 1d $\CN=2$ SQM on the defects. We check in an explicit example the matching between these indices and the expected Chern characters of the Schubert classes in the quantum K-theory ring of Fl$(3)$. To complement this section, we also added the details of the calculations for the Fl$(4)$ case in appendix~\ref{app:Fl(4) defects}.
Taking a strict 2d limit of the 3d GLSM, in section \ref{sec:0d-2d system} we discuss the resulting 0d-2d coupled system. We show in examples that the $\mathcal{N}=2$ supersymmetric matrix model realizes the Schubert classes in the equivariant quantum cohomology ring of the flag manifold. 
In appendix~\ref{app:Schubert classes}, we give a review of the definitions of the (quantum) Schubert and Grothendieck polynomials and discuss their geometric significance to our work.  In appendix~\ref{app:quiver-vs-math}, we provide some technical details relating the quiver construction on the defects to quiver Grassmannians and Bott-Samelson resolutions.

\medskip
\noindent
\textbf{A comment on conventions.}
 In this paper, we denote permutations using the `window' notation,
as in e.g.~\cite{Billey-Lakshmibai-2000}, in which the $i$-th entry is the image of $i$.
For example, $(3\,2\,4\,1)$ denotes the permutation that sends
\begin{displaymath}
    1 \,\mapsto\, 3, \: \: \:
    2 \,\mapsto\, 2, \: \: \:
    3 \,\mapsto \,4, \: \: \:
    4 \,\mapsto \, 1.
\end{displaymath}
This should be distinguished from the `cycle' notation, in which $(3\,2\,4\,1)$ denotes the permutation that maps
\begin{displaymath}
    3 \,\,\mapsto \,\,2 \,\,\mapsto\,\, 4 \,\,\mapsto\,\, 1\,\, \mapsto\,\, 3.
\end{displaymath}
To take another example, the identity permutation in window notation is $(1\,2\,3\,4)$, whereas in cycle notation it is $(1)\,(2)\,(3)\,(4)$.

\section{Defect line operators of 3d \texorpdfstring{$\CN=2$}{N=2} quivers as 1d--3d coupled quivers}\label{sec:defects and dualities}

In this section, we discuss general aspects of defect line operators in 3d $\CN=2$ supersymmetric gauge theories. We first discuss the coupling of 1d $\CN=2$ SQM to 3d gauge fields in some generality, and then specialize to the case of unitary quiver SQM.

\subsection{Twisted chiral ring of half-BPS lines and 1d-3d coupled defects}

The half-BPS lines of 3d $\CN=2$ supersymmetric field theories form an algebra $\CR_{\rm 3d}$ under fusion of parallel lines. Given any set of lines $\{\SL_i\}$ forming a basis of $\CR_{\rm 3d}$, the fusion ring takes the form (see~{\it e.g.}~\cite{Witten:1993xi, Nekrasov:2009uh, Closset:2016arn}):
\be\label{twisted chiral ring L}
\SL_i\,\star\,  \SL_j \,=\, {\CN_{ij}}^k(y,q)\, \SL_k~, \qquad\quad  {\CN_{ij}}^k(y)\,\in\,  \Z(y,q)~.
\ee
The fusion coefficients ${\CN_{ij}}^k$ generally depend on the flavor fugacities $y$ and $q$, where $y$ are ordinary flavor parameters while $q$ are parameters for topological symmetries (FI parameters) which give us the `quantum parameters' of the 3d ring in the geometric phase. This ring structure is best understood by considering the 3d theory on $\R^2 \times S^1_\beta$ and wrapping the lines along the circle $S^1_\beta$. Then, the operators $
\SL_i$ are interpreted as local operators in the 2d effective description along $\R^2$. The line operators are twisted chiral operators, {\it i.e.} they commute with the two supercharges $Q_-$ and $\overline{Q}_+$ of the 2d  $\CN=(2,2)$ supersymmetry algebra; as per usual, the twisted chiral ring~\eqref{twisted chiral ring L} is obtained in (joint) cohomology of these two supercharges.

A 3d GLSM is defined to be a 3d $\CN=2$ supersymmetric gauge theory with a Higgs phase, compactified on $S^1_\beta$. For simplicity, we assume that the Higgs branch is the only moduli space of vacua of the 3d theory --- {\it i.e.} that there are no additional Coulomb or hybrid vacua. The circle compactification allows us to understand the infrared dynamics in two-dimensional terms, as an NLSM with target space $X$ being the Higgs branch of the 3d gauge theory. The main effect of the circle compactification is that supersymmetric ground states correspond to the quantum K-theory of $X$, instead of the standard quantum cohomology in a 2d NLSM. In short, one identifies the twisted chiral ring geometrically, in 2d or 3d, respectively, as:
\be
\CR_{\rm 2d}\,\cong\, {\rm QH}^\bullet(X) \qquad \text{vs.} \qquad  \CR_{\rm 3d}\,\cong\, {\rm QK}(X)~.
\ee
In the 3d GLSM picture, the key question is to identify the quantum K-theory class to which a specific half-BPS line will flow in the infrared:
\be
\SL_i \qquad \rightsquigarrow\qquad  [\CE_i] \,\in \,{\rm QK}(X)~.
\ee
Here $\CE_i$ is a specific coherent sheaf over $X$ or, more generally, an element of the derived category of coherent sheaves of $X$ --- in (quantum, equivariant) K-theory, this amounts to a formal linear combination of coherent sheaves (with coefficients in $\Z(q, y)$).

Let us consider a 3d $\CN=2$ supersymmetric gauge theory with gauge group $G$ and chiral multiplets transforming in some linear representation $\FR\cong \C^{\text{dim}(\FR)}$ of $G$. The Higgs branch is then given, schematically, by a GIT quotient:
\be
X\,\cong \, [\FR \sslash G_\C ]~.
\ee
Note that $X$ has complex dimension 
\be
\text{dim}(X)\,= \,\text{dim}(\FR)\,-\, \text{dim}(G_\C)~. 
\ee
For instance, in the classic example of the complex Grassmannian $X={\rm Gr}(k,n)$, we take $G=U(k)$ and $\FR\cong\C^{kn}$ ($n$ copies of the fundamental representation), hence $\text{dim}(X)= k (n -k)$.

\medskip
\noindent
{\bf Wilson lines as locally free sheaves.} A typical half-BPS line operator in the 3d gauge theory is the Wilson loop in some representation $R$ of $G$~\cite{Closset:2019hyt}:
\be\label{Wline gen}
W_{R}\, = \,\Tr_{R}\, P  \exp{\left(-i \int_{S^1_\beta} \left(A - i \sigma d\tau\right) \right)}~.
\ee
In the geometric phase, this flows to the homogeneous vector bundle:
\be\label{def hom vec bundle}
\CE_R \,\equiv\, (\FR\times R)\sslash G_\C \,\longrightarrow \,X~,
\ee
where $G_\C$ acts on $R$ in the corresponding representation. 
In particular, the Chern character of $\CE_R$ can be directly recovered from~\eqref{Wline gen}.

\medskip
\noindent
{\bf Vortex lines as coherent sheaves.} A general {\it vortex line} or, more generally, {\it defect line}, can be defined as a 1d $\CN=2$ gauge theory coupled to the 3d bulk fields --- this is a standard way of constraining the 3d gauge and matter fields at the location of the line, thus defining a line defect after integrating out the worldline degrees of freedom.\footnote{For a more recent discussion in the 3d holomorphic-topological (HT) twist framework, see \cite{Dimofte:2025oqf}.} We refer to the recent work~\cite{Closset:2025zyl}, whose conventions we follow,  for a general discussion of 1d $\CN=2$ gauge theories.

The 1d $\CN=2$ SQM consists of a 1d $\CN=2$ vector multiplet for some 1d gauge group  $G_{\rm 1d}$, coupled to 1d $\CN=2$ chiral multiplets $\varphi$ in some representation $R_{\varphi}$ of $G_{\rm 1d}$ and some 1d $\CN=2$ Fermi multiplets $\Gamma$ in some representation $R_{\Gamma}$ of $G_{\rm 1d}$. The 1d chiral and Fermi multiplets also transform non-trivially under the 3d gauge field $G$, which enters as a flavor symmetry from the worldline perspective. Finally, we have the $E$- and $J$-terms that impose relations among chiral multiplets:
\be
E_\Gamma(\varphi,\phi)\,=\,0~, \qquad J_\Gamma(\varphi,\phi)\,=\,0~.
\ee
Note that they also couple together the 1d chiral multiplets $\varphi$ and the 3d chiral multiplets $\phi$. 
Here, for definiteness, let us assume that $J=0$ while we have a non-trivial $E$-term for each Fermi field $\Gamma$ --- this will be the case considered in this work. In appropriate circumstances, the defect line defines a coherent sheaf $\CO_Y$ which is simply the structure sheaf of a subvariety: 
\be
Y \,\cong\, \left(\FR\,\times\, R_\varphi \,| \, E_\Gamma\,=\,0 \right)\sslash G_\C\, \times\, (G_{\rm 1d})_\C~.
\ee
Namely, one constructs the zero locus of the section $E_\Gamma(\varphi,\phi)$ inside  $\FR\times R_\varphi$ before taking a GIT quotient by both the 1d and 3d gauge groups. The resulting coherent sheaf $\CO_Y$ over $X$ has support over the subvariety $Y$ of  dimension:
\be
 \text{dim}(Y)\,=\, \text{dim}(\FR)\,+\, \text{dim}(R_\varphi)\,-\,\text{dim}(R_\Gamma)\, -\, \text{dim}(G_\C)\,-  \,\text{dim}((G_{\rm 1d})_\C)~. 
\ee
This schematic description involves a number of subtleties which we shall not fully discuss, including the fact that the 1d Fermi multiplets must be in a phase of the $\CN=2$ SQM where they are all massive, and that we implicitly assumed specific choices of 1d Chern--Simons levels~\cite{Closset:2025zyl}. Our goal here is not to give a rigorous theory of 1d vortex lines in 3d GLSMs, but only to motivate the explicit construction of this paper in the special case of flag manifolds. Thus, let us simply illustrate the general discussion above with a very simple example.

\medskip
\noindent
{\bf Examples of defect lines on $X=\mathbb{P}^1 = {\rm Fl}(2)$.} Consider $G=U(1)$ coupled to two chiral multiplets $\phi_1$ and $\phi_2$ of unit charge. For a positive FI parameter, $\zeta>0$, the Higgs branch is the projective space:
\be
X\,=\, \C^2\sslash {GL}(1) \,\cong\, {\rm Fl}(2)\, =\, \mathbb{P}^1~,
\ee
with homogeneous coordinates $[\phi_1, \phi_2]$. First, consider a Wilson loop of charge $p\in \Z$,
\be
W_p \,= \,e^{- i p \int_{S^1_\beta} (A-i \sigma d\tau)} \,\cong\, x^p \,=\, {\rm ch}(\CO(-p))~,
\ee
where $x$ it the $U(1)$ gauge fugacity. At the location of the vortex line itself, we can interpret the charge $p$ as a 1d Chern--Simons level, so that the 1d ground states built out of the fields $\phi_1, \phi_2$ are precisely sections of the line bundle~\cite{Hori:2014tda,Closset:2025zyl}:
\be
\CO(-p) \,\cong\, (\C^2 \times \C)\sslash  { GL}(1)~,
\ee
as expected from~\eqref{def hom vec bundle}. Next, we would like to engineer a skyscraper sheaf $\CO_p$ with support at the point $p=[a,b]\in \mathbb{P}^1$. This is easily achieved by introducing a single 1d chiral multiplet $\varphi$ of charge $1$ under $G_{\rm 1d}=U(1)$ (with 1d bare CS level $Q=0$) and of charge $-1$ under $G=U(1)$. We also introduce a single Fermi multiplet $\Gamma$ of charge $1$ under $G_{\rm 1d}$ and which couples to the 1d and 3d fields by the $E$-term constraint:
\be\label{Egamma expl}
E_\Gamma \,=\, \varphi\, (b\, \phi_1\, -\, a\, \phi_2)\,=\,0~.
\ee
Finally, we need to pick the 1d FI parameter $\xi >0$. Then, the 1d and 1d-3d $D$-term equations on the worldline read:
\be
|\varphi|^2\,=\,\xi~, \qquad -|\varphi|^2\,+\, |\phi_1|^2\,+\,|\phi_2|^2\, =\, \zeta~.
\ee
We are forced to have $\varphi \neq 0$, which fully Higgses $G_{\rm 1d}$, and the $E$-term constraint then gives us $b \phi_1 - a \phi_2=0$. Up to gauge transformations, this fixes
\be
[\phi_1, \phi_2]\,=\, [a, b]~,
\ee
which is the expected support of the skyscraper sheaf. Indeed, we simply gave a physics construction to the resolution:
\be\label{res Op}
0\,\longrightarrow\, \CO(-1) \,\overset{E_\Gamma}{\longrightarrow}\, \CO\, \longrightarrow \,\CO_p \,\longrightarrow\, 0~,
\ee
as discussed in~\cite{Closset:2023bdr}. Finally, one can check that the Witten index of the 1d gauge theory reads:
\be
{\bf I}_{\rm W}\,=\, -\oint_{(z=x)} {dz\ov 2 \pi i z} \,{1\,-\,z\ov1\,-\, z\, x^{-1}} \,=\, 1\,-\,x\, =\, {\rm ch}(\CO_p)~,
\ee
as expected from~\eqref{res Op}.

\subsection{Unitary quivers and dualities on the worldline}  \label{sect:duality}

\begin{figure}[t!]
    \centering
    \includegraphics[width=0.25\linewidth]{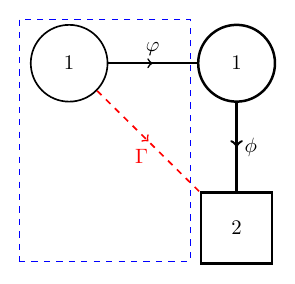}
    \caption{A simple example: the 3d quiver for $X={\rm Fl}(2)=\mathbb{P}^1$ coupled to a 1d $\CN=2$ SQM engineering the skyscraper sheaf $\CO_p$. The 1d quiver is indicated by the dashed blue rectangle. Note that the Fermi multiplet $\Gamma$ is not in the antifundamental of the $SU(2)$ global symmetry; instead, we have a single $\Gamma$ fields which breaks the flavor symmetry explicitly through the $E$-term~\protect\eqref{Egamma expl}.}
    \label{fig:simpleCase}
\end{figure}

Let us now specialize to the kind of 1d-3d coupled systems considered in this paper. Both the 3d and the 1d gauge groups are taken to be products of unitary gauge groups:
\be
G\,=\, \prod_{i}\, U(k_i)~, \qquad\quad  G_{\rm 1d}\, =\, \prod_\ell\, U({\rm r}_\ell)~.
\ee
The 3d chiral multiplets $\phi$ are in bifundamental or (anti)-fundamental representations of the $U(k_i)$'s, corresponding to a 3d quiver, and similarly for the 1d chiral multiplets $\varphi$, so that the 1d worldline theory is embedded into a larger quiver with gauge group $G\times G_{\rm 1d}$ --- we may also include flavor nodes, denoted by squares. 
Finally, we introduce bifundamental Fermi multiplets $\Gamma$, with corresponding $E$-terms constraints:
\be
E_\Gamma(\varphi, \phi)\,=\,0~,
\ee
which impose relations among the 1d-3d chiral-multiplet quiver arrows. 
A simple example, corresponding to the $X=\mathbb{P}^1$ discussed above, is shown in figure~\ref{fig:simpleCase}.

As part of the definition of the 1d defect, one needs to specify the 1d FI parameters $\xi_\ell$; they are chosen so that the 1d-3d quiver admits a non-trivial geometric phase. One also needs to specify the bare 1d Chern--Simons levels $Q_\ell\in \Z$, in the conventions of~\cite{Closset:2025zyl}. In this work, for simplicity, we choose
\be\label{Qellzero cond}
Q_\ell \,=\,0~, \qquad \xi_\ell\,>\,0~,
\ee
for every $U({\rm r}_\ell)$ gauge group. 
It is not difficult to relax this constraint, but this will not be necessary for our purpose.

\medskip
\noindent
{\bf Dualities in 1d $\CN=2$ SQM.}  The 1d unitary quivers of the type discussed here admit Seiberg-like dualities recently discussed in~\cite{Closset:2025zyl}. Here, we present some simple consequences of these dualities for our description of 1d defect lines. Indeed, using 1d dualities, complicated 1d quivers can often be simplified.  Here we will focus on duality moves that correspond to the `full confinement' of a 1d unitary gauge node, effectively removing the node and replacing it with some effective `mesonic' arrows. 

These duality operations are local on the quiver. Consider a  $U({\rm r})$ node connected to other nodes (gauge or flavor) by (anti)fundamental chiral and Fermi arrows, as follows:

\begin{equation}\label{Ur local in Q1d}
\raisebox{-2cm}{
\begin{tikzpicture}[>=stealth,
    every node/.style={circle, draw, minimum size=2em},->-/.style={decoration={
  markings,
  mark=at position #1 with {\arrow{>}}},postaction={decorate}}]
\pgfmathsetmacro{\a}{1.5}
\node (r) at (0,0) [circle,draw,thick]{\(\displaystyle {\rm r}\)};

\node (n2) at (-\a,0) [circle,draw]{\(\displaystyle n_2\)};
\node (n1) at (\a,0) [circle,draw]{\(\displaystyle n_1\)};

\node (n4) at (0,\a) [circle,draw]{\(\displaystyle n_4\)};
\node (n3) at (0,-\a) [circle,draw]{\(\displaystyle n_3\)};

\draw[->-=0.6,thick] (n2) -- (r);
\draw[->-=0.6,thick] (r) -- (n1);

\draw[->-=0.6,dashed,thick] (n4) -- (r);
\draw[->-=0.6,dashed,thick] (r) -- (n3);

\end{tikzpicture}}
\end{equation}
Here, $n_1$ and $n_3$ denote the total number of fundamental chiral and Fermi multiplets, respectively, while $n_2$ and $n_4$ denote the total number of antifundamental chiral and Fermi multiplets, respectively.%
\footnote{Hence, for instance, the node denoted by $n_2$ might not be a $U(n_2)$ node, because it could be split into several $U(m_i)$ nodes with $\sum_i m_i=n_2$, and similarly for the other  nodes surrounding $U({\rm r})$ in~\protect\eqref{Ur local in Q1d}.} In the following discussion, the condition~\eqref{Qellzero cond} is crucial. We then find three useful duality moves, as a direct application of the results of~\cite{Closset:2025zyl}.

\medskip
\noindent
{\bf Type I duality.}  The simplest confinement duality occurs if and only if $n_1={\rm r}$, in which case we have the following confinement duality:
\begin{equation}
\raisebox{-2cm}{
\begin{tikzpicture}[>=stealth,
    every node/.style={circle, draw, minimum size=2em},->-/.style={decoration={
  markings,
  mark=at position #1 with {\arrow{>}}},postaction={decorate}}]
\pgfmathsetmacro{\a}{1.5}
\node (r) at (0,0) [circle,draw,thick]{\(\displaystyle n_1\)};

\node (n2) at (-\a,0) [circle,draw]{\(\displaystyle n_2\)};
\node (n1) at (\a,0) [circle,draw]{\(\displaystyle n_1\)};

\node (n4) at (0,\a) [circle,draw]{\(\displaystyle n_4\)};
\node (n3) at (0,-\a) [circle,draw]{\(\displaystyle n_3\)};

\draw[->-=0.6,thick] (n2) -- (r);
\draw[->-=0.6,thick] (r) -- (n1);

\draw[->-=0.6,dashed,thick] (n4) -- (r);
\draw[->-=0.6,dashed,thick] (r) -- (n3);

\end{tikzpicture}}
\qquad
\longleftrightarrow
\qquad
\raisebox{-2cm}{
\begin{tikzpicture}[>=stealth,
    every node/.style={circle, draw, minimum size=2em},->-/.style={decoration={
  markings,
  mark=at position #1 with {\arrow{>}}},postaction={decorate}}]
\pgfmathsetmacro{\a}{1.5}
\node (n2) at (-\a,0) [circle,draw]{\(\displaystyle n_2\)};
\node (n1) at (\a,0) [circle,draw]{\(\displaystyle n_1\)};
\node (n4) at (0,\a) [circle,draw]{\(\displaystyle n_4\)};
\node (n3) at (0,-\a) [circle,draw]{\(\displaystyle n_3\)};

\draw[->-=0.6,thick] (n2) -- (n1);
\draw[->-=0.6,dashed,thick] (n4) -- (n1);
\draw[->-=0.6,dashed,thick] (n1) -- (n3);

\end{tikzpicture}
}
\end{equation}
In this work, we call this duality move a  `Type I' duality. In the language of the 1d $\CN=2$ Seiberg-like dualities~\cite{Closset:2025zyl}, this is a right-mutation at the $U({\rm r})$ node. 

\medskip
\noindent
{\bf Type II and Type III dualities.} 
The second and third duality moves, dubbed `Type II' and `Type III' dualities in the following, are more intricate and require one of the following two sets of conditions:
\bea
&\text{Type II:}\qquad  &&n_4\,=\,0~, \qquad   &&{\rm r}\,=
\,n_1\,-\,n_3~, \qquad &&{\rm r}\,\geq\, n_2~,\\
&\text{Type III:}\qquad  &&n_4\,=\,0~, \qquad   &&{\rm r}\,=\, n_2~, \qquad &&{\rm r}\,\leq\, n_1\,-\,n_3~.\\
\eea
If either of these conditions holds, we have the confinement duality:
\begin{equation}
\raisebox{-1cm}{
\begin{tikzpicture}[>=stealth,
    every node/.style={circle, draw, minimum size=2em},->-/.style={decoration={
  markings,
  mark=at position #1 with {\arrow{>}}},postaction={decorate}}]
\pgfmathsetmacro{\a}{1.5}
\node (r) at (0,0) [circle,draw,thick]{\(\displaystyle {\rm r}\)};

\node (n2) at (-\a,0) [circle,draw]{\(\displaystyle n_2\)};
\node (n1) at (\a,0) [circle,draw]{\(\displaystyle n_1\)};

\node (n3) at (0,-\a) [circle,draw]{\(\displaystyle n_3\)};

\draw[->-=0.6,thick] (n2) -- (r);
\draw[->-=0.6,thick] (r) -- (n1);

\draw[->-=0.6,dashed,thick] (r) -- (n3);

\end{tikzpicture}}
\qquad
\longleftrightarrow
\qquad
\raisebox{-1cm}{
\begin{tikzpicture}[>=stealth,
    every node/.style={circle, draw, minimum size=2em},->-/.style={decoration={
  markings,
  mark=at position #1 with {\arrow{>}}},postaction={decorate}}]
\pgfmathsetmacro{\a}{1.5}
\node (n2) at (-\a,0) [circle,draw]{\(\displaystyle n_2\)};
\node (n1) at (\a,0) [circle,draw]{\(\displaystyle n_1\)};
\node (n3) at (0,-\a) [circle,draw]{\(\displaystyle n_3\)};

\draw[->-=0.6,thick] (n2) -- (n1);
\draw[->-=0.6,dashed,thick] (n2) -- (n3);

\end{tikzpicture}
}
\end{equation}
In the language of Seiberg-like dualities, this is a right mutation followed by a trivial wall-crossing plus another right mutation (Type II) or a trivial wall-crossing followed by a left mutation (Type III)~\cite{Closset:2025zyl}.  
These three 1d duality moves often allow us to simplify the rectangular quivers significantly, to be introduced below, by removing `redundant' nodes.


\section{Review of GLSMs for complete flag manifolds} \label{sec:GLSM for flag}

\begin{figure}[t]
    \centering
    \includegraphics[scale=1.1]{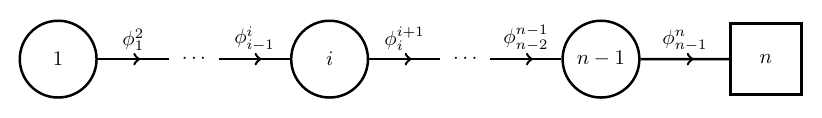}
    \caption{The 2d quiver gauge theory of interest. The circle nodes are gauge groups $U(i)$, and the square node denotes the flavor symmetry group $SU(n)$. 
    The fields $\phi_{i}^{i+1}$ are bifundamental scalars in chiral multiplets. The Higgs branch of this theory is the complete flag variety Fl$(n)$.}
    \label{fig:complete Flag quiver}
\end{figure}

In this section, we review the GLSM construction of the complete flag manifold Fl$(n)$:
\begin{equation}\label{Fl(n) defn}
    {\rm Fl}(n)\,=\,\{V_\bullet = (0\subset V_1\subset V_2\subset \cdots\subset V_{n-1}\subset\C^n)~|~\dim(V_i) = i\}~.
\end{equation}
In the first subsection, we review the classical and quantum cohomology relations for this manifold from the 2d GLSM perspective. In the second subsection, we consider the 3d (K-theoretic) uplift of the 2d theory, which gives us access to the quantum K-theory of Fl$(n)$. 

\subsection{2d GLSMs and quantum cohomology of complete flag manifolds}
The complete flag manifold Fl$(n)$ can be realized as the target space of a 2d GLSM by considering the 2d $\mathcal{N} = (2,2)$ quiver gauge theory shown in figure~\ref{fig:complete Flag quiver} ---
see~\cite{Witten:1993xi,Donagi:2007hi}. We have the gauge group:
\begin{equation}
    G \,=\, U(1)\,\times\,U(2)\,\times\,\cdots\,\times U(n-1)~. 
\end{equation}
Associated with each $U(i)$ gauge group we have the gauge vector multiplet $\mathcal{V}^{(i)} = (\widetilde{\sigma}^{(i)},\cdots)$, where $\widetilde{\sigma}^{(i)}$ is a complex adjoint scalar whose $i$ eigenvalues we denote by $\widetilde{\sigma}^{(i)}_{a_i}$. Moreover, between the quiver nodes $U(i)$ and $U(i+1)$, we have the 2d chiral multiplet $\Phi_i^{i+1} = (\phi^{i+1}_i,\cdots)$ in the bifundamental representation $(\square_i, \overline{\square}_{i+1})$.

Looking at the $D$-term equations of this theory, one can show that the complete flag manifold Fl$(n)$ is the Higgs branch of the theory, which can be written as the GIT quotient:
\begin{equation}
    {\rm Fl}(n) \,=\, \left(\bigtimes_{i=1}^{n-1} \C^{i(i+1)} \right)\sslash
    G_\C~.
\end{equation}
Here, 
$G_\C  = { GL}(1)\,\times \,\cdots\,\times\,{GL}(n-1)$ is the complexified gauge group. The complex dimension and topological Euler characteristic of this manifold are given by:
\begin{equation}\label{dim and chi}
    \dim({\rm Fl}(n)) \,=\, \frac{n\,(\,n\,-\,1\,)}{2}~, \qquad\quad \chi({\rm Fl}(n)) \,=\, n!~. 
\end{equation}

\medskip
\noindent
\textbf{Universal bundles and quantum cohomology.} Let us denote by $\CS_i$ the tautological vector bundle over Fl$(n)$ whose fiber is the vector space $V_i$ in~\eqref{Fl(n) defn}. The eigenvalues $\widetilde{\sigma}_{a_i}^{(i)}$ introduced above can be interpreted as the Chern roots of $\mathcal{S}_i$. The complete flag manifold also comes equipped with the universal quotient line bundles $\{\mathcal{Q}_1, \cdots, \mathcal{Q}_n\}$ such that:
\begin{equation}
    \mathcal{Q}_i \,\equiv\,\begin{cases}
        \mathcal{S}_1~, \qquad &i=1~, \\
        \mathcal{S}_i/\mathcal{S}_{i-1}~, \qquad &i=2, \cdots, n-1~, \\
        \C^n/\mathcal{S}_{n-1}~, \qquad &i = n~.
    \end{cases}
\end{equation}
Let us denote by $\sigma_i$ the Chern root of $\CQ_i$. Importantly, the line bundles~$\CQ_i$ generate the K-theory of the Fl$(n)$; indeed, they give us what is known as the Toda presentation of the quantum K-theory~\cite{ahkmox} --- see~\cite[section 4.1]{Closset:2026bnk} for a more detailed explanation of various presentations of the quantum K-theory ring.

They are ${n(n-1)/2}$ variables $\t\sigma^{(i)}$ given to us by the quiver gauge theory, while the K-theory or cohomology of the complete flag can be expressed in terms of the smaller set of $n$ variables $\sigma_i$. 
At the level of classical cohomology, the Chern roots of the two classes of universal bundles are related to each other via the exact sequence:
\begin{equation}\label{SES-i}
    0 \,\longrightarrow\, \mathcal{S}_{i}\, \longrightarrow\, \mathcal{S}_{i+1}\,\longrightarrow\, \mathcal{Q}_{i+1}\, \longrightarrow\, 0~.
\end{equation}
This implies that the total Chern classes of the involved bundles are related as:
\begin{equation}\label{CoH Fl(n) I}
    c(\mathcal{S}_i)\,\wedge\,c(\mathcal{Q}_{i+1})\,=\,c(\mathcal{S}_{i+1})~.
\end{equation}
Expanding both sides, one obtains:
\begin{equation}\label{CoH Fl(n) II}
    e_{b}(\widetilde{\sigma}^{(i+1)})\,=\, e_b(\widetilde{\sigma}^{(i)})\,+\,\sigma_{i+1}\,e_{b-1}(\widetilde{\sigma}^{(i)})~,\qquad b = 1, \cdots, i+1~,
\end{equation}
with the convention that $e_0(\widetilde{\sigma}^{(i)}) = 1$ for all $i$ and $e_b(\widetilde{\sigma}^{(i)}) = 0$ whenever $b>i$ and $b<0$.
In particular,
\begin{equation}\label{sigma1=sigma11}
    \sigma_1 = \widetilde{\sigma}_1^{(1)}~.
\end{equation}
For example, for Fl$(3)$, we have \eqref{sigma1=sigma11} together with:
\begin{equation}\label{CoH Fl(3)}
\begin{split}
        &e_1(\widetilde{\sigma}^{(2)})\,=\,e_1^2(\sigma)~,\qquad e_2(\widetilde{\sigma}^{(2)})\,=\,e_2^2(\sigma)~.
\end{split}
\end{equation}
Here, we are using the notation, $e_i^j(\sigma) \equiv e_i(\sigma_1, \cdots, \sigma_j)$. For Fl$(3)$, we have three variables $\t\sigma^{(i)}$ and three variables $\sigma_i$.

\medskip
\noindent
In the quantum cohomology, the classical relations \eqref{CoH Fl(n) I} get deformed as follows~\cite{Gu:2023tcv,Gu:2023fpw}:
\begin{equation}\label{QCoH Fl(n) I}
    c(\mathcal{S}_i)\,\wedge_\qcoh\,c(\mathcal{Q}_{i+1})\,=\,c(\mathcal{S}_{i+1})\,-\,\qcoh_i\,c(\mathcal{S}_{i-1})~.
\end{equation}
 More explicitly, one finds:
\begin{equation}\label{QCoH Fl(n) II}
    e_{b}(\widetilde{\sigma}^{(i+1)})\,=\, e_b(\widetilde{\sigma}^{(i)})\,+\,\sigma_{i+1}\,e_{b-1}(\widetilde{\sigma}^{(i)})\,+\,\qcoh_i\,e_{b-2}(\widetilde{\sigma}^{(i-1)})~,\qquad b = 1, \cdots, i+1~.
\end{equation}
From the 2d GLSM perspective, the quantum parameters $\qcoh_i$ are the exponentiated complexified FI parameters. Observe that the relation \eqref{sigma1=sigma11} is still valid at the quantum level; meanwhile, the higher-order relations are more involved. For example, for ${\rm QH}^{\bullet}$(Fl$(3)$) the classical relations \eqref{CoH Fl(3)} become:
\begin{equation}\label{QCoH Fl(3)}
\begin{split}
        &e_1(\widetilde{\sigma}^{(2)})\,=\,e_1^2(\sigma)~,\qquad e_2(\widetilde{\sigma}^{(2)})\,=\,e_2^2(\sigma) \,+\,\qcoh_1~.
\end{split}
\end{equation}

\medskip
\noindent
\textbf{Another example: relations in ${\rm QH}^\bullet({\rm Fl}(4))$.} Taking $n=4$, we find that, in addition to the relations \eqref{sigma1=sigma11} and \eqref{QCoH Fl(3)}, the quantum relations for Fl$(4)$ include:
\begin{equation}\label{QCoH Fl(4)}
    \begin{split}
        &e_1(\widetilde{\sigma}^{(3)})\,=\,e_1^3(\sigma)~,\\
        &e_2(\widetilde{\sigma}^{(3)})\,=\,\qcoh_1\,+\,\qcoh_2\,+\,e_2^3(\sigma)~,\\
        &e_3(\widetilde{\sigma}^{(3)})\,=\,\,e_3^3(\sigma)\,+\,\qcoh_2\,\sigma_1\,+\,\qcoh_1\,\sigma_3~.
    \end{split}
\end{equation}
This expresses the $6$ variables $\t\sigma^{(i)}$ in terms of the $4$ variables $\sigma_i$.

\subsection{3d GLSMs and quantum K-theory of complete flag manifolds}\label{subsec:3dGLSMQK}
Consider the 3d $\mathcal{N}=2$ uplift of the 2d GLSM considered in the previous subsection, where we take the 3d spacetime to be $\R^2\times S^1_\beta$. We will fix the gauge Chern--Simons (CS) levels to be such that:\footnote{Here we are using the conventions reviewed in subsection 2.2 of \cite{Closset:2023vos} for the CS levels.}
\begin{equation}
    \kk_i\,=\, 0~, \qquad \l_i \,=\,-1~, 
\end{equation}
where $\kk_i$ and $\l_i$ are the Chern--Simons levels of the 3d gauge group $U(i)_{\kk_i, \kk_i + i\,\l_i}$ for $i = 1, \cdots, n-1$. Moreover, there is a mixed  CS level $\kk_{i,i{+}1}=\frac{1}{2}$ between every two consecutive gauge groups $U(i)$ and $U(i+1)$. 

With this choice of CS levels, the 3d twisted chiral ring is known to be isomorphic to the quantum K-theory ring of the complete flag manifold. One way to see this is by looking at the Bethe ansatz equations (BAEs) coming from the 3d A-model subsector of the full theory. Motivated by BAE computations, the quantum K-theory relations can be compactly written in the following form \cite{Gu:2023tcv,Gu:2023fpw}:
\begin{equation}\label{QK lambda relations}
    \lambda_t(\mathcal{S}_i)\,\otimes_\qk\,\lambda_t(\mathcal{Q}_{i+1}) = \lambda_t(\mathcal{S}_{i+1})\,-\,t \,\frac{\qk_i}{1-\qk_i}\,\det(\mathcal{Q}_{i+1})\,\otimes_q\,(\lambda_t(\mathcal{S}_{i})-\lambda_t(\mathcal{S}_{i-1}))~,
\end{equation}
for $i=1, \cdots, n-1$, where $\otimes_q$ is the product operation of the quantum K-theory ring. Here, for any vector bundle $E$,
\begin{equation}
       \lambda_t(E) \equiv 1 \,+ \, t\,[E]\, + \,\cdots\, + \, t^{\rk (E)}\,[\wedge^{\rk (E)}E] ~,
\end{equation}
with $t$ a formal parameter. 
Taking $\qk_i = 0$ in the above equations, one obtains the classical K-theory ring relations of the complete flag manifold.
This presentation can also be related to the Toda presentation mentioned earlier, see
\cite[theorems 5.7, 5.11]{ahkmox}.

In the field-theory language, the K-theoretic Chern roots of the tautological bundle $\mathcal{S}_i$ are given by $x^{(i)}_{a_i} \sim e^{- \,\beta\, \widetilde{\sigma}^{(i)}_{a_i}}$ where $\widetilde{\sigma}^{(i)}$ is the 3d real adjoint scalar in the $U(i)$ 3d $\mathcal{N}=2$ vector multiplet. Moreover, the quantum parameters $\qk_i\sim e^{-\,\beta\,\zeta_i}$ are defined in terms of the 3d real FI parameters $\zeta_i$. With this in mind, and in terms of the corresponding Chern characters, the quantum relations \eqref{QK lambda relations} can be written more explicitly as:
\begin{equation}\label{QK relations of flag n}
    e_a(x^{(i+1)})\, =\, e_a(x^{(i)})\, +\,  \frac{1}{1\,-\,q_i}\,x_{i+1}\, \cdot\, e_{a-1}(x^{(i)}) \,-\, \frac{q_i}{1\,-\,q_i}\, x_{i+1} \,\cdot\, e_{a-1}(x^{(i-1)})~,
\end{equation}
for $a = 1, \cdots, i+1$. Here $x_i \sim e^{-\,\beta\,\sigma_i}$ is the K-theoretic Chern root of the quotient line bundle $\mathcal{Q}_i$. In the above equation~\eqref{QK relations of flag n}, we have adopted the convention that $q_0 = 0$, $e_0(x^{(i)}) = 1$ and $e_a(x^{(i)}) = 0$ when $a>i$ or $a<0$. In particular, looking at $i=0$, the above relations become:
\begin{equation}\label{x1=x11}
    x_1 \,=\,x_1^{(1)}~.
\end{equation}

\medskip
\noindent
\textbf{Example I: QK relations for Fl$(3)$.} For the case $n=3$, in addition to \eqref{x1=x11}, we find the following relations for the quantum K-theory ring of Fl$(3)$:
\begin{equation}\label{QK for Fl(3)}
    e_1(x^{(2)})\,=\,e_1^2(x)~, \qquad e_2(x^{(2)})\,=\,\frac{1}{1-\qk_1}\,e_2^2(x)~,
\end{equation}
where, $e_i^j(x)\equiv e_i(x_1, \cdots, x_j)$ are as defined in the previous subsection.

\medskip
\noindent
{\bf Example II: QK relations for Fl$(4)$.} In addition to \eqref{x1=x11} and \eqref{QK for Fl(3)}, we find the following relations for the quantum K-theory of Fl$(4)$:
\begin{equation}\label{QK for Fl(4)}
    \begin{split}
        &e_1(x^{(3)})\,=\,e_1^3(x)~,\\
        &e_2(x^{(3)})\,=\,\frac{1}{1-\qk_1}\,x_1\,x_2\,+\,\frac{1}{1-q_2}\,x_2\,x_3\,+\,x_1\,x_3~,\\
        &e_3(x^{(3)})\,=\,\frac{1}{1-\qk_1}\,\frac{1}{1-\qk_2}\,e_3^3(x)~.
    \end{split}
\end{equation}

\medskip
\noindent
\textbf{2d limit: back to quantum cohomology.} In the limit of small circle radius $\beta$, the quantum K-theory relations \eqref{QK lambda relations} reduce to the quantum cohomology relations \eqref{QCoH Fl(n) I}, keeping in mind that the quantum parameters are related by:
\begin{equation}
    \qk_i\,=\,\beta^2\,\qcoh_i~, \qquad i = 1, \cdots, n-1~.
\end{equation}
For example, the relations \eqref{QK for Fl(3)} reduce to \eqref{QCoH Fl(3)}, and \eqref{QK for Fl(4)} reduce to \eqref{QCoH Fl(4)} in this limit.


\section{Schubert line defects for complete flag manifolds}\label{sec:schubert line defects}
In this section, we first review the definition of the Schubert varieties in the complete flag manifold. We then construct and study the corresponding Schubert line defects in the 3d GLSM.

\subsection{Schubert varieties in complete flag manifolds}\label{sec:schubert in complete flag}

The Schubert varieties of the complete flag manifold Fl$(n)$~\eqref{Fl(n) defn} are indexed by the elements of the symmetric group $S_n$ --- that is, by permutations. There are thus $n!$ distinct Schubert varieties in Fl$(n)$, matching the topological Euler characteristic~\eqref{dim and chi}. Note that Schubert varieties can be defined in terms of fixed points of the $SL(n)$ action on Fl$(n)$ ---- we refer to~\cite{Anderson_Fulton_2023} for a detailed account. 

For a given permutation $w\in S_n$, we define the corresponding Schubert variety $X_w$ as follows~\cite[page 153]{Anderson_Fulton_2023}:
\begin{equation}\label{Xw defn}
     X_w \,:=\,  \left\{ \,F_{\bullet} \in {\rm Fl}(n) \, | \,
    \dim( F_{i} \cap E_j ) \geq {\rm r}^{w_0w}_{i, j}~,\quad  \forall \, 1\leq i, j\leq n
   \, \right\}~.
\end{equation}
Here, $w_0 = (n\,\cdots\,2\,1)$ is the longest permutation in $S_n$, and:
\begin{equation}\label{w0w}
    (w_0w)(i)\,=\, n\, +\, 1\, -\, w(i)~.
\end{equation}
Moreover, $E_\bullet$ in~\eqref{Xw defn} denotes the reference flag \cite[page 171]{Anderson_Fulton_2023}:
\begin{equation}\label{E flag}
  E_{\bullet} \: := \: \left( E_1\, \subset\, E_2 \,\subset\, \cdots \,\subset\, E_n \,\equiv\, {\mathbb C}^n \right)~,
  \end{equation}
with $E_i \;:=\; {\rm Span}_\C\{{\rm e}_{n+1-i}, \cdots, {\rm e}_n\}$. Here, $\{{\rm e}_{1}, \cdots, {\rm e}_n\}$ is the standard basis of $\C^n$. Finally, the rank matrix ${\rm r}^w$ associated with a permutation $w$ is defined as:
\begin{equation}\label{rw defn}
    {\rm r}^w_{i,j} \,:=\,\#\{\,l\,\leq\,i\,:\,w(l)\,\leq\,j\,\}~.
\end{equation}
The Schubert variety $X_w$ is the closure of the Schubert cell $X_w^\circ$, which is defined as in~\eqref{Xw defn} with equalities instead of inequalities. Note that the Schubert variety associated with the identity permutation $(\,1\, 2 \, \cdots n\,)$ is $X={\rm Fl}(n)$ itself, while the Schubert variety associated with the longest permutation $w_0$ is the point class. More generally, the codimension of a Schubert variety $X_w$ is given by the length of $w$:
\be
{\rm codim}(X_w) \,=\, \ell(w)~.
\ee
See~\eqref{def ellw} for the definition of $\ell(w)$. Inclusion relations of Schubert varieties follow the Bruhat order on $S_n$; in particular, $X_w \subset X_{w'}$ implies that $\ell(w)>\ell(w')$.


\medskip
\noindent
\textbf{Schubert classes.}  The cohomological or K-theoretic Schubert classes associated with the Schubert varieties form a basis for the cohomology or K-theory of Fl$(n)$. We refer to appendix~\ref{app:Schubert classes} for more details on the known presentations of the Schubert classes as polynomials. Briefly, in cohomology, the Schubert classes $[X_w]$ associated with the Schubert varieties form a basis of the cohomology ring. In (equivariant) classical cohomology H$^\bullet_T({\rm Fl}(n))$, these are represented by the  (double) Schubert polynomials $\mathfrak{S}_w$ \eqref{double-Schubert-defn}, and by quantum (double) Schubert polynomials $\mathfrak{S}_w^{(\qcoh)}$ \eqref{quantum-schub-poly-defn} in the (equivariant) quantum cohomology ring QH$_T^\bullet({\rm Fl}(n))$. See appendix~\ref{subsec: Schubert polys} for more details and explicit examples.

Meanwhile, in the K-theory, the Schubert classes  $[\CO_w]$ are the equivalence classes of the structure sheaves on the Schubert varieties. In classical (equivariant) K-theory K$_T({\rm Fl}(n))$, these are represented by the (double) Grothendieck polynomials $\mathfrak{G}_w$ \eqref{double-groth-polynomial-defn}, which are upgraded to quantum (double) Grothendieck polynomials $\mathfrak{G}^{(\qk)}_w$ \eqref{q-double-groth-polys-defn} in the (equivariant) quantum K-theory QK$_T({\rm Fl}(n))$. See appendix~\ref{subsec: Grothendeick polys} for more details and examples.

\subsection{Schubert line defects in the 3d GLSM}\label{sect:proposal}
Let us now describe a systematic construction of line defects in the 3d GLSM of the complete flag manifold that flow to the Schubert classes in the quantum K-theory. We call these defects the Schubert line defects. By a reasoning similar to the one in~\cite{Closset:2023bdr,Gu:2025tda}, we will show that the insertion of these line defects restricts the target geometry of the 3d GLSM at the point of insertion to the corresponding Schubert varieties.

The Schubert line defects are realized by coupling a $\mathcal{N}=2$ SQM quiver to the 3d GLSM described in section~\ref{sec:GLSM for flag}. In the following, we will first directly present the proposal, and then study its vacuum equations to show that the target geometry is indeed restricted to the Schubert varieties. We will actually present a stronger result, namely that the 1d-3d system provides a resolution of the (generally singular) Schubert variety. Furthermore, we will study the flavored Witten index of the 1d $\CN=2$ quiver. In all the examples we have checked, the index precisely agrees with the corresponding quantum Grothendieck polynomial once we use the quantum K-theory relations~\eqref{QK relations of flag n} to express them as polynomials in the $x_i$ variables.

\subsubsection{General proposal}\label{subsec:general proposal}

\begin{figure}[t]
    \centering
    \includegraphics[width=0.7\linewidth]{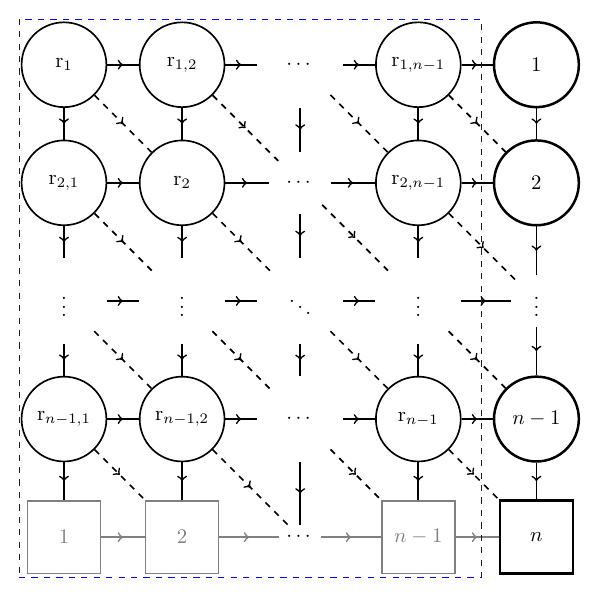}
    \caption{Quiver diagram of the 1d-3d coupled system realizing $X_w\subseteq {\rm Fl}(n)$. The last column represents the 3d GLSM to the complete flag manifold Fl$(n)$. The square quiver in the blue box is the 1d quiver gauge theory defining the defect. The circle nodes represent 1d gauge groups of the indicated ranks --- for compactness, we write ${\rm r}_{i,j}\equiv {\rm r}_{i,j}^{w_0w}$, and ${\rm r}_i \equiv {\rm r}_{i,i}$.  Horizontal and vertical arrows stand for 1d chiral matter multiplets in the bifundamental representation of the corresponding nodes. Moreover, diagonal arrows correspond to 1d Fermi multiplets in the bifundamental representation of the corresponding nodes. The gray squares in the last row represent fixed background data for the inclusion maps defined in~\protect\eqref{eqn:background field}.}
    \label{fig:general proposal}
\end{figure}

For the Schubert variety $X_w$ defined in equation \eqref{Xw defn}, we construct the corresponding Schubert defects by coupling a 1d $\mathcal{N}=2$ SQM to the 3d GLSM described in section~\ref{sec:GLSM for flag}. The resulting 1d-3d coupled theory can be realized as the quiver diagram displayed in figure~\ref{fig:complete Flag quiver}. The last column alone gives the 3d GLSM for the complete flag manifold, and we use a blue box to encircle the 1d theory of the Schubert defect.

In the quiver, circle nodes denote unitary gauge groups of the indicated ranks, while the square nodes denote unitary flavor groups. Solid arrows represent chiral multiplets, and dashed arrows correspond to Fermi multiplets. Both types of multiplets live in the bifundamental representations of the corresponding groups. We adopt the convention that each multiplet lives in the fundamental representation under the group of the source node, and the anti-fundamental representation under the group of the target node.

More specifically, the circle node at position $(i,j)$ (that is, the $i$-th row and the $j$-th column) for $i,j=1,2,\cdots,n-1$, corresponds to a 1d gauge group factor $U({\rm r}_{i,j})$, where $\text{r}_{i,j}$ is a short hand notation for $\text{r}_{i,j}^{w_0 w}$ defined as in~\eqref{rw defn}, namely:
\be\label{rij def}
\text{r}_{i,j} \equiv  \#\{l \le i \mid n+1-w(l) \leq j\}~.
\ee
We use the following notational conventions for the 1d-3d quiver theory. Bulk 3d chiral multiplets are denoted by $\phi$, while defect chiral multiplets are denoted by $\varphi$, and defect Fermi multiplets by $\Gamma$. For bulk chiral multiplets, we use the notation $\phi_\ell^{\ell+1}$ to indicate the bifundamental between the gauge groups $U(\ell)$ and $U(\ell+1)$ in the last column of the quiver. For defect fields, superscripts and subscripts are used to indicate the direction of the corresponding arrow: a field $\varphi_{i,j}^{k,l}$ transforms in the fundamental representation of the unitary group at the source node $(i,j)$ and in the anti-fundamental of the group at the target node $(k,l)$, and similarly for the Fermi multiplets $\Gamma_{i,j}^{k,l}$.

Therefore, for $i,j=1,2,\cdots, n-1$, we have chiral multiplets $\varphi_{i,j}^{i,j+1}$ corresponding to horizontal arrows, and $\varphi_{i,j}^{i+1,j}$ corresponding to vertical arrows. We also have Fermi multiplets $\Gamma_{i,j}^{i+1,j+1}$ corresponding to the diagonal arrows in each block. 

The last row in the quiver of figure~\ref{fig:general proposal} needs special treatment. Instead of the product structure $U(1) \times U(2) \times \cdots \times U(n)$ that is common for quiver gauge theories, what we really mean here is that we chose a flavor symmetry group structure
\be
U(1) \subset U(2) \subset \cdots \subset U(n)~,
\ee
which corresponds to a choice of reference flag $E_\bullet$. 
Hence, the horizontal arrows in the last row are not dynamical fields, but represent fixed background data. For the arrow between flavor $U(i)$ and $U(i+1)$, we will denote the background field by $\iota^{(i)}$, and it has the following form as an $i\times (i+1)$ matrix:
\begin{equation}\label{eqn:background field}
    \iota^{(i)} \,=\, \begin{pmatrix}
        \mathbb{I}_{i\times i} & 0_{i\times 1}
    \end{pmatrix}~.
\end{equation}
In the quiver diagram, we indicate these background arrows in gray.

To describe the interaction terms, we need to specify the $E$-term and $J$-term for each Fermi multiplet $\Gamma_{i,j}^{i+1,j+1}$. In our construction, the $J$-terms are set to be zero, while the $E$-terms are given as follows:%
\footnote{
This can also be formally encoded in a gauge-invariant superpotential (see \protect\cite[equation (2.12)]{Closset:2025zyl})
\protect\begin{equation}
 W= \text{Tr}\,\left(\,\overline{\Gamma}_{i,j}^{i+1,j+1}\, (\,\varphi_{i,j}^{i,j+1}\, \cdot\, \varphi_{i,j+1}^{i+1,j+1}\, -\, \varphi_{i,j}^{i+1,j} \,\cdot\, \varphi_{i+1,j}^{i+1,j+1})\,\right)~.
\end{equation}
It should be noted, however, that this expression is solely formal. It is used in the mathematical description of graded quivers with potential to encode relations in a convenient manner --- see {\it e.g.}~\cite{Franco:2017lpa, Closset:2018axq}.
}%
\begin{equation}\label{eqn:E-term}   
    \overline{D}_+ \,\Gamma_{i,j}^{i+1,j+1}
    \, = \,
    E_{i,j}^{i+1,j+1}
    \, = \,
    \varphi_{i,j}^{i,j+1}\, \cdot\, \varphi_{i,j+1}^{i+1,j+1}\, -\, \varphi_{i,j}^{i+1,j}\, \cdot\, \varphi_{i+1,j}^{i+1,j+1}~,
\end{equation}
where $\cdot$ denotes matrix multiplication, and $\varphi_{n,j}^{n,j+1} \equiv \iota^{(j)}$, $\varphi_{i,n}^{i+1,n} \equiv \phi_i^{i+1}$ are understood.

\subsubsection{Vacuum equations}
To uncover the underlying geometry, let us first study the vacuum equations. We choose a phase of the $\CN=2$ SQM where all the 1d FI parameters are taken to be positive. Let us denote the 1d FI parameter associated with $U(1)\subseteq U({\rm r}_{i,j})$ by $\xi_{i,j}$.

\medskip
\noindent \textbf{$D$-term equations.}
The $D$-term equation for the gauge group $U(\text{r}_{i,j})$ is given by:
\begin{equation}
    \varphi_{i,j}^{i+1,j} \cdot (\varphi_{i,j}^{i+1,j})^\dagger 
    + \varphi_{i,j}^{i,j+1} \cdot (\varphi_{i,j}^{i,j+1})^\dagger 
    - (\varphi_{i-1,j}^{i,j})^\dagger \cdot \varphi_{i-1,j}^{i,j}
    - (\varphi_{i,j-1}^{i,j})^\dagger \cdot \varphi_{i,j-1}^{i,j} = \xi_{i,j} \,\mathbb{I}_{\text{r}_{i,j}}~.
\end{equation}
Note that, for certain edge cases, some of these fields may be absent. The positive terms on the LHS correspond to chiral multiplets transforming in the fundamental representation of the gauge group, while the negative terms correspond to those transforming in the anti-fundamental representation.

Each term of the schematic form $AA^\dagger$ or $A^\dagger A$ is a Hermitian matrix that is positive semi-definite, while $\xi_{i,j} \mathbb{I}_{\text{r}_{i,j}}$ is positive definite for $\xi_{i,j}>0$. Hence, we can conclude that:
\begin{equation} \label{eqn:reduced D-term condition}
    \varphi_{i,j}^{i+1,j}\, \cdot\, (\varphi_{i,j}^{i+1,j})^\dagger 
    \,+\, \varphi_{i,j}^{i,j+1} \,\cdot\, (\varphi_{i,j}^{i,j+1})^\dagger
   \, \in \,\mathbb{H}^+_{\,\text{r}_{i,j}}~.
\end{equation}
Here, we denote the set of positive definite Hermitian matrices of size $\text{r}_{i,j}\times \text{r}_{i,j}$ by $\mathbb{H}^+_{\,\text{r}_{i,j}}$.
If one of $\varphi_{i,j}^{i+1,j}$ and $\varphi_{i,j}^{i,j+1}$ is absent, we can conclude that the remaining one corresponds to a matrix of full rank. But, when both of them are present (which is our current case for defect gauge groups), we cannot conclude that both of them correspond to full rank matrices. We have to resort to the $E$-term equations.

\medskip
\noindent \textbf{$E$-term equations.}
The supersymmetric ground state conditions require the vanishing of the $E$-terms.
For the $E$ term given in equation \eqref{eqn:E-term}, we then have
\begin{equation}\label{eqn:E-term equation}
    \varphi_{i,j}^{i,j+1}\, \cdot\, \varphi_{i,j+1}^{i+1,j+1}\,-\,\varphi_{i,j}^{i+1,j} \,\cdot \,\varphi_{i+1,j}^{i+1,j+1} \,=\, 0~.
\end{equation}

\medskip
\noindent \textbf{Rank constraints from the vacuum equations.}
Combining the $D$-term equations \eqref{eqn:reduced D-term condition} with the $E$-term equations \eqref{eqn:E-term equation}, we see that, if $\varphi_{i,j+1}^{i+1,j+1}$ is a matrix of full rank, then so is $\varphi_{i,j}^{i+1,j}$; similarly if $\varphi_{i+1,j}^{i+1,j+1}$ is a matrix of full rank, then so is $\varphi_{i,j}^{i,j+1}$. To make the notation more intuitive, one can think in terms of a square in the quiver: combining the $D$-term equation for the top-left gauge group, and the $E$-term equation for the diagonal Fermi, we see that, if the bottom arrow corresponds to a full rank matrix, so does the top arrow; if the right arrow corresponds to a full rank matrix, so does the left arrow.

In what follows, we prove the first statement, and the second statement can be verified using an analogous argument. Indeed, suppose $\varphi_{i,j+1}^{i+1,j+1}$ corresponds to a matrix of full rank, then we have:
\begin{equation}
    A\,\equiv\, \varphi_{i,j+1}^{i+1,j+1}\,\cdot\, (\varphi_{i,j+1}^{i+1,j+1})^\dagger\, \in \,\mathbb{H}_{\,\text{r}_{i,j+1}}^+~,
\end{equation}
and $A$ is invertible. Multiplying both sides of the $E$-term equation \eqref{eqn:E-term equation} on the right by $(\varphi_{i,j+1}^{i+1,j+1})^\dagger \cdot A^{-1}$, we can solve for $\varphi_{i,j}^{i,j+1}$ to obtain:
\begin{equation}
    \varphi_{i,j}^{i,j+1} \,=\, \varphi_{i,j}^{i+1,j} \,\cdot\, \varphi_{i+1,j}^{i+1,j+1} \,\cdot\, (\varphi_{i,j+1}^{i+1,j+1})^\dagger \,\cdot\, A^{-1} \,\equiv\, \varphi_{i,j}^{i+1,j}\,\cdot\, B~.
\end{equation}
Plugging this back into equation \eqref{eqn:reduced D-term condition}, we have:
\begin{equation} \label{eqn:D-term E-term combined}
    \varphi_{i,j}^{i+1,j}\, \cdot\, (\,\mathbb{I}_{\text{r}_{i+1,j}}\, +\, B\, \cdot \,B^\dagger\,)\,\cdot\,(\varphi_{i,j}^{i+1,j})^\dagger\, \in \,\mathbb{H}^+_{\,\text{r}_{i,j}}~.
\end{equation}
Hence $\varphi_{i,j}^{i+1,j}$ is an $\text{r}_{i,j}\times \text{r}_{i+1,j}$ matrix of full rank. Otherwise there would be a non-zero vector $v$ of size $1 \times \text{r}_{i,j}$, such that $v \cdot \varphi_{i,j}^{i+1,j} = 0$, contradicting equation \eqref{eqn:D-term E-term combined}. This concludes our proof.

With this result in hand, we can show by induction that all the chiral multiplets in the quiver correspond to full rank matrices.
For our quiver in figure~\ref{fig:general proposal}, the gray arrows in the last row correspond to full rank matrices by construction, as given in equation \eqref{eqn:background field}, while the arrows in the last column correspond to the bulk fields and we can use the bulk $D$-term equations to show that they all correspond to full rank matrices (since each bulk gauge group only has one set of chiral multiplets in the fundamental representation). These provide the base cases for the induction.
Applying the argument above to each square of the quiver then propagates the full rank condition inward, and we conclude that all chiral multiplets in the quiver correspond to full rank matrices.

\subsubsection{The underlying geometry}
To understand why the bulk geometry gets restricted to the corresponding Schubert varieties, there are two complementary perspectives. First, one can analyze the geometry of the full 1d-3d coupled system and then project onto the bulk direction. Second, one can instead focus directly on the bulk fields and study how the defect restricts the bulk fields through the $E$-term relations. We now describe both viewpoints.

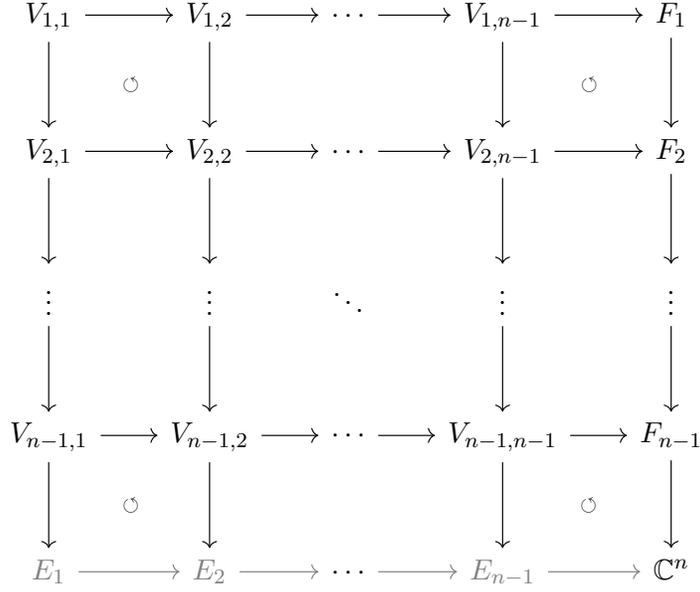
\begin{figure}[t]
\centering
\begin{tikzcd}[column sep=2em, row sep=3em]
    {V_{1,1}} & {V_{1,2}} & \cdots & {V_{1,n-1}} & {F_1} \\
    {V_{2,1}} & {V_{2,2}} & \cdots & {V_{2,n-1}} & {F_2} \\
    \vdots & \vdots & \ddots & \vdots & \vdots \\
    {V_{n-1,1}} & {V_{n-1,2}} & \cdots & {V_{n-1,n-1}} & {F_{n-1}} \\
    {\textcolor{gray}{E_1}} & \textcolor{gray}{E_2} & \cdots & \textcolor{gray}{E_{n-1}} & {\mathbb{C}^n}
    \arrow[from=1-1, to=1-2]
    \arrow[from=1-1, to=2-1]
    \arrow["\circlearrowleft"{description}, draw=none, from=1-1, to=2-2]
    \arrow[from=1-2, to=1-3]
    \arrow[from=1-2, to=2-2]
    \arrow[from=1-3, to=1-4]
    \arrow[from=1-4, to=1-5]
    \arrow[from=1-4, to=2-4]
    \arrow["\circlearrowleft"{description}, draw=none, from=1-4, to=2-5]
    \arrow[from=1-5, to=2-5]
    \arrow[from=2-1, to=2-2]
    \arrow[from=2-1, to=3-1]
    \arrow[from=2-2, to=2-3]
    \arrow[from=2-2, to=3-2]
    \arrow[from=2-3, to=2-4]
    \arrow[from=2-4, to=2-5]
    \arrow[from=2-4, to=3-4]
    \arrow[from=2-5, to=3-5]
    \arrow[from=3-1, to=4-1]
    \arrow[from=3-2, to=4-2]
    \arrow[from=3-4, to=4-4]
    \arrow[from=3-5, to=4-5]
    \arrow[from=4-1, to=4-2]
    \arrow[from=4-1, to=5-1]
    \arrow["\circlearrowleft"{description}, draw=none, from=4-1, to=5-2]
    \arrow[from=4-2, to=4-3]
    \arrow[from=4-2, to=5-2]
    \arrow[from=4-3, to=4-4]
    \arrow[from=4-4, to=4-5]
    \arrow[from=4-4, to=5-4]
    \arrow["\circlearrowleft"{description}, draw=none, from=4-4, to=5-5]
    \arrow[from=4-5, to=5-5]
    \arrow[from=5-1, to=5-2, gray]
    \arrow[from=5-2, to=5-3, gray]
    \arrow[from=5-3, to=5-4, gray]
    \arrow[from=5-4, to=5-5, gray]
\end{tikzcd}

\caption{Diagram of nested vector spaces in $\mathbb{C}^n$, with a fixed reference flag $E_\bullet$ shown in gray. The arrows indicate the inclusions of subspaces, while the circular arrows indicate that the squares commute. Here, $\dim(V_{i,j}) = \text{r}_{i,j}$ and $\dim(F_i) = i$.}
\label{fig:quiver geometry}
\end{figure}

\medskip
\noindent \textbf{Geometry of the coupled 1d-3d system.} We have already shown that all the chiral multiplets correspond to full rank matrices. Thus, they can be interpreted as inclusions of vector spaces. For example, the chiral field $\varphi_{i,j}^{i,j+1}$ can be viewed as an $\text{r}_{i,j} \times \text{r}_{i,j+1}$ matrix of full rank, describing the inclusion of an $\text{r}_{i,j}$-dimensional vector space into an $\text{r}_{i,j+1}$-dimensional vector space, with the gauge symmetries acting by rotations of the basis in the corresponding vector spaces. The $E$-term equations can be understood as commutativity relations, which ensure that these inclusions are compatible: the composition of maps along different paths connecting the same nodes gives the same subspace in the target vector space. Finally, the chiral multiplets in the last row are not dynamical, and they describe a fixed complete flag $E_1 \subset E_2 \subset \cdots \subset E_n$. Therefore, the full collection of chiral multiplets and their relations describes all possible consistent configurations of nested subspaces of the form shown in figure~\ref{fig:quiver geometry}, where $\dim(V_{i,j}) = \text{r}_{i,j}$ and $\dim(F_i) = i$.

In passing, this setup can be seen as a generalization of Grassmannian $\text{Gr}(k,n)$, which describes all possible $k$-dimensional subspaces of an $n$-dimensional vector space. Mathematically, this geometry of our 1d-3d system corresponds to a particular quiver Grassmannian \cite{iezzi2025quivergrassmanniansbottsamelsonresolution} or, equivalently, to bioriented flags \cite{cibotaru2020bioriented}. Both of these constructions have been proven to be isomorphic to Bott--Samelson resolutions \cite{BottSamelsonResolution} of Schubert varieties in Fl$(n)$ (for a particular presentation of each permutation) --- we discuss these mathematical constructions in more detail in appendix~\ref{app:quiver-vs-math}.
The 1d-3d target geometry $\widetilde{X}_w$ therefore serves as a resolution of the singular Schubert variety $X_w$. The projection map 
\begin{equation}\label{res Xw}
    \pi\,:\, \widetilde{X}_w\, \to\, X_w~,
\end{equation}
sends each configuration in figure~\ref{fig:quiver geometry} to its embedding into the last column, which corresponds to the projection onto the Schubert variety $X_w\subseteq X$ inside the bulk geometry.

Indeed, projecting to the bulk geometry, one can see that the complete flag $F_1 \subset F_2 \subset \cdots \subset F_{n-1} \subset \mathbb{C}^n$ satisfies the incidence relations defining Schubert varieties as in equation~\eqref{Xw defn}:
\begin{equation}
    \dim(F_i \,\cap\, E_j) \,\geq\, \dim(V_{i,j}\, \cap\, E_j) \,=\, \dim(V_{i,j}) \,=\, \text{r}_{i,j}~,
\end{equation}
where in the first step, we used the fact that $V_{i,j} \subset F_i$, and in the second step, we used the fact that $V_{i,j} \subset E_j$, as shown in figure~\ref{fig:quiver geometry}. 
For this check, $i$ and $j$ range from $1$ to $n-1$. Note that the condition $\dim(F_i \cap E_j) \geq {\rm r}_{i,j}$ is automatically satisfied whenever $i=n$ or $j=n$.
Therefore, the bulk geometry is given by the corresponding Schubert variety $X_w$.

\medskip
\noindent \textbf{Bulk geometry via $E$-term constraints.} 
One can also study the bulk geometry directly by considering all the constraints on the bulk fields $\phi_\ell^{\ell+1}$ arising from the $E$-term equations. Individually, the $E$-term equations~\eqref{eqn:E-term equation} do not impose constraints on the bulk fields. We have to combine them so that both the fixed background $\iota^{(i)}$ and the bulk fields $\phi_\ell^{\ell+1}$ are involved simultaneously. 

To that end, consider two paths in the quiver connecting the $U(\text{r}_{i,j})$ node and the flavor $U(n)$ node. The first path runs horizontally and then vertically, leading to a composed field:
\begin{equation}
\varphi_{i,j}^{i,j+1}\, \cdot\, \varphi_{i,j+1}^{i,j+2}  \,\cdot\,\cdots \,\cdot\, \varphi_{i,n-1}^{i,n}\, \cdot \,\phi_i^{i+1} \,\cdot\, \phi_{i+1}^{i+2} \,\cdot\,\cdots  \,\cdot\,\phi_{n-1}^n~.
\end{equation}
The second path runs vertically and then horizontally, leading to another composed field:
\begin{equation}\label{eqn:second-composed-field}
    \varphi_{i,j}^{i+1,j}\, \cdot \varphi_{i+1,j}^{i+2,j} \,\cdot\, \cdots \,\cdot\, \varphi_{n-1,j}^{n,j}\,\cdot \,\iota^{(j)} \,\cdot\, \iota^{(j+1)} \,\cdot\, \cdots  \,\cdot\,\iota^{(n-1)}~,
\end{equation}
where the background fields $\iota$'s are given in equation~\eqref{eqn:background field}. Commutativity of the quiver implies that these two compositions must coincide; equivalently, one can obtain their equality by repeatedly applying the $E$-term equations.

Now, due to the special form of the fixed background $\iota$'s, we have:
\begin{equation}
    \varphi_{i,j}^{i,j+1}\, \cdot\, \varphi_{i,j+1}^{i,j+2} \,\cdot\, \cdots \,\cdot\, \varphi_{i,n-1}^{i,n}\, \cdot \,\phi_i^{i+1} \,\cdot\, \phi_{i+1}^{i+2} \,\cdot\,\cdots \,\cdot\, \phi_{n-1}^n\, = \,\begin{pmatrix}
        *_{j\times j} & 0_{j\times (n-j)}
    \end{pmatrix}~,
\end{equation}
where the block form on the RHS arises from plugging $\iota$'s into the composed field~\eqref{eqn:second-composed-field}.

The composed field $\phi_i^{i+1}\,\cdot\, \phi_{i+1}^{i+2} \,\cdot\, \cdots \,\cdot\, \phi_{n-1}^n$ is an $i \times n$ full rank matrix, describing $F_i \subset \mathbb{C}^n$, and we consider the submatrix $A_{i \times (n-j)}$ formed by its last $(n-j)$ columns, which is not necessarily of full rank. Then the above equation implies that:
\begin{equation}
    \varphi_{i,j}^{i,j+1}\, \cdot\, \varphi_{i,j+1}^{i,j+2} \,\cdot\, \cdots \,\cdot\, \varphi_{i,n-1}^{i,n} \,\cdot\, A_{i \times (n-j)}\,  = \,0~.
\end{equation}
The composed field $\varphi_{i,j}^{i,j+1}\, \cdot\, \varphi_{i,j+1}^{i,j+2} \,\cdot\, \cdots  \,\cdot\,\varphi_{i,n-1}^{i,n}$ is an $\text{r}_{i,j} \times i$ full-rank matrix, and consequently,
\begin{equation}\label{eqn:rank-constraint}
    \text{rk}(A_{i\times (n-j)}) \,\leq\, i-\text{r}_{i,j}~.
\end{equation}
This condition is actually equivalent to the defining incidence relation $\dim(F_i \cap E_j) \geq \text{r}_{i,j}$. The vector space $F_i$ in $\mathbb{C}^n$ is represented by an $i\times n$ full rank matrix:
\begin{center}
    $\phi_i^{i+1}\,\cdot \,\phi_{i+1}^{i+2} \,\cdot\,\cdots\,\cdot\, \phi_{n-1}^n$~,
\end{center}
whose rows can be viewed as $n$-vectors, spanning $F_i$.
Then any $n$-vector $v_{1\times n} \in F_i$ can be written as a linear combination of these basis vectors:
\begin{equation}
    v_{1\times n}  \,=\, w_{1\times i} \,\cdot\, \left(\,\phi_i^{i+1}\,\cdot \,\phi_{i+1}^{i+2}\,\cdot\, \cdots\,\cdot\, \phi_{n-1}^n\,\right)~,
\end{equation}
where $w$ is an $i$-vector of the coefficients.
If we further want this $n$-vector $v$ to lie in $F_i \cap E_j$, its last $n-j$ entries need to vanish:
\begin{equation}
    w_{1\times i} \,\cdot\, A_{i\times (n-j)} \,=\, 0~.
\end{equation}
That is, $w_{1\times i}$ has to lie in the kernel of $A_{i\times (n-j)}$. The set of all $n$-vectors $v$ in $F_i \cap E_j$ is in one-to-one correspondence with the set of $i$-vectors $w$ satisfying $w \cdot A = 0$. Therefore, we have:
\begin{equation}
    \dim(F_i\, \cap\, E_j)\, =\, \dim(\ker(A)) \,=\, i\, -\, \rk(A)\, \geq\, \text{r}_{i,j}~,
\end{equation}
where in the second step we used the rank-nullity theorem, and in the last step we used the rank constraint~\eqref{eqn:rank-constraint}.
Again, we see that the bulk geometry gets restricted to the Schubert variety.

\medskip
\noindent\textbf{Further comment.} Although both perspectives show that the bulk geometry is restricted to the corresponding Schubert variety $X_w$, the first approach also reveals something stronger: the underlying geometry of the full 1d-3d system is a resolution~\eqref{res Xw} of $X_w$.
Thus, the coupled system contains additional geometric structure beyond simply requiring the bulk fields to lie in $X_w$. The additional 1d degrees of freedom on the line defect provide us with precisely the ingredients needed to engineer the smooth resolution $\widetilde{X}_w$ as a target space. 

As we will show in subsection~\ref{subsec:1dindexOw}, the flavored Witten index of the 1d gauge theory produces the Chern character ch$(\mathcal{O}_{X_w})$ of the structure sheaf on $X_w$. This is consistent with the fact that the pushforward of the structure sheaf on $\widetilde{X}_w$ reproduces the structure sheaf on $X_w$ \cite[Proposition 2.2.5]{brion-lec}, \cite[section 3.4]{brion-kumar}:
\begin{equation}
    \pi_* \mathcal{O}_{\widetilde{X}_w} \,=\, \mathcal{O}_{X_w}~,
\end{equation}
so that their K-theory classes, and their Chern characters, coincide. Note that what we previously called $\mathcal{O}_w$ is a shorthand notation for $\mathcal{O}_{X_w}$ here.

\subsubsection{Mass terms induced by background fields}
\label{subsec:integrating-out}
Due to the existence of the non-dynamical fields, certain chiral fields and Fermi fields are massive and can be integrated out to obtain the infrared physics.
To see this in general, consider a Fermi multiplet $\Gamma= (\eta, ...)$ whose $E$-term contains a term proportional to a single chiral multiplet $\Phi= (\phi, \psi)$, say $E(\Phi)\, =\, \mu \,\Phi$. Then the Lagrangian contains the following term:
\begin{equation}
    \overline{\eta}\,\frac{\partial E(\phi)}{\partial \phi}\, \psi \,=\, \mu\, \overline{\eta}\, \psi~,
\end{equation}
which generates a complex mass $\mu$ for the pair $(\Gamma, \Phi)$. For a detailed discussion of 1d $\mathcal{N}=2$ supersymmetry, see, for example, section 2 of \cite{Closset:2025zyl}.

Consider the block in the quiver whose diagonal arrow is the Fermi multiplet $\Gamma_{n-1,j}^{n,j+1}$ for $j=1,2, \cdots, n-1$. For simplicity, let us denote it by $\Gamma^{(j)}$. Its corresponding 
$E$-term interaction is given by:
\begin{equation}   \label{eqn:superpotential-last-row}
\overline{D}_+ \Gamma^{(j)} \: = \:
\varphi_{n-1,j}^{n,j}\, \cdot\, \iota^{(j)}\, -\, \varphi_{n-1,j}^{n-1,j+1}\,\cdot\, \varphi_{n-1,j+1}^{n, j+1}~.
\end{equation}
Here, $\Gamma^{(j)}$ is viewed as an $\text{r}_{n-1,j}\times (j+1)$ matrix.
We denote the $(j+1)$ columns of $\Gamma^{(j)}$ by:
\begin{equation}
\Gamma_1^{(j)}~, \, \Gamma_2^{(j)}~, \, \cdots\, , \, \Gamma_{j+1}^{(j)}~,
\end{equation}
each transforming in the fundamental representation of $U(\text{r}_{n-1,j})$.

The chiral multiplet $\varphi_{n-1,j}^{n,j}$ can be viewed as an $\text{r}_{n-1,j} \times j$ matrix, whose columns will be denoted by:
\begin{equation}
    \varphi^{(j)}_1\, , \, \varphi^{(j)}_2\, , \, \cdots \, , \varphi^{(j)}_j~,
\end{equation}
each transforming in the fundamental representation of $U(\text{r}_{n-1,j})$.

Since the background field $\iota^{(j)}$ takes a special form $\iota^{(j)} = (\mathbb{I}_{j \times j} \quad 0_{j \times 1})$, the first term in 
\eqref{eqn:superpotential-last-row} 
reduces to
mass terms for the pairs $(\Gamma^{(j)}_i, \varphi^{(j)}_i)$ for $i$ from $1$ to $j$. Then, in the infrared, integrating out these massive fields, the vertical chiral multiplets are absent, and we are left with only one Fermi multiplet $\Gamma^{(j)}_{j+1}$ in the fundamental representation of $U(\text{r}_{n-1,j})$.

\begin{figure}[t]
    \centering
    \includegraphics[width=0.8\linewidth]{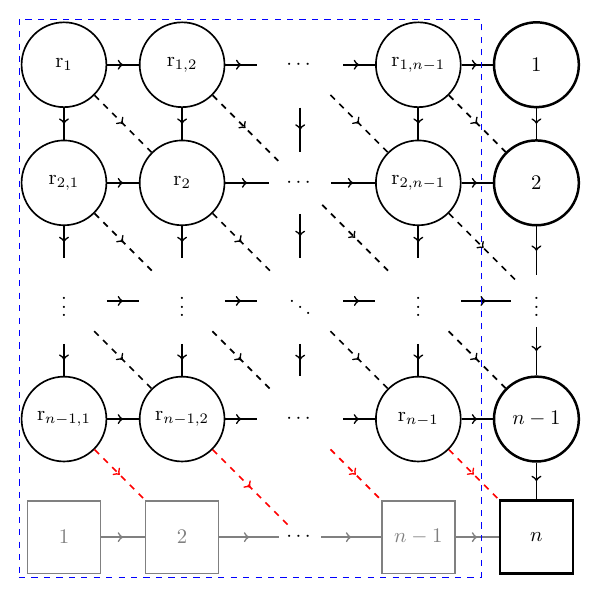}
    \caption{The 1d-3d coupled system after integrating out the massive 1d fermionic and bosonic degrees of freedom from the defect given in figure \ref{fig:general proposal}. In the last row, the red diagonal arrows now represent Fermi multiplets in the fundamental representation of the gauge group represented by the circle node and have a charge $-1$ under the $U(1)$ subgroup of the bulk flavor symmetry $SU(n)$ indexed by the square node.}
    \label{fig:last-row-reduced}
\end{figure}

Therefore, the last row in figure~\ref{fig:general proposal} can be reduced, and we obtain the quiver shown in figure~\ref{fig:last-row-reduced}. The Fermi multiplets in the last row are no longer bifundamentals, but instead reduce to single components $\Gamma^{(j)}_{j+1}$, each coupled to a distinct $U(1)\subset U(1)^n \subset U(n)$. We use the color red to make this distinction that they are only charged under one $U(1)$ factor in the flavor group associated with the target node. In order to keep track of these charges, we assign the flavor fugacities as discussed next.

\medskip
\noindent
\textbf{Flavor fugacity assignment: a convention.}
In the computation of the flavored Witten index of the 1d theory (discussed in the next subsection), one can turn on flavor fugacities associated with the flavor symmetry groups --- this corresponds to probing the equivariant quantum K-theory of the complete flag manifold. Due to the inclusion maps, and as part of our conventions, the flavor fugacities associated with $U(j)$ in the last row of the quiver are $y_n, \cdots,y_{n-j+1}$. The surviving Fermi multiplet $\Gamma^{(j)}_{j+1}$ charged under $U(1) \subset U(1)^{j+1} \subset U(j+1)$ corresponds to the fugacity $y_{n-j}$. The pair $(\Gamma^{(j)}_i, \varphi^{(j)}_i)$ carries fugacities $y_{n+1-i}$, and their contributions cancel out in the flavored Witten index, providing a consistency check that integrating out these fields is valid and does not affect the counting of supersymmetric ground states.

The bulk gauge groups are also treated as flavor symmetry groups from the 1d perspective. For each $U(i)$ in the last column of the quiver, we denote the flavor fugacities by $x^{(i)}_1, x^{(i)}_2, \cdots x^{(i)}_i$ for $i$ from $1$ to $n-1$, which, as we discussed in subsection \ref{subsec:3dGLSMQK}, are the K-theoretic Chern roots of the tautological bundle $\mathcal{S}_i$ on the flag manifold.

\subsection{1d indices as Chern characters of the Schubert classes}\label{subsec:1dindexOw}
In the above discussion, we argued that, for a permutation $w\in S_n$, the proposed 1d line defect $\SL_w$ restricts the bulk scalar fields $\phi_i^{i+1}$ at the point of insertion to land onto the Schubert variety $X_w$ inside the complete flag. On general ground, the Witten index of the defining quiver gauge theory should give us the Chern character of the Schubert class $[\mathcal{O}_w]$ in the equivariant quantum K-theory of the flag manifold. These characters are given in terms of the quantum double Grothendieck polynomials $\mathfrak{G}_w^{(q)}(x,y)$. See appendix~\ref{subsec: Grothendeick polys} for further details. Checking this relationship would constitute a very strong check of our proposal. 

Following the supersymmetric localization results of \cite{Hori:2014tda}, and using the conventions spelled out in~\cite{Closset:2023bdr, Closset:2025zyl}, the 1d index of the quiver in figure \ref{fig:last-row-reduced} can be written as a so-called Jeffrey--Kirwan (JK) residue:
\begin{equation}\label{1d index JK}
    \mathcal{I}^{({\rm 1d})}_{w}(x,y) = \oint_{\rm JK}\; ({\rm dM})\;{\rm Z}_{\rm chiral}^{\rm ver}\;{\rm Z}_{\rm chiral}^{\rm hor}\;{\rm Z}_{\rm Fermi}^{\rm black}\;{\rm Z}_{\rm Fermi}^{\rm red}~.
\end{equation}
The measure $({\rm dM})$ is given by:
\begin{equation}  \label{eq:measure}
    ({\rm dM})\,:=\,\prod_{i=1}^{n-1} \,\prod_{j = 1}^{n-1}\,\left[\,\Delta^{(i,j)}(z)\,\frac{1}{{\rm r}_{i,j}\,!}\,\prod_{\alpha=1}^{{\rm r}_{i,j}}\, \frac{-\,dz_\alpha^{(i,j)}}{2\pi i \,z_\alpha^{(i,j)}}\,\right]~,
\end{equation}
with the 1d ranks defined in~\eqref{rij def}, and $\Delta^{(i,j)}$ the Vandermonde determinant factor coming from the 1d $U({\rm r}_{i,j})$ gauge W-bosons:
\begin{equation}  \label{eq:vandermonde}
    \Delta^{(i,j)}(z)\,:=\,\prod_{1\leq \alpha\neq\beta\leq{\rm r}_{i,j}}\left(1-\frac{z^{(i,j)}_\alpha}{z^{(i,j)}_\beta}\right)~.
\end{equation}
As for the different components appearing in the integrand of \eqref{1d index JK}, these are the 1-loop contributions of the chiral and Fermi multiplets. More explicitly: 
\begin{align}\label{matter contr 1d index}
    \begin{split}
        &{\rm Z}_{\rm chiral}^{\rm ver} \,:=\,\prod_{i=1}^{n-2}\,\prod_{j=1}^{n-1}\,\prod_{\alpha=1}^{{\rm r}_{i,j}}\,\prod_{\beta=1}^{{\rm r}_{i+1,j}}\,\left(\,1\,-\,\frac{z^{(i,j)}_{\alpha}}{z^{(i+1,j)}_{\beta}}\,\right)^{-1} ~,\\
        &{\rm Z}_{\rm chiral}^{\rm hor} \,:=\,\prod_{i=1}^{n-1}\,\left[\,\prod_{\alpha=1}^{{\rm r}_{i,n-1}}\,\prod_{a = 1}^{i}\,\left(\,1\,-\,\frac{z_{\alpha}^{(i,n-1)}}{x_{a}^{(i)}}\,\right)^{-1}\,\prod_{j=1}^{n-2}\,\prod_{\beta=1}^{{\rm r}_{i,j}}\,\prod_{\gamma=1}^{{\rm r}_{i,j+1}}\,\left(\,1\,-\,\frac{z^{(i,j)}_{\beta}}{z^{(i,j+1)}_{\gamma}}\,\right)^{-1}\,\right] ~,\\ 
        &{\rm Z}^{\rm black}_{\rm Fermi} \,:=\, \prod_{i=1}^{n-2}\,\left[\,\prod_{\gamma=1}^{{\rm r}_{i,n-1}}\prod_{a=1}^{i+1}\left(\,1\,-\,\frac{z^{(i,n-1)}_\gamma}{x_{a}^{(i+1)}}\,\right)\prod_{j = 1}^{n-2}\,\prod_{\alpha=1}^{{\rm r}_{i,j}}\,\prod_{\beta = 1}^{{\rm r}_{i+1,j+1}}\,\left(\,1\,-\,\frac{z^{(i,j)}_\alpha}{z_\beta^{(i+1,j+1)}}\,\right)\,\right]\,~,\\
        &{\rm Z}^{\rm red}_{\rm Fermi} \,:=\,\prod_{i=1}^{n-1}\,\prod_{\alpha=1}^{{\rm r}_{n-1,i}}\,\left(\,1\,-\,\frac{z^{(n-1,i)}_\alpha}{y_{n-i}}\,\right) ~.\\
    \end{split}
\end{align}
Let us make the following remarks:
\begin{itemize}
    \item The form of the final contribution in \eqref{matter contr 1d index} is obtained after the cancellation that occurs between the massive chiral and Fermi multiplets in the last row in figure~\ref{fig:general proposal}, leading to figure~\ref{fig:last-row-reduced}, as explained in subsection \ref{subsec:integrating-out}.
    \item Due to the fact that all 1d real FI parameters are taken to be positive, the JK prescription instructs us to pick only the poles coming from the chiral fields in the fundamental representation. That is, those in ${\rm Z}_{\rm chiral}^{\rm ver}$ and ${\rm Z}_{\rm chiral}^{\rm hor}$, but not the poles coming from the measure $({\rm dM})$. One should take the JK residue iteratively, starting from the first row and working from left to right, then proceeding to the second row in the same way, and continuing row by row. (Equivalently, one can proceed column by column from left to right.)
    \item When working out the residue integrals in \eqref{1d index JK}, we end up with polynomial expressions in terms of the K-theoretic Chern roots $x^{(\bullet)}_\bullet$ of the tautological vector bundles $\mathcal{S}_\bullet$. Since the Grothendieck polynomials defined in appendix \ref{subsec: Grothendeick polys} are given in terms of the roots of the quotient bundles $\mathcal{Q}_\bullet$, we further need to use the quantum relations \eqref{QK relations of flag n}.%
    \footnote{In fact, if we instead use the classical ring relations obtained by taking $\qk_i = 0$ in \protect\eqref{QK lambda relations}, we end up with the classical Grothendieck polynomials $\mathfrak{G}_{w}(x,y)$. These are the Chern characters of the Schubert classes in the classical K-theory ring of Fl$(n)$. See appendix~\protect\ref{subsec: Grothendeick polys} for the definition and further properties of these polynomials. The original index in terms of $x^{(\bullet)}_\bullet$ is believed to coincide with the universal Grothendieck polynomial \protect\cite{buch2005grothendieck}.} 
    More specifically, we can start from the index, which is symmetric in the K-theoretic Chern roots of $\mathcal{S}_i$ (by symmetry, from the 1d perspective). Using the quantum K-theory ring relation \eqref{QK relations of flag n}, we may eliminate the elementary symmetric polynomials in $x^{(i+1)}$, in terms of those in $x^{(i)}$ and $x^{(i-1)}$, and in terms of the quotient bundle Chern roots $x_{i+1}$. 
    Iterating this procedure, we eventually reach the base case in which $x^{(1)}_1$ is identified with $x_1$, leaving an expression entirely in terms of the quotient bundle Chern roots, which should give us the expected quantum Grothendieck polynomial. (One can systematize this procedure on a computer using Gr\"obner bases methods, as in~\cite{Closset:2023vos}.)
    Note that one cannot go the other way around, hence the Witten indices carry strictly more information than the quantum Grothendieck polynomials. (If we were to start from the latter, given in terms of the quotient-bundle Chern roots, one can eliminate $x_{i+1}$ using equation~\eqref{QK relations of flag n}. But this elimination is not unique as equation~\eqref{QK relations of flag n} contains a family of identities (labeled by the index $a$), and different choices of $a$ lead to different ways of eliminating $x_{i+1}$.)
    \item To speed up the calculation of the Witten indices, one can locally apply the 1d Seiberg-like dualities that we reviewed in section \ref{sec:defects and dualities}. As mentioned earlier, this amounts to `confining' some of the 1d gauge nodes that, from the geometric perspective discussed above, lead to redundant constraints. 
\end{itemize}

\subsection{Example: Schubert defects for Fl\texorpdfstring{$(3)$}{(3)}}

\begin{figure}[t]
    \centering
    \includegraphics[width=0.77\linewidth]{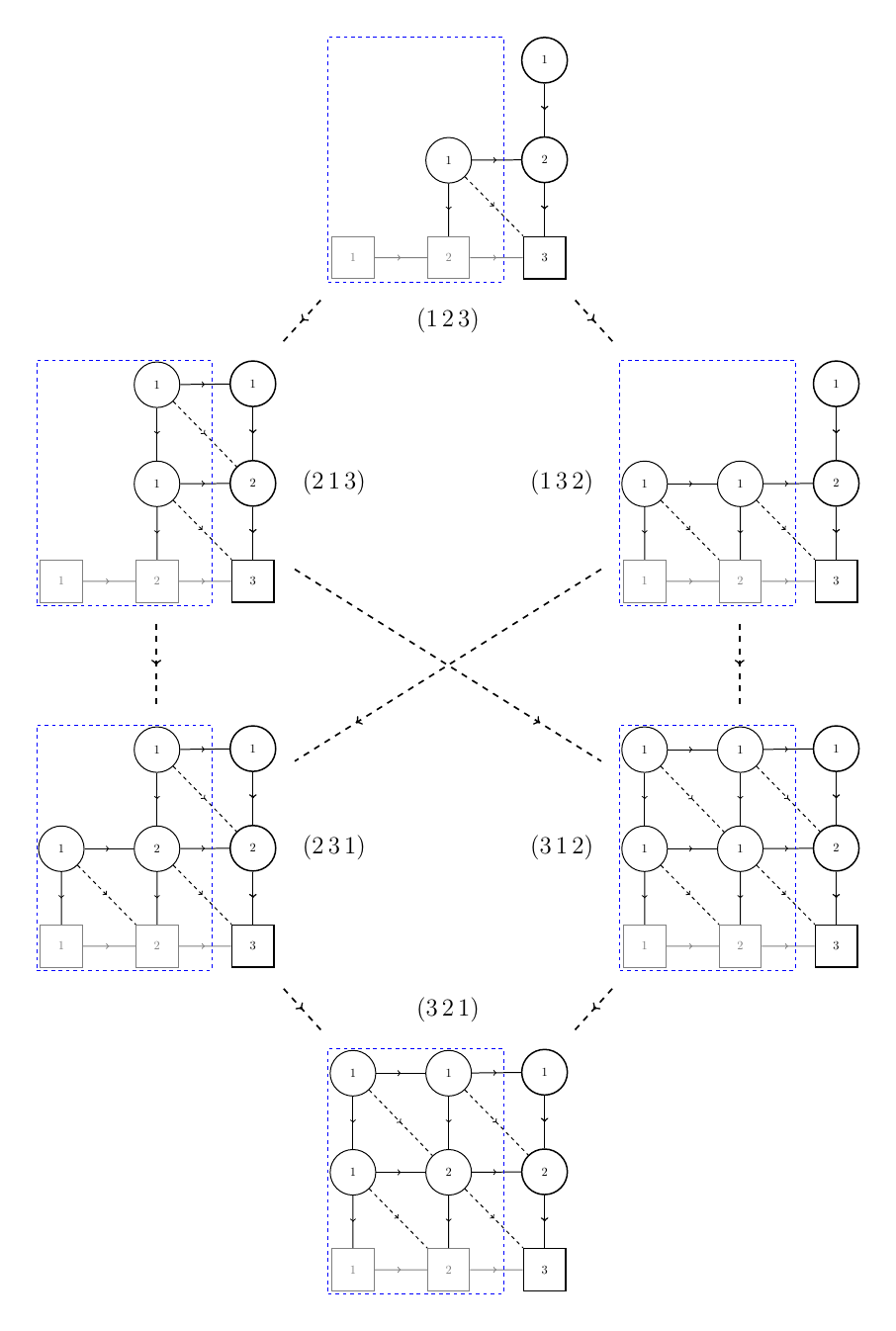}
    \caption{Hasse diagram for the Schubert line defects of Fl$(3)$ 
    (compare {\it e.g.} to~\protect\cite[example 4.8.6]{Billey-Lakshmibai-2000}). 
    For each one of the 1d quiver defects, the associated $S_3$ permutation is displayed next to it. The arrows connecting different defects correspond to a transposition connecting the indexing permutations. In our notation, $(1\,2\,3)$ is the identity, corresponding to $X={\rm Fl}(3)$.
    }
    \label{fig:Hasse S3}
\end{figure}

\begin{figure}[t!]
    \centering
    \includegraphics[width=0.8\linewidth]{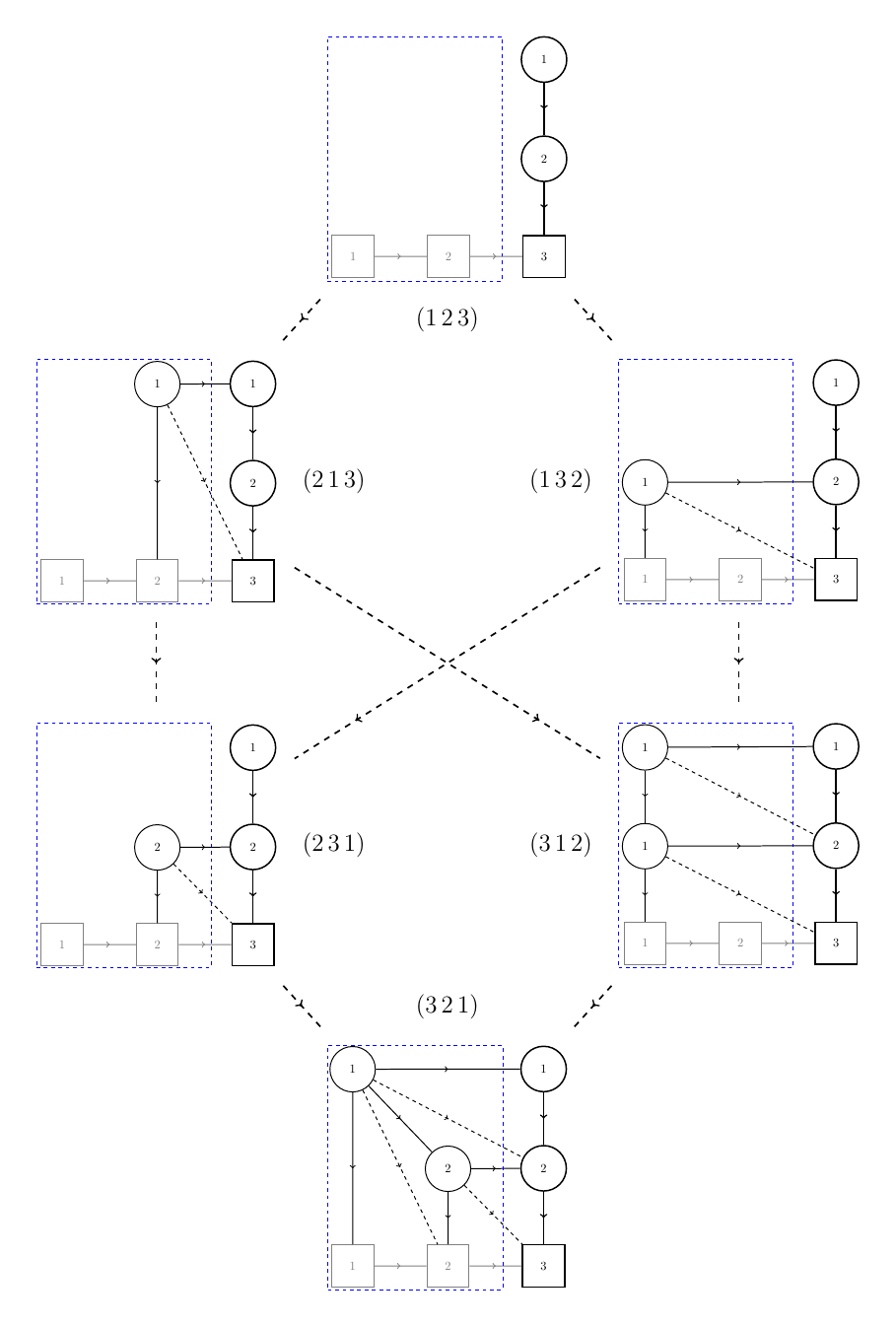}
    \caption{Hasse diagram for the defects in Fl$(3)$ after applying the 1d duality moves discussed in section \ref{sec:defects and dualities}. For example, note that for the trivial permutation, the quiver defect reduces to the trivial defect line.}
    \label{fig:Hasse S3 FR}
\end{figure}

As a pedagogical example, let us look at the case $n=3$, in which case we have six Schubert varieties indexed by $w\in S_3$. For each permutation, the corresponding line defect is shown in figure~\ref{fig:Hasse S3}. In figure \ref{fig:Hasse S3 FR}, we show the simplified 1d quivers obtained after using the 1d duality moves discussed in section~\ref{sect:duality}. Note that the arrows in the Hasse diagram denote the projections of the Schubert varieties (that is, $X_w\rightarrow X_{w'}$ if $X_{w'}\subset X_w$).
Looking at these defects as 1d-3d coupled systems, one can compute the indices of the defect SQM using the JK residue formula~\eqref{1d index JK}.

\begin{itemize}
    \item \underline{For $w \,=\, (1\,2\,3)$.}
Here, the permutation $w = (1\, 2\, 3)$ is the identity permutation.  The quiver is slightly nontrivial, but the polynomial is trivial.  As a warm-up exercise, we will walk through this case systematically, and in later examples, we will only give the highlights.

First, we compute the ${\rm r}_{i,j}$ matrix.  For this permutation, $w_0 w = (3\, 2\, 1)$,
and one quickly finds:
\begin{align}\label{123rMatrix}
\begin{split}
    {\rm r}_{1,1}^{w_0w} & \,=\,  \#\,\{\,\ell \,\leq \,1 \, | \, (w_0w)(\ell)\, \leq\, 1 \,\} \,= \, 0~,
    \\
    {\rm r}_{1,2}^{w_0w} & \,=\,  \#\,\{\,\ell \,\leq\, 1 \, | \, (w_0w)(\ell)\, \leq\, 2 \,\} \, = \,0~,
    \\
    {\rm r}_{2,1}^{w_0w} & \,=\,  \#\,\{\,\ell\,\leq\, 2 \, | \, (w_0w)(\ell)\, \leq\, 1 \,\} \, = \,0~,
    \\
    {\rm r}_{2,2}^{w_0w} & \,=\,  \#\,\{\, \ell \,\leq \,2 \, | \, (w_0w)(\ell)\, \leq \,2\, \} \, = \, 1~,
\end{split}
\end{align}
thus recovering the quiver in figure~\ref{fig:Hasse S3}.

From equation~(\ref{eq:measure}), the measure is given in this case by:
\begin{eqnarray*}
    ({\rm dM}) & \,=\, &
    \prod_{i=1}^2 \,\prod_{j=1}^2\, \left(\, \Delta^{(i,j)}(z) \, \frac{1}{{\rm r}_{i,j}!} \,
    \prod_{\alpha=1}^{{\rm r}_{i,j}}\, \left(\, \frac{ - 1}{2\,\pi \,i}\,\right) \,\left( \,\frac{dz_{\alpha}^{(i,j)} }{z_{\alpha}^{(i,j)}}\, \right)
    \right)~,
    \\
    & \,=\, &
     \Delta^{(1,1)}(z) \,
     \Delta^{(1,2)}(z)
    \,
     \Delta^{(2,1)}(z)\,
   \Delta^{(2,2)}(z)\, \left(\, -\, \frac{1}{2\,\pi\, i}\,\frac{ dz_1^{(2,2)}}{z_1^{(2,2)}}
   \, \right)~,
\end{eqnarray*}
with, from~(\ref{eq:vandermonde}),
\begin{equation}
    \Delta^{(1,1)}(z) \, = \, 1 \, = \, \Delta^{(1,2)}(z) \, = \,
    \Delta^{(2,1)}(z) \, = \, \Delta^{(2,2)}(z)~.
\end{equation}
From~(\ref{matter contr 1d index}),
\begin{eqnarray*}
    Z^{\rm hor}_{\rm chiral} & \,=\, & 
    \left(\, 1\, -\, \frac{ z_1^{(2,2)}}{x_1^{(2)}} \,\right)^{-1}\,
    \left( \,1\, -\, \frac{ z_1^{(2,2)}}{x_2^{(2)}}\, \right)^{-1}~,
    \\
    Z^{\rm red}_{\rm Fermi} & \,=\, & \left(\, 1 \,-\, \frac{z_1^{(2,2)}}{y_1} \,\right)~.
\end{eqnarray*}
For notational simplicity, define $z \,\equiv\, z_1^{(2,2)}$, then our index for the case $w = (1\, 2\, 3)$ is:
\begin{displaymath}
\mathcal{I}^{\rm (1d)}_{(1\, 2\, 3)}(x,y) \, = \,
    -\, \frac{1}{2\,\pi \,i} \,\oint_{\rm JK} \frac{dz}{z} \,\left(\, 1\, -\, \frac{z}{x_1^{(2)}}\, \right)^{-1}\,
    \left( \,1\, -\, \frac{z}{x_2^{(2)}} \,\right)^{-1}\,
    \left( \,1\, - \,\frac{z}{y_1}\, \right)~,
\end{displaymath}
and the contour integral encloses the poles at $z \,=\, x_{1,2}^{(2)}$, but not the pole at $z=0$.
It is straightforward to evaluate the contour integral to find:
\begin{equation}\label{I123=1}
    \mathcal{I}^{\rm (1d)}_{(1\, 2 \,3)}(x,y) \: = \: 1~.
\end{equation}

    \item \underline{For $w \,=\, (2\,1\,3)$.} In this case, we have the following integrand:
    \begin{equation*}
        \frac{\left(1-\frac{z^{(1,2)}_{1}}{x_1^{(2)}}\right)\left(1-\frac{z^{(1,2)}_{1}}{x_2^{(2)}}\right)
   \left(1-\frac{z^{(2,2)}_{1}}{y_1}\right)}{\left(1\,-\,\frac{z_1^{(1,2)}}{x_1^{(1)}}\right)
   \left(1-\frac{z^{(1,2)}_{1}}{z^{(2,2)}_{1}}\right) 
   \left(1-\frac{z^{(2,2)}_{1}}{x^{(2)}_{1}}\right)
   \left(1-\frac{z^{(2,2)}_{1}}{x^{(2)}_{2}}\right)}~.
    \end{equation*}
  Computing the corresponding residue integrals, we get the following polynomial:
    \begin{equation} \label{I213}
        \CI^{(\rm 1d)}_{(2\,1\,3)} (x,y) \,=\, 1\,-\,\frac{x^{(1)}_{1}}{y_1} \quad\implies\quad \mathcal{I}_{(2\,1\,3)}^{(\rm 1d)}(x,y;\qk)\,=\,1\,-\,\frac{x_{1}}{y_1}  ~.
    \end{equation}
    Here, `$\implies$' means that the we used \eqref{x1=x11} to get the quantum version $\mathcal{I}_{(2,1,3)}^{(\rm 1d)}(x,y;\qk)$.
    \item \underline{For $w \,=\, (1\,3\,2)$.} In this case, the integrand is of the form:
    \begin{equation*}
        \frac{\left(1-\frac{z^{(2,1)}_{1}}{y_2}\right)
   \left(1-\frac{z^{(2,2)}_{1}}{y_1}\right)}{
   \left(1-\frac{z^{(2,1)}_{1}}{z^{(2,2)}_{1}}\right) 
   \left(1-\frac{z^{(2,2)}_{1}}{x^{(2)}_{1}}\right)
   \left(1-\frac{z^{(2,2)}_{1}}{x^{(2)}_{2}}\right)}~.
    \end{equation*}
    Taking the residue integrals for this expression, we get:
    \begin{equation}
        \CI_{(1\,3\,2)}^{(\rm 1d)}(x,y)\, =\,  1\,-\,\frac{x^{(2)}_{1}\, x^{(2)}_{2}}{y_1\, y_2} \quad\implies\quad\mathcal{I}^{(\rm 1d)}_{(1\,3\,2)}(x,y;\qk) \, = \,1\,-\,\frac{1}{1\,-\,q_1}\,\frac{x_{1} \,x_{2}}{y_1\, y_2}~.
    \end{equation}
    In the last equality here, we used the second relation in \eqref{QK for Fl(3)}.
    \item \underline{For $w \,= \,(2\,3\,1)$.} In this case, the 1d quiver has the following contributions:
    \begin{equation*}
    \frac{\left(1-\frac{z^{(2,2)}_{1}}{z^{(2,2)}_{2}}\right) \left(1-\frac{z^{(2,2)}_{2}}{z^{(2,2)}_{1}}\right)\left(1-\frac{z^{(2,1)}_{1}}{y_2}\right) \prod_{a=1}^2\left(1-\frac{z^{(1,2)}_{1}}{x^{(2)}_{a}}\right)\left(1-\frac{z^{(2,2)}_a}{y_1}\right)}{ \left(1-\frac{z^{(1,2)}_{1}}{x^{(1)}_{1}}\right)\prod_{a=1}^2\left(1-\frac{z^{(1,2)}_{1}}{z^{(2,2)}_{a}}\right) \left(1-\frac{z^{(2,1)}_{1}}{z^{(2,2)}_{a}}\right)\prod_{a,b=1}^2\left(1-\frac{z^{(2,2)}_{a}}{x^{(2)}_{b}}\right) }~.
    \end{equation*}

    The associated residue integrals yield the following polynomial:
    \begin{equation}\label{I231}
        \CI^{(\rm 1d)}_{(2\,3\,1)} (x,y)\, =\,1\,-\,\frac{x^{(2)}_{1}}{y_1}\,-\,\frac{x^{(2)}_{2}}{y_1}\,+\, \frac{x^{(2)}_{1}\,x^{(2)}_{2} }{y_1^2}~.
    \end{equation}
Using the two relations in \eqref{QK for Fl(3)}, we see that this index can be written as:
\begin{equation}
    \CI^{(\rm 1d)}_{(2\,3\,1)} (x,y;\qk) \,=\,1\,-\,\frac{x_{1}}{y_1}\,-\,\frac{x_{2}}{y_1}\,+\, \frac{1}{1\,-\,q_1}\,\frac{x_{1}\,x_{2} }{y_1^2}~.
\end{equation}
    \item \underline{For $w \,=\, (3\,1\,2)$.} We have the following integrand:
    \begin{equation*}
        \frac{\left(1-\frac{z^{(1,1)}_{1}}{z^{(2,2)}_{1}}\right)  \left(1-\frac{z^{(2,1)}_{1}}{y_2}\right)
   \left(1-\frac{z^{(2,2)}_{1}}{y_1}\right)\prod_{a=1}^2\left(1-\frac{z^{(1,2)}_{1}}{x^{2}_{a}}\right)}{\left(1-\frac{z^{(1,1)}_{1}}{z^{(1,2)}_{1}}\right) \left(1-\frac{z^{(1,1)}_{1}}{z^{(2,1)}_{1}}\right) 
   \left(1-\frac{z^{(1,2)}_{1}}{z^{(2,2)}_{1}}\right) \left(1-\frac{z^{(2,1)}_{1}}{z^{(2,2)}_{1}}\right)  \left(1-\frac{z^{(1,2)}_{1}}{x^{(1)}_{1}}\right)
   \prod_{a=1}^2\left(1-\frac{z^{(2,2)}_{1}}{x^{(2)}_{a}}\right) }~.
    \end{equation*}

    Doing the residue integral for this, we get the following polynomial:
    \begin{align}
    \begin{split}
        \CI_{(3\,1\,2)}^{(\rm 1d)}(x,y) \,=\, 1\,-\,\frac{x^{(1)}_{1}}{y_1}\,-\,\frac{x^{(1)}_{1}}{y_2}\,+\, \frac{x^{(1)}_{1}\,x^{(2)}_{1} }{y_1 \,y_2}\,+\,\frac{x^{(1)}_{1}\,x^{(2)}_{2} }{y_1\, y_2}\,-\,\frac{x^{(2)}_{1}\, x^{(2)}_{2}}{y_1 \,y_2}~.\\
        \end{split}
    \end{align}
    Using both \eqref{x1=x11} and \eqref{QK for Fl(3)}, we obtain:
    \begin{equation}
       \CI_{(3\,1\,2)}^{(\rm 1d)}(x,y;\qk) \,=\, 1\,-\,\frac{x_{1}}{y_1}\,-\,\frac{x_{1}}{y_2}\,+\, \frac{x_{1}^2}{y_1 y_2}\,+\,\frac{x_{1}\,x_{2} }{y_1 \,y_2}\,-\,\frac{1}{1\,-\,q_1}\,\frac{x_{1}\, x_{2}}{y_1 \,y_2}~.
    \end{equation}
    \item \underline{For $w \,=\, (3\,2\,1)$.} In this case, the integrand has the following form:
    \begin{equation*}
        \frac{\left(1-\frac{z_{2}^{(2,1)}}{z_{2}^{(2,2)}}\right)
   \left(1-\frac{z_{2}^{(2,2)}}{z_{2}^{(2,1)}}\right) \prod_{a=1}^2\left(1-\frac{z^{(1,1)}_{1}}{z^{(2,2)}_{a}}\right)\left(1-\frac{z^{(1,2)}_{1}}{x^{(2)}_{a}}\right)  \left(1-\frac{z^{(2,2)}_{a}}{y_1}\right)  \left(1-\frac{z^{(2,1)}_{1}}{y_2}\right)
   }{ \left(1-\frac{z^{(1,1)}_{1}}{z^{(1,2)}_{1}}\right) 
   \left(1-\frac{z^{(1,1)}_{1}}{z^{(2,1)}_{1}}\right) 
 \left(1-\frac{z^{(1,2)}_{1}}{x^{(1)}_{1}}\right)
\prod_{a=1}^2 \left(1-\frac{z^{(1,2)}_{1}}{z^{(2,2)}_{a}}\right)   \left(1-\frac{z^{(2,1)}_{1}}{z^{(2,2)}_{a}}\right)\prod_{b=1}^2  \left(1-\frac{z^{(2,2)}_{a}}{x^{(2)}_{b}}\right) }~. 
\end{equation*}
Computing the associated residue integral, we get:
\begin{align}
\begin{split}
        \CI_{(3\,2\,1)}^{(\rm 1d)}(x,y) \,&=\, 1\,-\,\frac{x^{(1)}_{1}}{y_2}\,-\,\frac{x^{(2)}_{1}}{y_1}\,-\,\frac{x^{(2)}_{2}}{y_1}\,+\,\frac{x^{(1)}_{1}\,x^{(2)}_{1} }{y_1 \,y_2}\,+\,\frac{ x^{(1)}_{1}\,x^{(2)}_{2}}{y_1 \,y_2}\\
        &\quad+\,\frac{x^{(2)}_{1}\,
   x^{(2)}_{2}}{y_1^2}\,-\,\frac{x^{(1)}_{1}\,x^{(2)}_{1} \,x^{(2)}_{2} }{y_1^2\, y_2}~,\\
   &= \left(1\,-\,\frac{x_1^{(1)}}{y_2}\right)\,\left(1\,-\,\frac{x_1^{(2)}}{y_1}\right)\,\left(1\,-\,\frac{x_2^{(2)}}{y_1}\right)~.
   \end{split}
\end{align}
This indeed matches with \cite[Theorem 1.2]{ahkmox} upon identifying the parameters $\epsilon_i$ in that reference with our 2d twisted masses.\footnote{As another explicit example, see the 1d index for $w_0 \,=\, (\,4\,3\,\,2\,1\,)\in S_4$ in the corresponding table in appendix \ref{app:Fl(4) defects}.} We explain this correspondence to the results of~\cite{ahkmox} in part II~\cite{Closset:2026bnk}.

As before, using the relations \eqref{x1=x11} and \eqref{QK for Fl(3)}, we get:
\begin{equation}\label{321index}
    \begin{split}
    \CI_{(3\,2\,1)}^{(\rm 1d)}(x,y;\qk)\,&=\,1\,-\,\frac{x_{1}}{y_2}\,-\,\frac{x_{1}}{y_1}\,-\,\frac{x_{2}}{y_1}\,+\,\frac{x_{1}^2 }{y_1 \,y_2}\,+\,\frac{ x_{1}\,x_{2}}{y_1\, y_2}\,+\,\frac{1}{1\,-\,q_1}\,\frac{x_{1}
   \,x_{2}}{y_1^2}\\
   &\quad \,-\,\frac{1}{1\,-\,q_1}\,\frac{x^2_{1}\, x_{2} }{y_1^2\, y_2}~.
   \end{split}
\end{equation}
\end{itemize}
Note that these 1d indices match exactly with the corresponding quantum Grothendieck polynomials $\mathfrak{G}_w^{(\qk)}(x,y)$ listed in \eqref{q-double-groth-polys-S3}. Moreover, the polynomial expression for the longest permutation \eqref{321index} can be factorized as:
\begin{equation}
     \CI_{(3\,2\,1)}^{(\rm 1d)}(x,y;\qk)\,=\,\left(1\,-\,\frac{x_1}{y_2}\right) \left(1\,+\,\frac{x_1 \,x_2}{\left(1\,-\,q_1\right) \,y_1^2}\,-\,\frac{x_1\,+\,x_2}{y_1}\right)~.
\end{equation}
This indeed matches with \eqref{q-double-groth-poly-longest-perm}. 

\medskip
\noindent
In appendix \ref{app:Fl(4) defects}, we work the $4!=24$ 1d indices for the case $n=4$ --- the complete flag manifold Fl$(4)$ --- and show that they match with the corresponding quantum double Grothendieck polynomials.

\medskip
\noindent
\textbf{\underline{Conjecture I:}}  For the general Fl$(n)$ case, we conjecture that the formula \eqref{1d index JK} is a JK residue representation of the quantum Grothendieck polynomials. Namely:
\begin{equation}
    \mathcal{I}_{w}^{({\rm 1d})}(x,y;\qk)\,=\,\mathfrak{G}_w^{(\qk)}(x,y)~,
\end{equation}
for any $w\in S_n$, where it is understood here that the equality is upon using the relations \eqref{QK relations of flag n} to transform $ \mathcal{I}_{w}^{({\rm 1d})}(x,y)$ into $ \mathcal{I}_{w}^{({\rm 1d})}(x,y;\qk)$.

\subsubsection{Index calculation for the reduced quivers}
For completeness, in this subsection, we exhibit by examples how the dualities discussed in section \ref{sec:defects and dualities} can speed up the calculation of the 1d indices. For that goal, let us look again at the example of Fl$(3)$. The result of applying these dualities to the six possible line defects is shown in figure \ref{fig:Hasse S3 FR}. Let us consider the following three cases in more detail:

\begin{itemize}
    \item \underline{For $w\,=\,(1\,2\,3)$.} As expected from our calculations for this case around \eqref{I123=1}, upon using the duality moves, the Schubert defect trivializes, as one can see in figure \ref{fig:Hasse S3 FR}. And indeed, the 1d Witten index in this case is equal to $1$.
    \item \underline{For $w\,=\,(2\,1\,3)$.}
     For this reduced quiver, we have the following simpler integrand:
    \begin{equation*}
       \left(1\,-\,\frac{z^{(1,2)}_{1}}{y_1}\right)
   \left(1\,-\,\frac{z^{(1,2)}_{1}}{x^{(1)}_{1}}\right)^{-1}.
    \end{equation*}
    The residue computation is much simpler but reproduces the same result as in~\eqref{I213}. 
    \item \underline{For $w\,=\,(2\,3\,1)$.}
This 1d reduced quiver gives the following integrand:
    \begin{equation*}
    \left(1\,-\,\frac{z^{(2,2)}_{1}}{z^{(2,2)}_{2}}\right) \left(1\,-\,\frac{z^{(2,2)}_{2}}{z^{(2,2)}_{1}}\right)\prod_{a=1}^2\left(1\,-\,\frac{z^{(2,2)}_a}{y_1}\right) \prod_{a,b=1}^2\left(1\,-\,\frac{z^{(2,2)}_{a}}{x^{(2)}_{b}}\right)^{-1}~.
    \end{equation*}
This much simpler residue yields the same result as in~\eqref{I231}.
\end{itemize}


\section{Schubert point defects in the 2d GLSM}\label{sec:0d-2d system}

In the previous section, we gave a UV construction of line defects in the 3d GLSM that, in the IR, flow to the Schubert classes in QK$_T$(Fl$(n)$). In this section, we discuss the cohomological limit of this construction. That is, we view the quiver in figure~\ref{fig:general proposal} as a 0d-2d coupled system and propose that the corresponding half-BPS point defects flow to the Schubert classes in QH$^{\bullet}_T$(Fl$(n)$). In what follows, we will refer to these defects as \textit{Schubert point defects}.

For a permutation $w\in S_n$, the Poincar\'e dual to the Schubert variety $[X_w]$ is $\Omega_w\in {\rm H}^{2\ell(w)}({\rm Fl}(n))$. In the $T$-equivariant case,  this cohomology class is written in terms of the quantum double Schubert polynomial $\mathfrak{S}_w^{(\qcoh)}(\sigma, m)$ \cite{fomin1997quantum,ciocan1997quantum,kirillov2000quantum}. For definition and examples, see appendix~\ref{subsec: Schubert polys}.

Viewing the quiver defect as a 0d $\mathcal{N}=2$ supersymmetric matrix model (SMM), we will argue in examples that its partition function is none other than the corresponding quantum Schubert polynomial. Like in the 3d case, we conjecture that this matching holds in full generality.

\subsection{Partition functions of the matrix model}

In the 2d limit $\beta \rightarrow 0$, the figure~\ref{fig:general proposal} is now interpreted as a 0d $\mathcal{N}=2$ SMM with gauge group $G_{\rm 0d} = \times_{i,j}\, U({\rm r}_{i,j})$ and bifundamental chiral and Fermi matter multiplets as discussed above. Following \cite{Franco:2016tcm, Closset:2017yte, Closset:2023bdr}, the partition function of this quiver SMM can be written in the following JK residue form:
\begin{equation}\label{0d JK par func}
    \mathcal{I}_{w}^{({\rm 0d})}(\sigma, m)\,=\,\oint_{\rm JK} \;(dM) \;Z_{\rm chiral}^{\rm vec}\;Z_{\rm chiral}^{\rm hor}\;Z_{\rm Fermi}^{\rm black}\;Z_{\rm Fermi}^{\rm red}~.
\end{equation}
In the integrand, we have the contributions coming from the bifundamental matter multiplets:
\begin{align}
    \begin{split}
        &{Z}_{\rm chiral}^{\rm ver} \,:=\,\prod_{i=1}^{n-2}\,\prod_{j=1}^{n-1}\,\prod_{\alpha=1}^{{\rm r}_{i,j}}\,\prod_{\beta=1}^{{\rm r}_{i+1,j}}\,\left(\,{u^{(i,j)}_{\alpha}}-{u^{(i+1,j)}_{\beta}}\,\right)^{-1} ~,\\
        &{Z}_{\rm chiral}^{\rm hor} \,:=\,\prod_{i=1}^{n-1}\,\left[\,\prod_{\alpha=1}^{{\rm r}_{i,n-1}}\,\prod_{a = 1}^{i}\,\left(\,{u_{\alpha}^{(i,n-1)}}-{\widetilde{\sigma}_{a}^{(i)}}\,\right)^{-1}\,\prod_{j=1}^{n-2}\,\prod_{\beta=1}^{{\rm r}_{i,j}}\,\prod_{\gamma=1}^{{\rm r}_{i,j+1}}\,\left(\,{u^{(i,j)}_{\beta}}-{u^{(i,j+1)}_{\gamma}}\,\right)^{-1}\,\right] ~,\\ 
        &{Z}^{\rm black}_{\rm Fermi} \,:=\, \prod_{i=1}^{n-2}\,\left[\,\prod_{\gamma=1}^{{\rm r}_{i,n-1}}\prod_{a=1}^{i+1}\left(\,{u^{(i,n-1)}_\gamma}-{\widetilde{\sigma}_{a}^{(i+1)}}\,\right)\prod_{j = 1}^{n-1}\,\prod_{\alpha=1}^{{\rm r}_{i,j}}\,\prod_{\beta = 1}^{{\rm r}_{i+1,j+1}}\,\left(\,{u^{(i,j)}_\alpha}-{u_\beta^{(i+1,j+1)}}\,\right)\,\right]\,~,\\
        &{Z}^{\rm red}_{\rm Fermi} \,:=\,\prod_{i=1}^{n-1}\,\prod_{\alpha=1}^{{\rm r}_{n-1,i}}\,\left(\,{u^{(n-1,i)}_\alpha}-{m_{n-i}}\,\right) ~.
    \end{split}
\end{align}
As for the measure $(dM)$ in \eqref{0d JK par func}, it is given more explicitly by:
\begin{equation}
    (dM) \,=\, \prod_{i=1}^{n-1}\,\prod_{j=1}^{n-1}\,\left[\,\Delta^{(i,j)}_{\rm 0d}(u)\,\frac{1}{{\rm r}_{i,j}\,!}\,\prod_{\alpha=1}^{{\rm r}_{i,j}}\,\frac{du_\alpha^{(i,j)}}{2\pi i }\,\right]~,
\end{equation}
with $\Delta_{\rm 0d}^{(i,j)}$ being the Vandermonde determinant factor:
\begin{equation}
    \Delta_{\rm 0d}^{(i,j)}(u)\,:=\,\prod_{1\leq \alpha\neq \beta\leq {\rm r}_{i,j}}\,\left(u^{(i,j)}_\alpha - u_\beta^{(i,j)}\right)~.
\end{equation}
Note that the poles in \eqref{0d JK par func} come from the chiral matter fields, and the JK prescription in this case is simply to consider the contributions coming from all of them.

Analogously to the remarks we made in the Schubert line defects case in subsection \ref{subsec:general proposal}, to obtain the final expression $Z_{\rm Fermi}^{\rm red}(\sigma,m)$, one needs to cancel the contributions coming from the massive chiral fields with their massive Fermi counterparts in the last row in figure~\ref{fig:general proposal}. Additionally, to match with the quantum double Schubert polynomials discussed in appendix \ref{subsec: Schubert polys}, we use the quantum relations \eqref{QCoH Fl(n) II} after performing the residue calculation.\footnote{Similarly to the comment we made earlier for the Schubert line defects, using the classical cohomology relations \protect\eqref{CoH Fl(n) II} rather than the quantum ones, we obtain the corresponding double Schubert polynomials which are the Poincar\'e duals to the Schubert classes in the classical cohomology ring of Fl$(n)$. For definitions and examples, see appendix \protect\ref{subsec: Schubert polys}.}

\subsection{Example: Schubert defects for Fl\texorpdfstring{$(3)$}{3}}
As a first example, let us consider the case with $n=3$. The quiver defects for each one of the six permutations in $S_3$ are given in figure~\ref{fig:Hasse S3}. From these, one can compute the corresponding partition function. Let us go through these six cases:
\begin{itemize}
    \item \underline{For $w \,=\, (1\,2\,3)$.} Following our discussion around \eqref{123rMatrix}, and using the general form \eqref{0d JK par func}, we see that, in this case, the 0d partition function is:
    \begin{equation}
        \mathcal{I}_{(1\,2\,3)}^{(\rm 0d)} (\widetilde{\sigma}, m)\,=\,\oint\frac{du}{2\pi i}\,\frac{u\,-\,m_1}{(\,u\,-\,\sigma_1^{(2)}\,)\,(\,u\,-\,\sigma_2^{(2)}\,)} \,=\,\frac{\sigma_1^{(2)}\,-\,m_1}{\sigma_1^{(2)}\,-\,\sigma_2^{(2)}}\,+\,\frac{\sigma_2^{(2)}\,-\,m_1}{\sigma_2^{(2)}\,-\,\sigma_1^{(2)}}\,=\,1~.
    \end{equation}
where, for brevity, we introduced $u\equiv u_1^{(2,2)}$. The two terms appearing in the second equality are the contributions of the two simple poles in $u$ coming from the 1d bifundamental chiral multiplet.

    \item \underline{For $w \,=\, (2\,1\,3)$.}
    The integrand in this case is:
    \begin{equation*}
 \frac{(u_1^{(2,2)}-m_1)\prod_{a=1}^2(u_{1}^{(1,2)}-\widetilde{\sigma}_{a}^{(2)})}{(u_1^{(1,2)}-u_1^{(2,2)})(u_1^{(1,2)}-\widetilde{\sigma}_1^{(1)})\prod_{b=1}^2(u_{1}^{(2,2)}-\widetilde{\sigma}_b^{(2)})} ~.
    \end{equation*}
    Performing the residue integral, we get:
    \begin{equation}
   \mathcal{I}^{(\rm 0d)}_{(2\,1\,3)} (\widetilde{\sigma}, m)\, =\, \widetilde{\sigma}^{(1)}_1 \,-\, m_1 \quad \implies \quad \mathcal{I}^{(\rm 0d)}_{(2\,1\,3)} (\sigma, m;\qcoh)\, =\, \sigma_1\, - \,m_1~.
    \end{equation}
    To obtain the second result, one also needs to use the relation \eqref{sigma1=sigma11}. 
\item \underline{For $w \,= \,(1\,3\,2)$.} In this case, the integrand is of the form:
    \begin{equation*}
       \frac{(u_1^{(2,1)}-m_1)(u_1^{(2,2)}-m_2)}{(u_1^{(2,1)}-u_1^{(2,2)})(u_1^{(2,2)}-\widetilde{\sigma}_1^{(2)})(u_1^{(2,2)} - \widetilde{\sigma}_2^{(2)})}~.
    \end{equation*}
    Similarly to the above, this residue can be computed easily. Upon using the first relation in \eqref{QCoH Fl(3)}, one obtains:
    \begin{equation}
        \CI_{(1\,3\,2)}^{(\rm 0d)}({\sigma}, m;\qcoh) \,=\, \sigma_1 \,+\, \sigma_2\, -\, m_1\, -\,m_2~.
    \end{equation}
    \item \underline{For $w \,= \,(2\,3\,1)$.} In this case, we have that the integrand of the 0d partition function is given by:
    \begin{align*}
        \frac{\Delta_{\rm 0d}^{(2,2)}(u)\,(u^{(2,1)}_{1}-m_2)\prod_{a=1}^2(u^{(2,2)}_{1}-m_1)\prod_{b=1}^2(u^{(1,2)}_1-\widetilde{\sigma}^{(2)}_{b})}{(u^{(1,2)}_1 - \widetilde{\sigma}^{(1)}_1)\prod_{a=1}^2(u^{(2,1)}_1-u^{(2,2)}_a)\prod_{a=1}^2(u_1^{(1,2)}-u_a^{(2,2)})\prod_{a,b=1}^2(u^{(2,2)}_a-\widetilde{\sigma}^{(2)}_b)}~.
    \end{align*}

    Taking the residue of this, we obtain the following form for the 0d index:
    \begin{equation}
        \CI^{(\rm 0d)}_{(2\,3\,1)}(\widetilde{\sigma},m)\, =\, (\widetilde{\sigma}_1^{(2)}\,-\,m_1)\,(\widetilde{\sigma}_2^{(2)}\,-\,m_1)~.
    \end{equation}
    Using the two relations in \eqref{QCoH Fl(3)}, we obtain:
    \begin{equation}
         \CI^{(\rm 0d)}_{(2\,3\,1)}(\sigma,m;\qcoh) \,=\, (\sigma_1\,-\,m_1)(\sigma_2\,-\,m_1)\,-\,\qcoh_1~,
    \end{equation}
\item \underline{For $w\, =\, (3\,1\,2)$.} 
    The partition function for this quiver gauge theory has the following integrand:
    \begin{equation*}
        \frac{(u_1-u_4)(u_4-m_1)(u_3-m_2)\prod_{a=1}^2(u_2-\widetilde{\sigma}_a^{(2)})}{(u_1-u_2)(u_1-u_3)(u_3-u_4)(u_2-\widetilde{\sigma}_1^{(1)})(u_2-u_4)\prod_{a=1}^2(u_4-\widetilde{\sigma}_a^{(2)})}~,
    \end{equation*}
    where here, for compactness, we use the notation:
    \begin{equation}\label{u notation}
        u_1 \equiv u_{1}^{(1,1)}~, \quad u_2\equiv u_1^{(1,2)}~, \quad u_3\equiv u_1^{(2,1)}~, \quad u_4\equiv u_1^{(2,2)}~.
    \end{equation}
   The residue integral with this integrand can be performed recursively, starting with $u_1$ and ending with $u_4$. Doing so, one obtains:
    \begin{equation}
        \CI_{(3\,1\,2)}^{(\rm 0d)}(\widetilde{\sigma}, m)\, =\, -\,\widetilde{\sigma}_1^{(2)}\,\widetilde{\sigma}_{2}^{(2)} \,+\, \widetilde{\sigma}_1^{(1)}\, (\widetilde{\sigma}_1^{(2)}\, +\, \widetilde{\sigma}_2^{(2)}\,-\,m_1\,-\,m_2) \,+\,m_1\,m_2~,
    \end{equation}
    which, upon using \eqref{sigma1=sigma11} and \eqref{QCoH Fl(3)}, gives us:
    \begin{equation}
        \CI_{(3\,1\,2)}^{(\rm 0d)}(\sigma, m;\qcoh) \,=\, (\sigma_1\,-\,m_1)(\sigma_1\,-\,m_2)\,-\,\qcoh_1~.
    \end{equation}
    \item \underline{For $w \,=\,(3\,2\,1) $}. In this case, the integrand has the following form:
    {\begin{equation*}
      {  \frac{
   \Delta_{\rm 0d}^{(2,2)}(u)\,
   (u_{3}-m_2)
  \prod_{a=4}^5(u_{a}-m_1)(u_{1}-u_{a})(u_{2}-\widetilde{\sigma}
   ^{(2)}_{a})}{(u_{1}-u_{2}) (u_{1}-u_{3})
   (u_{2}-\widetilde{\sigma}^{(1)}_{1}) \prod_{a=4}^5   (u_{2}-u_{a}) (u_{3}-u_{a})\prod_{b=1}^2  (u_{a}-\widetilde{\sigma}^{(2)}_{b})}}~.
    \end{equation*}}
Here, we used the notation \eqref{u notation} along with $u_5\equiv u_2^{(2,2)}$.

    Computing the corresponding residue, we obtain the following expression for the 0d partition function:
    \begin{equation}
        \mathcal{I}_{(3\,2\,1)}^{(\rm 0d)}(\widetilde{\sigma}, m) \,=\,(\widetilde{\sigma}^{(1)}_{1}\,-\,m_2) (\widetilde{\sigma}^{(2)}_{1}\,-\,m_1)
   (\widetilde{\sigma}^{(2)}_{2}\,-\,m_1)~.
    \end{equation}
    Upon using the quantum cohomology relations \eqref{sigma1=sigma11} and \eqref{QCoH Fl(3)}, one finds that the above expressions can be written as follows:
    \begin{equation}
        \CI_{(3\,2\,1)}^{(\rm 0d)}(\sigma, m;\qcoh) \,=\, (\sigma_1\,-\,m_1)(\sigma_1\,-\,m_2)(\sigma_2\,-\,m_1) \,+\,\qcoh_1 \,\left(\sigma _1\,-\,m_2\right)~.
    \end{equation}
\end{itemize}
Note that, after using the quantum ring relations, these partition functions match exactly with the corresponding quantum double Schubert polynomials $\mathfrak{S}_{w}^{(\qcoh)}(\sigma, m)$ listed in \eqref{quantum-doub-schub-S3}.

\medskip
\noindent
As another example, in appendix~\ref{app:Fl(4) defects} we work out the partition functions of the SMM defects in Fl$(4)$ and check that they match with the quantum double Schubert polynomials upon using the quantum relations  \eqref{sigma1=sigma11} and \eqref{QCoH Fl(4)}.

\medskip
\noindent
\textbf{\underline{Conjecture II:}} We conjecture that, for any $w\in S_n$, the 0d partition function \eqref{0d JK par func} is a residue formula representation for the quantum double Schubert polynomial. Namely:
\begin{equation}
    \mathcal{I}^{(\rm 0d)}_w(\sigma, m;\qcoh)\,=\,\mathfrak{S}_w^{(\qcoh)}(\sigma, m)~.
\end{equation}
It is understood here that $ \mathcal{I}^{(\rm 0d)}_w(\sigma, m;\qcoh)$ is the final form of $ \mathcal{I}^{(\rm 0d)}_w(\sigma, m)$ upon using the quantum relations \eqref{QCoH Fl(n) II}.

\section{Conclusions}

In this work, we studied 3d $\mathcal{N}=2$ supersymmetric gauged linear sigma models on $\mathbb{R}^2\times S^1_\beta$ whose target space is the complete flag manifold $X = $ Fl$(n)$. We proposed a description of half-BPS line operators wrapping ${S}_\beta^1$ in terms of 1d $\mathcal{N}=2$ supersymmetric quiver gauge theories. We argued that, in the IR, these line defects flow to the Schubert classes in the quantum K-theory ring of Fl$(n)$. We provided two strong pieces of evidence for our proposal:
\begin{itemize}
    \item Solving the 1d vacuum equations, we find that they restrict the target $X$, at the point of insertion of the line, to a particular Schubert variety $X_w \subseteq X$ indexed by the permutation $w$ in terms of which the quiver quantum mechanical system is defined. Furthermore, we found that the 1d-3d system actually engineers a resolution $\t X_w\rightarrow X_w$ of the Schubert variety.
    \item Computing the 1d index of the quiver gauge theories indexed by $w\in S_n$, we argued that one obtains the quantum Grothendieck polynomials $\mathfrak{G}^{(\qk)}_w(x,y)$, which are none other than the Chern characters of the structure sheaf classes $[\mathcal{O}_w]\in {\rm QK}(X)$ of $X_w$.
\end{itemize}
Taking the small circle limit $\beta \rightarrow 0$, we also proposed a coupled 0d-2d system, where now the bulk theory is the 2d GLSM with target $X$ and the point defect is defined in terms of a $\CN=2$ supersymmetric quiver matrix model. We showed that, computing the 0d partition function, one obtains the quantum Schubert polynomials $\mathfrak{S}_w^{(\qcoh)}(\sigma,m)$. These represent the Schubert classes $[X_w]\in{\rm QH}^\bullet(X)$.

In the companion paper~\cite{Closset:2026bnk}, we generalize this construction of Schubert line defects to arbitrary partial flag manifolds. More generally, it would be very interesting to construct even more general line defects that would directly engineer arbitrary coherent sheaves over some target space $X$ of a general 3d GLSM. In addition, already in the case of the complete flag manifold $X={\rm Fl}(n)$, it would be interesting to compute the quantum K-theory ring coefficients in~\eqref{twisted chiral ring L} directly in the Schubert line basis, using the 3-point functions in the 3d $A$-model. While the necessary physics formalism is well developed~\cite{Closset:2023bdr}, carrying out such computations efficiently for $n>4$ might require clever tricks or advances in computational algebraic geometry. We leave these questions as challenges for future work.

\section*{Acknowledgments}
We would like to thank I.~Melnikov, L.~Mihalcea, J.~Wynne, and W.~Xu for useful discussions.
C.~Closset is a Royal Society
University Research Fellow.  The research of W.~Gu is funded by the National Natural Science Foundation of China (NSFC) with Grant
No.12575077. O.~Khlaif is a Junior Research Fellow at the Philippe Meyer Institute, and, in the early stages of this project, he was supported by the School of Mathematics at the University of Birmingham.
E.~Sharpe and H.~Zhang were partially supported by NSF
grant PHY-2310588. H.~Zou was partially supported by the National Natural Science Foundation of China (NSFC) with Grant No.~12405083 and~12475005 and the Shanghai Pujiang Program with Grant No.~24PJA119.



\appendix

\section{Schubert classes for complete flag manifolds}\label{app:Schubert classes}
In this appendix, we review the (equivariant) cohomological and K-theoretical Schubert classes in the complete flag variety Fl$(n)$. These are represented, respectively, by the (double) Schubert and Grothendieck polynomials. Moreover, we discuss the quantum version of these classes, which are represented by the quantum double Schubert polynomials in the quantum cohomology case, and by the quantum double Grothendieck polynomials in the quantum K-theory case. 

\medskip
\noindent
\textbf{A comment on permutations.} Let us recall here that a special type of permutations are the \textit{transpositions} $s_{ij}$. These exchange the elements at positions $i$ and $j$ while keeping the other elements fixed. For the case where $i$ and $j$ are consecutive, we denote the transposition by $s_i \equiv s_{i, i+1}$. The permutation group $S_{n}$ is generated by these transpositions. A reduced word of the permutation $w\in S_n$ is of the form $w = s_{i_1} \cdots s_{i_{\ell(w)}}$, where $\ell(w)$ is the length of the permutation. The length can also be defined as
\be\label{def ellw}
\ell(w)\,:=\, \#\{\,i\,<\,j \,:\,w(i)\, >\, w(j)\,\}~.
\ee
In particular, we have $\ell({\rm id})=0$ for the identity permutation ${\rm id}=(\,1\,2\,\cdots\,n\,)$ and $\ell(w_0)=n(n-1)/2$ for the longest permutation $w_0=(\,n\, \cdots\, 2\,1\,)$.

\subsection{Schubert classes in the cohomology of the complete flag}\label{subsec: Schubert polys}
In this section, we review the Chern characters associated with the Schubert classes in the (quantum) cohomology of the complete flag manifold Fl$(n)$. These are called the (quantum) double Schubert polynomials, which are polynomials in the Chern roots of the universal quotient bundles $\mathcal{Q}_i$ and in the equivariant parameters of the $T = U(1)^{n-1}\subset SU(n)$ torus action. As discussed in the bulk of this paper, we denote the equivariant parameters by $m_i\in\mathbb{C}$ for $i=1, \cdots, n$ with the constraint that $\sum_{i=1}^{n}m_i=0$. From the 2d perspective, these correspond to twisted masses for the fundamental matter multiplets $\phi_{n-1}^{n}$ --- see the discussion around figure \ref{fig:complete Flag quiver} for more details.

To define these polynomials, let us start by defining the \textit{divided difference operators} $\partial_i$ associated with transposition $s_i$ as follows~\cite{lascoux1982structure}:
\begin{equation}\label{divided-difference}
	\partial^{(\sigma)}_{i}\, f(\sigma, m) \,=\, \frac{f(\sigma, m)\, - s_i^{(\sigma)}\,\cdot\, f(\sigma, m)}{\sigma_i \,-\, \sigma_{i+1}}~,
\end{equation}
for some polynomial $f(\sigma, m)$. It is understood here that the superscripts indicate with respect to which of the variables ($\sigma$ or $m$) we are deriving. The action of the transposition in the numerator is understood as follows:
\begin{equation}\label{transp-action}
	s_{i}^{(\sigma)} \,\cdot\, f(\sigma_1, \cdots,  \sigma_{i}, \sigma_{i+1}, \cdots, \sigma_{n}, m) \,=\,  f(\sigma_1, \cdots,  \sigma_{i+1}, \sigma_{i}, \cdots, \sigma_{n}, m) ~.
\end{equation}
That is, we swap $\sigma_i$ and $\sigma_{i+1}$. 
For a permutation $w\in S_{n}$ with a reduced word $w = s_{i_1}\cdots s_{i_{\ell(w)}}$,
 the associated divided difference operator $\partial_w$ is defined as:
\begin{equation}\label{partial-w}
	\partial_w \,\equiv\, \partial_{i_1}\, \cdots \,\partial_{i_{\ell(w)}}~.
\end{equation}

\subsubsection{Double Schubert polynomials}
Now we come to defining the \textit{double Schubert polynomial} $\mathfrak{S}_{w}(\sigma, m)$ \cite{lascoux1982structure, LS82b, MacDonald1991NotesOS, Fulton92, lascoux2006symmetry}. For the longest permutation $w_0\in S_{n}$, {\it i.e.}~$w_0 = (n\; \cdots\;2 \; 1)$, we have:
\begin{equation}\label{double schubert w0}
    \mathfrak{S}_{w_0}(\sigma, m) \,= \,\prod_{j=1}^{n-1}\, \det\, \left(\,D_j\, -\, m_{n-j}\, \mathbb{I}_j\,\right)~,
\end{equation}
where, the matrix $D_{n}$ is defined as the $n\times n$ matrix with $\sigma_1, \cdots, \sigma_{n}$ along the diagonal and $-1$ on the upper-diagonal. $D_j$ is the upper-left $j\times j$ sub-matrix of $D_{n}$. Expanding the determinant factor in \eqref{double schubert w0}, we find that:
\begin{equation}\label{double schubert w0 exp}
    \mathfrak{S}_{w_0}(\sigma, m) \,=\, \prod_{j=1}^{n-1}\, \prod_{i=1}^{j}\, \left(\,\sigma_i \,-\, m_{n-j}\,\right)\, =\, \prod_{j=1}^{n-1}\, \left[\,\sum_{i=0}^{j}\, (-1)^{j-i}\, m_{n-j}^{j-i}\, e_{i}^{j}(\sigma)\,\right]~,
\end{equation}
where
\begin{equation}\label{elementary-polynomials}
    e_{i}^j(\sigma) \,:=\, e_i(\sigma_1, \cdots, \sigma_j)~,
\end{equation}
is the $i$-th symmetric polynomial in the $j$ variables $\sigma_1, \cdots, \sigma_j$. 
For a permutation $w\in S_{n}$, we define the corresponding double Schubert polynomial as:
\begin{equation}\label{double-Schubert-defn}
    \mathfrak{S}_{w}(\sigma, m)  \,=\,
   (-1)^{\ell(w^{-1} w_0)} \,\partial_{w^{-1} w_0}^{(m)}\, \mathfrak{S}_{w_0}(\sigma, m)~. 
\end{equation}

\medskip
\noindent
{\bf Example I: double Schubert polynomials in $S_3$.} Let us take the case $n=3$. In this case, we have $6$ possible permutations, and the corresponding double Schubert polynomials, apart from the one associated with the trivial permutation, have the following explicit form:
\begin{align}\label{double-schubert-polys-S3}
    \begin{split}
        &\mathfrak{S}_{(2\,1\,3)}(\sigma, m)  \,=\,\sigma_1 \,-\, m_1~,\\
        &\mathfrak{S}_{(1\,3\,2)}(\sigma, m)  \,=\, \sigma_1 \,+\,\sigma_2 \,- \,m_1\, -\,m_2~,\\
        &\mathfrak{S}_{(2\,3\,1)}(\sigma, m)  \,=\, (\,\sigma_1 \,-\, m_1\,)\,(\,\sigma_2\,-\,m_1\,)~,\\
        &\mathfrak{S}_{(3\,1\,2)}(\sigma, m)  \,=\,(\,\sigma_1\,-\,m_1\,)\,(\,\sigma_1\,-\,m_2\,)~,\\
        &\mathfrak{S}_{(3\,2\,1)}(\sigma, m)  \,=\,(\,\sigma_1\,-\,m_1\,)(\,\sigma_1\,-\,m_2\,)\,(\,\sigma_2\,-\,m_1\,)~.\\
    \end{split}
\end{align}

\medskip
\noindent
{\bf Example II: double Schubert polynomials in $S_4$.} For the case $n=4$ we have $24$ possible permutations. For instance, for four randomly selected permutations, we find:
\begin{align}\label{double-schubert-polys-S4}
    \begin{split}
        &\mathfrak{S}_{(1\,2\,4\,3)}(\sigma, m) \,=\, \sigma _1\,+\,\sigma _2\,+\,\sigma _3\,-\,m_1\,-\,m_2\,-\,m_3~,\\
        &\mathfrak{S}_{(3\,1\,4\,2)}(\sigma, m) \,=\, (\,\sigma _1\, -\,m_1\,) \,(\,\sigma _1\,-\,m_2\,)\,(\,\sigma _2\,+\,\sigma_3\, -\, m_1\,-\, m_2\,)~,\\
        &\mathfrak{S}_{(3\,1\,2\,4)}(\sigma, m) \,= \,(\,\sigma _1\,-\,m_1\,)\,(\,\sigma _1\,-\,m_2\,)\,(\,\sigma _2\,+\,\sigma _3\,-\,m_1\,-\,m_2\,)~,\\
         &\mathfrak{S}_{(2\,4\,1\,3)}(\sigma, m) \,=\, (\,\sigma _1\,-\,m_1\,) \,(\,\sigma _2\,-\,m_1\,) \,(\,\sigma _1\,+\,\sigma _2\,-\,m_2\,-\,m_3\,)~,\\
         &\mathfrak{S}_{(2\,3\,1\,4)}(\sigma, m) \,=\, (\,\sigma_1\,-\,m_1\,)\,(\,\sigma_2\,-\,m_1\,)~.
    \end{split}
\end{align}
Geometrically, the double Schubert polynomial $\mathfrak{S}_w(\sigma, m)$ is known to represent the Schubert class $[X_w]\in {\rm H}^\bullet_{T}({\rm Fl}(n);\Z)$ in the classical $T$-equivariant cohomology ring of the complete flag Fl$(n)$.

\subsubsection{Quantum double Schubert polynomials}
Let us now review the quantum version of \eqref{double-Schubert-defn}: the \textit{quantum double Schubert polynomial} $\mathfrak{S}^{(\qcoh)}_{w} (\sigma, m)$ \cite{kirillov2000quantum,fomin1997quantum,LS14} associated with a permutation $w\in S_{n}$. To define this, let us first define the quantum version of \eqref{double schubert w0} for the longest permutation. We have:
\begin{equation}\label{longest-perm-quantum-double-schubert}
    \mathfrak{S}_{w_0}^{(\qcoh)} (\sigma, m)\, =\, \prod_{j=1}^{n-1} \, \det\,\left(\,D^{(\qcoh)}_{j} \,-\,m_{n-j}\, \mathbb{I}_{j}\,\right)~,
\end{equation}
where the matrix $D^{(\qcoh)}_j$ is the upper-left $j\times j$ submatrix of the $n\times n$ matrix $D^{(\qcoh)}_{n}$ whose entries are: 
\begin{equation}\label{qD-matrix}
    (D^{(\qcoh)}_{n})_{i,j} \,=\, \sigma_{i}\,\delta_{i,j}\, +\, \qcoh_{j}\,\delta_{i, j+1}\, -\, \delta_{i+1,j}~, \qquad i, j \,=\, 1, \cdots, n~.
\end{equation}
One can easily see that setting the quantum parameters $\qcoh_i$ to zero recovers the classical case:
\begin{equation}
    \mathfrak{S}^{(\qcoh=0)}_{w_0}(\sigma, m)\, = \,\mathfrak{S}_{w_0}(\sigma, m)~.
\end{equation}
Similarly to~\eqref{double schubert w0 exp}, one can expand the determinant factor in \eqref{longest-perm-quantum-double-schubert} as:
\begin{equation}\label{quantum-elementary-polys-defn}
    \det\,\left(\,D_j^{(\qcoh)} \,-\, m_{n-j}\,\mathbb{I}_j\,\right) \,=\, \sum_{i=0}^{j}\,(-1)^{j-i}\, m_{n-j}^{j-i}\,{\rm E}_{i}^{j} (\sigma; \qcoh)~,
\end{equation}
where we introduced the \textit{quantum elementary symmetric polynomials} ${\rm E}_i^{j}(\sigma; \qcoh)$, which are usually defined as the coefficients of the expansion of the above determinant \cite{fomin1997quantum}. These polynomials have the obvious property that they reduce to the symmetric polynomials \eqref{elementary-polynomials} in the classical limit:
\begin{equation}
    {\rm E}_{i}^{j}(\sigma; \qcoh=0)\, =\, e_{i}^{j}(\sigma)~.
\end{equation}

\medskip
\noindent
{\bf Example III: Quantum symmetric polynomials for $n=4$.} As an example of these polynomials, we have:
\begin{align}
    \begin{split}
 &{\rm E}_1^4(\sigma;\qcoh) \,=\, e_1^4(\sigma)~,\\
 &{\rm E}^4_2(\sigma;\qcoh) \,=\, e_2^4(\sigma) \,+\, \qcoh_1\,+\,\qcoh_2\,+\,\qcoh_3 ~,\\
 &{\rm E}_3^4(\sigma;\qcoh) \,=\, e_3^4(\sigma) \,+\, \qcoh_2\, \sigma _1\,+\,\qcoh_3\, \sigma _1\,+\,\qcoh_3\, \sigma _2\,+\,\qcoh_1\, \sigma _3\,+\,\qcoh_1\, \sigma _4\,+\,\qcoh_2\, \sigma _4~, \\
 &{\rm E}_4^4(\sigma;\qcoh)\, =\,e_4^4(\sigma)\,+\, \qcoh_3\, \sigma _1 \,\sigma _2\,+\,\qcoh_2\, \sigma _1\, \sigma _4\,+\,\qcoh_1 \,\sigma _3\, \sigma _4\,+\,\qcoh_1\, \qcoh_3~. \\
    \end{split}
\end{align}
Looking at the form of the determinant $D^{(\qcoh)}_{j}$ in \eqref{qD-matrix}, it is not difficult to deduce that the quantum elementary polynomials satisfy the following recurrence relation:
\begin{equation}\label{recurrence-for-quantum-elementary-polys}
    {\rm E}_{i}^j(\sigma; \qcoh) \,=\, {\rm E}_i^{j-1}(\sigma; \qcoh)\, +\, \sigma_j\, {\rm E}_{i-1}^{j-1} (\sigma; \qcoh)\,+\, \qcoh_{j-1} \,{\rm E}_{i-2}^{j-2}(\sigma; \qcoh)~.
\end{equation}
Note the similarity between these equations and the quantum relations \eqref{QCoH Fl(n) II}.

\medskip
\noindent
Given \eqref{longest-perm-quantum-double-schubert}, one defines the quantum double Schubert polynomial for a permutation $w\in S_{n}$ as follows \cite{ciocan1997quantum,kirillov2000quantum}:
\begin{equation}\label{quantum-schub-poly-defn}
{\mathfrak{S}}_{w}^{(\qcoh)} (\sigma, m)\,  =\, (-1)^{\ell(w^{-1}w_0)} \,\partial^{(m)}_{w^{-1}w_0}\, \mathfrak{S}^{(\qcoh)}_{w_0}(\sigma, m)~.
\end{equation}

\medskip
\noindent
{\bf Example IV: quantum double Schubert polynomials for $S_3$.} As we did for the double Schubert case in \eqref{double-schubert-polys-S3} for $n=3$, we have the following quantum version:
\begin{align}\label{quantum-doub-schub-S3}
    \begin{split}
        &\mathfrak{S}^{(\qcoh)}_{(1\,3\,2)}(\sigma, m) \,=\, \mathfrak{S}_{(1\,3\,2)}(\sigma, m)~,\\
        &\mathfrak{S}^{(\qcoh)}_{(2\,1\,3)}(\sigma, m) \,=\, \mathfrak{S}_{(2\,1\,3)}(\sigma, m)~,\\
        &\mathfrak{S}^{(\qcoh)}_{(2\,3\,1)}(\sigma, m) \,=\,\mathfrak{S}_{(2\,3\,1)}(\sigma, m)\,+\, \qcoh_1 ~,\\
        &\mathfrak{S}^{(\qcoh)}_{(3\,1\,2)}(\sigma, m) \,=\, \mathfrak{S}_{(3\,1\,2)}(\sigma, m)\,-\,\qcoh_1~,\\
        &\mathfrak{S}^{(\qcoh)}_{(3\,2\,1)}(\sigma, m) \,=\,\mathfrak{S}_{(3\,2\,1)}(\sigma, m) \,+\,\qcoh_1 \,(\,\sigma _1\,-\,m_2\,) ~.\\
    \end{split}
\end{align}

\medskip
\noindent
{\bf Example V: quantum double Schubert polynomials for $S_4$.} Let us consider the case with $n=4$, for the same four random permutations as in~\eqref{double-schubert-polys-S4}. One finds:
\begin{align}\label{q-double-schubert-polys-S4}
    \begin{split}
        &\mathfrak{S}^{(\qcoh)}_{(1\,2\,4\,3)}(\sigma, m)  \,=\, \mathfrak{S}_{(1\,2\,4\,3)}(\sigma, m) ~,\\
        &\mathfrak{S}^{(\qcoh)}_{(3\,1\,4\,2)}(\sigma, m) \,=\, \mathfrak{S}_{(3\,1\,4\,2)}(\sigma, m) \,+\, \qcoh_1\,(\,\sigma_1\,-\,\sigma_3\,) ~,\\
        &\mathfrak{S}^{(\qcoh)}_{(3\,1\,2\,4)}(\sigma, m) \,=\, \mathfrak{S}_{(3\,1\,2\,4)}(\sigma, m)\,-\,\qcoh_1~,\\
        &\mathfrak{S}_{(2\,4\,1\,3)}^{(\qcoh)} (\sigma, m) \,=\,\mathfrak{S}_{(2\,4\,1\,3)}(\sigma, m)\,+\,\qcoh_1\,(\sigma _1\,+\,\sigma _2)\,-\, \qcoh_2\,(\sigma_1\,-\,m_1)\, \\
        &\qquad\qquad\qquad\qquad\qquad\qquad\qquad-\,m_2 \,\qcoh_1\,-\,m_3 \, \qcoh_1~,\\
        &\mathfrak{S}_{(2\,3\,1\,4)}^{(\qcoh)}(\sigma, m) \,=\, \mathfrak{S}_{(2\,3\,1\,4)}(\sigma, m)\,+\,\qcoh_1~.
    \end{split}
\end{align}
In the classical limit, these clearly reduce to the double Schubert polynomials in \eqref{double-schubert-polys-S4}.

\medskip
\noindent
{\bf Geometric interpretation.} From a geometric point of view, 
the quantum double Schubert polynomial $\mathfrak{S}^{(\qcoh)}_{w}(\sigma, m)$ represents the Schubert class of $[X_w]\in {\rm QH}^\bullet_T({\rm Fl}(n);\Z)$ in the $T$-equivariant quantum cohomology ring of Fl$(n)$.

\subsection{Schubert classes in the K-theory of the complete flag}\label{subsec: Grothendeick polys} 
In this subsection, we uplift the Schubert polynomials discussed above from the cohomology to the K-theory. In this case, the Schubert classes $[\mathcal{O}_w]$ will be the (quantum) K-theory classes of the structure sheaves of the Schubert variety $X_w$. The relevant polynomials will be the equivariant Chern characters of these sheaves. Recall that the equivariant Chern character provides a map:
\begin{equation}
    {\rm ch}_T\, :\, {\rm K}_{T}(X)\, \longrightarrow \,{\rm H}^\bullet_{T}(X)~.
\end{equation}

\medskip
\noindent
To define these polynomials, we start by defining the \textit{Demazure operator} (or isobaric divided difference operator). For a transposition $s_i\in S_{n}$, we denote the corresponding Demazure operator by $\pi_i$ and we define it explicitly as:
\begin{align}\label{pi x and pi y}
\begin{split}
        &\pi^{(x)}_i  \,:=\, 1\,-\,x_i\,\partial^{(x)}_i \, =\, \frac{x_{i+1}\,-\,x_{i}\, s^{(x)}_i}{x_{i+1}\, -\, x_{i}} ~, \\
        &\pi^{(y)}_i  \,:= \, 1\,+\,y_{i+1}\,\partial_i^{(y)}  \,=\, \frac{y_i\,-\,y_{i+1}\,s^{(y)}_i}{y_i\,-\,y_{i+1}}~.
        \end{split}
\end{align}
For a permutation $w\in S_{n}$ with a reduced word $w = s_{i_1}\cdots s_{i_{\ell(w)}}$, we define:
\begin{equation}
    \pi^{(x)}_{w} \,:=\, \pi^{(x)}_{i_1}\, \cdots\, \pi^{(x)}_{i_{\ell (w)}}~, \qquad\qquad \pi^{(y)}_w \,:=\,\pi^{(y)}_{i_1}\, \cdots\, \pi^{(y)}_{i_{\ell (w)}}~.
\end{equation}

\medskip
\noindent
\textbf{A comment on the physics conventions.} Note that, in the mathematical literature, the Demazure operator with respect to a variable $z$ is usually defined as:
\protect\begin{equation}
    \pi_i^{(z)}\, :=\, \partial_{i}^{(z)}\, \left(\,1\,-\,z_{i+1}\,\right) ~.
\end{equation}
To compare with our QFT calculations in the main text of this paper, we are here defining the operators $\pi_i^{(x)}$ and $\pi^{(y)}_i$ in \eqref{pi x and pi y} in terms of the `shifted Wilson line operators' \cite{Gu:2022yvj}
$z_i =1-x_i$ or $z_i=1-y_i^{-1}$, respectively. 

\subsubsection{Double Grothendieck polynomials}
To define the double Grothendieck polynomial for a permutation $w\in S_n$, we start by defining the double Grothendieck polynomial associated with the longest permutation $w_0\in S_{n}$ as \cite[(2.1)]{lascoux1982structure} (see also \cite[section 8]{lenart2006quantum}):
\begin{equation}\label{double-groth-longest-perm}
    \mathfrak{G}_{w_0}(x,y) \,:=\, \prod_{j=1}^{n-1}\, \prod_{i=1}^{j}\, \left(\,1\,-\,\frac{x_i}{y_{n-j}}\,\right)\,=\, \prod_{j=1}^{n-1}\, \left[\,1\, +\, \sum_{i=1}^{j} \,\frac{(-1)^{i}}{y^i_{n-j}} \,e_{i}^j(x)\,\right]~.
\end{equation}
The polynomials $e_i^j(x)$ are defined in \eqref{elementary-polynomials}. For a permutation $w\in S_{n}$, we define the associated double Grothendieck polynomial as:
\begin{equation}\label{double-groth-polynomial-defn}
    \mathfrak{G}_{w}(x, y)\,  :=\, \pi^{(y)}_{w^{-1} w_0}\,\mathfrak{G}_{w_0} (x,y)~.
\end{equation}

\medskip
\noindent
{\bf Example VIII: double Grothendieck polynomials in $S_3$.} For $n = 3$, we have the following list of double Grothendieck polynomials:
\begin{align}\label{double-groth-polys-S3}
    \begin{split}
        &\mathfrak{G}_{(2\,1\,3)}(x,y)\, =\, 1\,-\,\frac{x_1}{y_1}~,\\
        &\mathfrak{G}_{(1\,3\,2)}(x,y) \,=\,1\,-\,\frac{x_1 \,x_2}{y_1\, y_2} ~,\\
        &\mathfrak{G}_{(2\,3\,1)}(x,y) \,=\,1\,-\,\frac{x_1}{y_1}\,-\,\frac{x_2}{y_1}\,+\,\frac{x_1\, x_2}{y_1^2}~,\\
        &\mathfrak{G}_{(3\,1\,2)}(x,y) \,=\, 1\,-\,\frac{x_1}{y_1}\,-\,\frac{x_1}{y_2}\,+\,\frac{x_1^2}{y_1 \,y_2}~,\\
        &\mathfrak{G}_{(3\,2\,1)}(x,y) \,=\,1\,-\,\frac{x_1}{y_1}\,-\,\frac{x_1}{y_2}\,-\,\frac{x_2}{y_1}\,+\,\frac{x_1^2}{y_1 \,y_2}\,+\,\frac{x_1\,x_2 }{y_1\, y_2}\,+\,\frac{ x_1\,x_2}{y_1^2}\,-\,\frac{ x_1^2\,x_2}{y_1^2\,y_2}~.\\
    \end{split}
\end{align}

\medskip
\noindent
{\bf Example IX: double Grothendieck polynomials in $S_4$.} Let us now take $n=4$. From the definition \eqref{double-groth-polynomial-defn} we find the following polynomials:

\begin{align}\label{double-groth-polys-S4}
    \begin{split}
        &\mathfrak{G}_{(1\,2\,4\,3)}(x,y) \,=\,1\,-\,\frac{x_1 \,x_2 \,x_3}{y_1\, y_2\, y_3} ~,\\
        &\mathfrak{G}_{(3\,1\,4\,2)}(x,y) \,=\,1\,-\,\frac{x_1}{y_1}\,-\,\frac{x_1}{y_2}\,+\,\frac{x_1^2}{y_1\, y_2}\, -\,\frac{x_2\, x_3}{y_1\,y_2}\,+\,\frac{x_1\, x_2 \,x_3}{y_1^2\, y_2}\,+\,\frac{x_1\, x_2\, x_3}{y_1\, y_2^2}\, -\,\frac{x_1^2\, x_2\, x_3}{y_1^2\, y_2^2}~,\\
        &\mathfrak{G}_{(3\,1\,2\,4)}(x,y) \,=\, 1\,-\,\frac{x_1}{y_1}\,-\,\frac{x_1}{y_2}\,+\, \frac{x_1^2}{y_1\, y_2}~,\\
        &\mathfrak{G}_{(2\,4\,1\,3)}(x,y) \,=\, 1\,-\,\frac{x_1}{y_1}\,-\,\frac{x_2}{y_1}\,+\,\frac{x_1\, x_2}{y_1^2}\,-\,\frac{x_1\, x_2}{y_2\, y_3}\,+\,\frac{ x_1^2\, x_2}{y_1\, y_2 \,y_3}\,+\,\frac{x_1\, x_2^2 }{y_1\, y_2\, y_3}\,-\,\frac{x_1^2\, x_2^2 }{y_1^2 \,y_2\, y_3}~,\\
        &\mathfrak{G}_{(2\,3\,1\,4)}(x, y) \,=\, 1\,-\,\frac{x_1}{y_1}\,-\,\frac{x_2}{y_1}\,+\,\frac{x_1\, x_2}{y_1^2}~.
    \end{split}
\end{align}

\medskip
\noindent
\textbf{Geometric interpretation.} From a geometric point of view, the double Grothendieck polynomials are known to represent the equivariant Chern characters of the Schubert classes $[\mathcal{O}_w]\in {\rm K}_T(\text{Fl}(n))$, see {\it e.g.}~\cite{FL94}.

\medskip
\noindent
{\bf 2d limit and cohomology classes.}
To reduce these polynomials to their cohomological counterparts, recall the definition of the K-theoretic roots $x_i = e^{-\beta \sigma_i}$ for the quotient bundle and  $y_i = e^{-\beta m_i}$ for the weights of the $U(1)^{n-1}$ torus action. At the level of the double Grothendieck polynomial \eqref{double-groth-polynomial-defn}, we find that:
\begin{equation}\label{2d-limit-double-groth}
    \lim_{\beta\rightarrow 0}\,\left( \,\beta^{-\ell(w)}\, \mathfrak{G}_{w}\left(e^{-\beta \sigma_i},e^{-\beta m_i}\right)\,\right)\, =\, \mathfrak{S}_{w} (\sigma, m) ~.
\end{equation}
As an example of this, one can indeed check that, in this limit, the polynomials \eqref{double-groth-polys-S3} reduce to \eqref{double-schubert-polys-S3}, and the polynomials \eqref{double-groth-polys-S4} reduce to \eqref{double-schubert-polys-S4}.

\subsubsection{Quantum double Grothendieck polynomials}
Motivated by the form of the double Grothendieck polynomial in \eqref{double-groth-longest-perm}, and following the steps of \cite{lenart2006quantum}, let us ``quantize'' it in the following way:\footnote{Our convention for the quantum Grothendieck polynomials differs slightly from that of \protect\cite{lenart2006quantum} (see Example~3.19 there). 
Starting from our variables, one may perform the substitution $x_i \,\mapsto\, (1\,-\,q_i)\,(1\,-\,x_i)$ and $y_i\,\mapsto\,(1\,-\,y_i)^{-1}$ to match their expressions. Our notation appears more natural from the point of view of the 3d GLSM.}
\begin{equation}\label{q-double-groth-poly-longest-perm}
    \mathfrak{G}_{w_0}^{(q)} (x,y) \,:=\, \prod_{j=1}^{n-1}\, \left[\,1\, +\, \sum_{i=1}^{j}\, \frac{(-1)^{i}}{y^i_{n-j}}\, {\rm F}_{i}^j(x; q)\,\right]~,
\end{equation}
where ${\rm F}_i^j$ are the quantum Lenart--Maeno (LM) polynomials defined explicitly as:
\begin{equation}\label{LM polynomials}
    {\rm F}_{i}^j(x;\qk)\,:=\, \sum_{\substack{I\subseteq [j]\\|I| = i}}\, 
    \left[\,\prod_{k\in I}\,{x_k} \,\prod_{\substack{k\in I\\k+1\in I}}\,\frac{1}{1\,-\,q_k}\,\right]~,
\end{equation}
where here $[j]:= \{1, 2, \cdots, j\}$. In the classical limit $q_i \rightarrow 0$, this reduces to the elementary symmetric polynomials $e_{i}^j(x)$ that we used in defining the double Grothendieck polynomials \eqref{double schubert w0}.
With the expression~\eqref{q-double-groth-poly-longest-perm} in hand, we define the quantum double Grothendieck polynomial $\mathfrak{G}^{(q)}_{w}(x,y)$ for any permutation $w\in S_{n}$ as follows:
\begin{equation}\label{q-double-groth-polys-defn}
    \mathfrak{G}_w^{(q)} (x,y) \,:=\, \pi_{w^{-1} w_0}^{(y)}\,  \mathfrak{G}_{w_0}^{(q)} (x,y)~.
\end{equation}
Similar to the symmetric polynomials that we studied earlier in this appendix, the LM polynomials \eqref{LM polynomials} can be generated using the following $j\times j$ matrix:
\begin{equation}\label{G matrix}
    [{\rm G}_j(x;\qk)]_{k,l} \,=\, \begin{cases}
      x_k\,\prod_{s = l}^{k-1}\,q_s\,x_s~, \qquad &l\,<\,k~,\\
        x_k~, \qquad &l\,=\, k ~,\\
        -\,\frac{1}{1\,-\,q_k}~,\qquad &l \,=\, k\,+\,1~,\\
        0~,\qquad &{\rm otherwise}~,
    \end{cases}
\end{equation}
for $k,l=1, \cdots, j$. For example, for $j = 4$, the $4\times 4$ generating matrix G$_4$ has the following explicit form:
\begin{equation}
   {\rm G}_4(x;\qk) =  \left(
\begin{array}{cccc}
 x_1 ~~& ~~-\,\frac{1}{1\,-\,q_1} ~~&~~ 0 ~~&~~ 0 \\
 q_1 \,x_1\, x_2 ~~&~~ x_2 ~~&~~ -\,\frac{1}{1\,-\,q_2} ~~& ~~0 \\
 q_1 \,q_2 \,x_1 \,x_2\, x_3 ~~& ~~q_2 \,x_2 \,x_3~~ & ~~x_3~~ & ~~-\,\frac{1}{1\,-\,q_3} \\
 q_1\, q_2 \,q_3 \,x_1\, x_2\, x_3\, x_4~~ &~~ q_2 \,q_3\, x_2\, x_3\, x_4~~ &~~ q_3\, x_3\, x_4~~ & ~~x_4 \\
\end{array}
\right)~.
\end{equation}
Indeed, one can easily check that:
\begin{equation}
    \det\,\left(\,{\rm G}_j(x;q)\,-\,\lambda\, \mathbb{I}_{j\times j}\,\right)\, = \,\sum_{i=0}^j\, (-1)^{j-i}\, \lambda^{j-i} \,{\rm F}_{i}^j(x;\qk)~.
\end{equation}
Moreover, we observe that, in the classical $q_i\rightarrow 0$ limit, the matrix \eqref{G matrix} reduces to the matrix $D_j$ appearing in \eqref{double schubert w0}. 

Similarly to the recurrence relations \eqref{recurrence-for-quantum-elementary-polys}, one can show that the LM polynomials F$_i^j(x;\qk)$ satisfy the following recurrence relation:
\begin{equation}
    {\rm F}_i^j(x;\qk)\, =\, {\rm F}_i^{j-1} (x;\qk) \,+\, \frac{1}{1\,-\,\qk_{j-1}}\,x_j \,{\rm F}_{i-1}^{j-1}(x; \qk)\, -\, \frac{\qk_{j-1}}{1\,-\,\qk_{j-1}} \,x_j\, {\rm F}_{i-1}^{j-2}(x;\qk)~.  
\end{equation}
Note that this recurrence relation is similar to the QK relations of Fl$(n)$ \eqref{QK relations of flag n}. This has to do with the fact that these polynomials form a basis of the QK$({\rm Fl}(n))$ \cite{lenart2006quantum}. 

\medskip
\noindent
{\bf Example X: quantum double Grothendieck polynomials in $S_3$.} For $n = 3$, we have the following quantum double Grothendieck polynomials:
\begin{align}\label{q-double-groth-polys-S3}
    \begin{split}
        &\mathfrak{G}^{(q)}_{(2\,1\,3)}(x,y) \,=\, \mathfrak{G}_{(2\,1\,3)}(x,y)~,\\
        &\mathfrak{G}^{(q)}_{(1\,3\,2)}(x,y) \,=\, 1 \,-\, \frac{1}{1\,-\,q_1}\,\frac{x_1\, x_2}{ y_1\, y_2},\\
        &\mathfrak{G}^{(q)}_{(2\,3\,1)}(x,y) \,=\,1\,-\,\frac{x_1}{y_1}\,-\,\frac{x_2}{y_1}\,+\,\frac{1}{1\,-\,q_1}\,\frac{ x_1 \,x_2}{
   y_1^2}~,\\
        &\mathfrak{G}^{(q)}_{(3\,1\,2)}(x,y) \,=\,\left(\,1\,-\,\frac{x_1}{y_1}\,\right)\,\left(\,1\,-\,\frac{x_1}{y_2}\,\right)\, -\, \frac{q_1}{1\,-\,q_1}\, \frac{x_1 \,x_2}{y_1\,y_2},\\
        &\mathfrak{G}^{(q)}_{(3\,2\,1)}(x,y) \,=\,1\,-\,\frac{x_1}{y_1}\,-\,\frac{x_1}{y_2}\,-\,\frac{x_2}{y_1}\,+\,\frac{x_1^2}{y_1\, y_2}\,+\,\frac{x_1\,x_2 }{y_1 \,y_2}\,+\,\frac{1}{1\,-\,q_1}\,\frac{ x_1\,x_2}{y_1^2}\,-\,\frac{1}{1\,-\,q_1}\,\frac{ x_1^2\,x_2}{y_1^2\, y_2}~.\\
    \end{split}
\end{align}
In the limit $q_i\rightarrow0$, these reduce to the corresponding double Grothendieck polynomials.

\medskip
\noindent
\textbf{Example XI: quantum double Grothendieck polynomials in $S_4$.} Let us look at the case of $n=4$. We have the following quantum Grothendieck polynomials for some of the possible permutations:
\begin{align}\label{q-double-groth-S4}
    \begin{split}
        &\mathfrak{G}^{(\qk)}_{(1\,2\,4\,3)}(x,y) \,=\,1\,-\,\frac{1}{(1\,-\,q_1)(1\,-\,q_2)}\frac{x_1\, x_2 \,x_3}{y_1\, y_2\, y_3}~,\\
        &\mathfrak{G}^{(\qk)}_{(3\,1\,4\,2)}(x,y) \,=\,1\,-\,\frac{x_1}{y_1}\,-\,\frac{x_1}{y_2}\,+\,\frac{x_1^2}{y_1\, y_2} \,-\,\frac{1}{1\,-\,q_2}\frac{x_2\, x_3}{y_1\,y_2}\,+\,\frac{1}{(1\,-\,q_1)(1\,-\,q_2)}\frac{x_1\, x_2\, x_3}{y_1^2\, y_2}\\
        &\qquad \qquad \qquad+\,\frac{1}{(1\,-\,q_1)(1\,-\,q_2)}\frac{x_1\, x_2\, x_3}{y_1 \,y_2^2} -\frac{1}{(1\,-\,q_1)(1\,-\,q_2)}\frac{x_1^2\, x_2\, x_3}{y_1^2\, y_2^2}~,\\
        &\mathfrak{G}^{(\qk)}_{(3\,1\,2\,4)}(x,y) \,=\, \mathfrak{G}^{(\qk)}_{(3\,1\,2)}(x,y) ~,\\
        &\mathfrak{G}^{(\qk)}_{(2\,4\,1\,3)}(x,y) \,=\,1\,-\,\frac{x_1}{y_1}\,-\,\frac{x_2}{y_1}\,+\,\frac{1}{1\,-\,q_1}\frac{x_1\, x_2}{y_1^2}\,-\,\frac{1}{1\,-\,q_1}\,\frac{x_1\, x_2}{y_2\, y_3}\,+\,\frac{1}{1\,-\,q_1}\,\frac{ x_1^2\, x_2}{y_1\, y_2 \,y_3}\\
        &\qquad \qquad \qquad +\frac{1}{1\,-\,q_1}\,\frac{x_1\, x_2^2 }{y_1\, y_2\, y_3}-\frac{1}{(1\,-\,q_1)^2}\frac{x_1^2 \,x_2^2 }{y_1^2 \,y_2 \,y_3}~,\\
        &\mathfrak{G}^{(\qk)}_{(2\,3\,1\,4)}(x,y)\, =\, \mathfrak{G}_{(2\,3\,1)}(x,y)~,
    \end{split}
\end{align}
where, $\mathfrak{G}^{(q)}_{(3\,1\,2)}(x,y)$ and $\mathfrak{G}^{(q)}_{(2\,3\,1)}(x,y)$ are given in \eqref{q-double-groth-polys-S3}. Clearly, these reduce to \eqref{double-groth-polys-S4} in the classical limit.

\medskip
\noindent
\textbf{Geometric interpretation.} The quantum double Grothendieck polynomials are the $T$-equivariant Chern characters of the Schubert classes $[\mathcal{O}_w]$ in the $T$-equivariant quantum K-theory ring of the complete flag Fl$(n)$, see {\it e.g.}~\cite{lenart2006quantum}.

\medskip
\noindent 
{\bf 2d limit and quantum cohomology.} Similarly to what we did for the double Grothendieck polynomials earlier in \eqref{2d-limit-double-groth}, let us consider the 2d limit of the quantum double Grothendieck polynomials \eqref{q-double-groth-polys-defn}. 
As reviewed in the main text, we need to keep in mind that the quantum cohomology and quantum K-theory deformation parameters $\qcoh_i$ and $\qk_i$ are related via:
\begin{equation}\label{2d-limit-full-flag}
    {\qk}_i \,\equiv\, \beta^{2} \,\qcoh_i~, \qquad i\, =\, 1, \cdots, n-1~.
\end{equation}
Therefore, taking $x_i\,\equiv\,e^{-\beta \sigma_i}$ and $y_i \equiv e^{-\beta m_i}$, we find the following limit:
\begin{equation}\label{2d-limit-q-double-groth}
    \lim_{\beta\rightarrow 0}\,\left(\, \beta^{-\ell(w)}\, \mathfrak{G}^{({\qk})}_{w}\left(e^{-\beta \sigma_i},e^{-\beta m_i}\right)\,\right)\, =\, \mathfrak{S}^{({\qcoh})}_{w} (\sigma, m) ~,
\end{equation}
where the quantum double Schubert polynomial $\mathfrak{S}_{w}^{(\qcoh)}(\sigma, m)$ is defined in \eqref{quantum-schub-poly-defn}. 
For instance, in this limit, \eqref{q-double-groth-polys-S3} reduces to \eqref{quantum-doub-schub-S3}. Similarly, the polynomials \eqref{q-double-groth-S4} reduce to \eqref{q-double-schubert-polys-S4}.

\section{Resolution of Schubert varieties from mathematics}  \label{app:quiver-vs-math}

In this appendix, we explain how the geometry arising from our 1d-3d GLSM corresponds to two known constructions of resolutions of Schubert varieties: via bioriented flags and via quiver Grassmannians. 
The relation to bioriented flags is relatively direct, while for quiver Grassmannians we briefly unpack the relevant definitions before showing the equivalence.

\subsection{Resolved Schubert variety as a bioriented flag}
Following \cite{cibotaru2020bioriented}, the Schubert varieties with respect to a reference flag $E_\bullet$, are defined as:
\begin{equation}
    X^w(E_\bullet) \,:= \,\{F_\bullet\, \in\, \text{Fl}(n) \,\mid\, \dim(F_i \,\cap\, E_j)\, \geq\, \text{r}_{i,j}^w\}~, \quad {\rm r}^w_{i,j} \,=\,\#\{\,l\,\leq\,i\mid w(l)\,\leq\,j\,\}~.
\end{equation}
This definition differs slightly from our convention for $X_w$ in equation \eqref{Xw defn} --- the two conventions are equivalent up to relabeling: $X_{w} = X^{w_0w}$.
One introduces the product space:
\begin{equation}
    \mathcal{P}^w\, :=\, \prod_{i,j=1}^n \,\text{Gr}({\text{r}_{i,j}^w} , \mathbb{C}^n)~,
\end{equation}
whose points consist of products of subspaces $F_{i,j} \subset \mathbb{C}^n$ of dimension $\text{r}_{i,j}^w$. The \textit{bioriented flag} is defined as:
\begin{equation}
    \text{Fl}^w := \{F_{\bullet,\bullet} \in \mathcal{P}^w \mid F_{i,j} \subset F_{i,j+1}, ~F_{i,j} \subset F_{i+1,j}\}~.
\end{equation}
The authors of~\cite{cibotaru2020bioriented} further define: 
\begin{equation}
    \widehat{X}^w(E_\bullet)\, :=\, \{F_{\bullet, \bullet}\, \in\, \text{Fl}^w \,\mid\, F_{n,i}\, = \,E_i,~ i \,=\,1, 2,\cdots,n\}~.
\end{equation}
Our physical construction realizes precisely $\widehat{X}^{w_0 w}(E_\bullet)$, as is clear from figure~\ref{fig:quiver geometry}. Moreover, it is proved in \cite{cibotaru2020bioriented} that the projection:
\begin{equation}
    \widehat{X}^w (E_\bullet) \to \text{Fl}(n)~, \quad F_{\bullet,\bullet} \mapsto F_{\bullet, n}~,
\end{equation}
is a resolution of $X_w$, which is isomorphic to the Bott--Samelson resolution of $X_w$ corresponding to a certain word representation of $w$.

\subsection{Resolved Schubert variety as a quiver Grassmannian}
In this appendix, we would like to explain why the target space of our 1d-3d GLSM is also precisely a quiver Grassmannian $\text{Gr}_{{\bf r}^w}(M)$ in the sense of~\cite{iezzi2025quivergrassmanniansbottsamelsonresolution}.

To define the quiver Grassmannian, one starts with a specific quiver representation $M$ (with relations), then $\text{Gr}_{{\bf r}^w}(M)$ is the collection of all the quiver subrepresentations $N$ of $M$, whose dimension vectors are specified by ${\bf r}^w$. For the full definitions and technical details, we refer the reader to \cite{iezzi2025quivergrassmanniansbottsamelsonresolution}.
For our purposes, it will be sufficient to specify the quiver representation $M$ directly, without introducing the underlying quiver and its arrows explicitly. The quiver representation $M$ is given by:
\begin{equation*}
\begin{tikzcd}[sep=large]
\overset{E_1}{\bullet} \ar[r, "\text{id}"] \ar[d, "\iota_{2,1}"]  & \overset{E_1}{\bullet} \ar[r, "\text{id}"] \ar[d, "\iota_{2,1}"] & ...\ar[r, "\text{id}"] \ar[d, "\iota_{2,1}"] & \overset{E_1}{\bullet} \ar[d, "\iota_{2,1}"] \\
\overset{E_2}{\bullet} \ar[ur, phantom, "\scalebox{1.5}{$\circlearrowleft$}"] \ar[r, "\text{id}"] \ar[d, "\iota_{3,2}"] & \overset{E_2}{\bullet} \ar[ur, phantom, "\scalebox{1.5}{$\circlearrowleft$}"] \ar[r, "\text{id}"] \ar[d, "\iota_{3,2}"]  & ... \ar[ur, phantom,xshift=5, "\scalebox{1.5}{$\circlearrowleft$}"] \ar[r, "\text{id}"] \ar[d, "\iota_{3,2}"] & \overset{E_2}{\bullet} \ar[d, "\iota_{3,2}"] \\
... \ar[ur, phantom,xshift=5, "\scalebox{1.5}{$\circlearrowleft$}"] \ar[r, "\text{id}"] \ar[d, "\iota_{n,n\!-\!1}"] & ... \ar[ur, phantom,xshift=5, "\scalebox{1.5}{$\circlearrowleft$}"] \ar[r, "\text{id}"] \ar[d, "\iota_{n,n\!-\!1}"] & ... \ar[ur, phantom, xshift=5, "\scalebox{1.5}{$\circlearrowleft$}"] \ar[r, "\text{id}"] \ar[d, "\iota_{n,n\!-\!1}"] & ... \ar[d, "\iota_{n,n\!-\!1}"]\\
\overset{E_n}{\bullet} \ar[ur, phantom, "\scalebox{1.5}{$\circlearrowleft$}"] \ar[r, "\text{id}"] & \overset{E_n}{\bullet} \ar[ur, phantom, "\scalebox{1.5}{$\circlearrowleft$}"] \ar[r, "\text{id}"] & ... \ar[ur, phantom, xshift=5, "\scalebox{1.5}{$\circlearrowleft$}"] \ar[r, "\text{id}"] & \overset{E_n}{\bullet}
\end{tikzcd}.
\end{equation*}
Here each node is a vector space $M_{i,j} = E_j$, where $( 0 \subset E_1 \subset E_2 \subset \cdots \subset E_n = \mathbb{C}^n )$ forms a fixed full flag in $\mathbb{C}^n$. The horizontal arrows are identity maps, while the vertical arrows are inclusions $\iota_{i+1,i}: E_i \hookrightarrow E_{i+1}$.
By construction, all squares in the diagram commute. This commutativity realizes quiver relations that are otherwise specified as part of the quiver data (and follow from the $E$-term equations in our GLSM).

The quiver Grassmannian $\text{Gr}_{{\bf r}^w} (M)$ consists of subrepresentations $N$ of $M$. For each subrepresentation $N$, it consists of the following data:
\begin{itemize}
    \item We have subspaces $N_{i,j} \subseteq M_{i,j} = E_i$ and $\dim(N_{i,j}) = \text{r}^w_{i,j}$.
    \item For each arrow in $M$, the induced arrow in $N$ is the unique inclusion map between the corresponding subspaces, of the form $N_{i,j} \hookrightarrow N_{i+1,j}$ or $N_{i,j} \hookrightarrow N_{i,j+1}$.
    \item The commutativity relation
    $(N_{i,j} \hookrightarrow N_{i,j+1} \hookrightarrow N_{i+1,j+1}) = (N_{i,j} \hookrightarrow N_{i+1,j} \hookrightarrow N_{i+1,j+1})$
    is trivially satisfied.
\end{itemize}
Therefore, we have the representation $N$:
\begin{equation*}
\begin{tikzcd}[sep=large]
\overset{N_{1,1}}{\bullet} \ar[r, hook] \ar[d, hook]  & \overset{N_{1,2}}{\bullet} \ar[r, hook] \ar[d, hook] & ...\ar[r, hook] \ar[d, hook] & \overset{N_{1,n-1}}{\bullet} \ar[d, hook] \\
\overset{N_{2,1}}{\bullet} \ar[ur, phantom, "\scalebox{1.5}{$\circlearrowleft$}"] \ar[r, hook] \ar[d, hook] & \overset{N_{2,2}}{\bullet} \ar[ur, phantom, "\scalebox{1.5}{$\circlearrowleft$}"] \ar[r, hook] \ar[d, hook]  & ... \ar[ur, phantom,xshift=5, "\scalebox{1.5}{$\circlearrowleft$}"] \ar[r, hook] \ar[d, hook] & \overset{N_{2,n-1}}{\bullet} \ar[d, hook] \\
... \ar[ur, phantom,xshift=5, "\scalebox{1.5}{$\circlearrowleft$}"] \ar[r, hook] \ar[d, hook] & ... \ar[ur, phantom,xshift=5, "\scalebox{1.5}{$\circlearrowleft$}"] \ar[r, hook] \ar[d, hook] & ... \ar[ur, phantom, xshift=5, "\scalebox{1.5}{$\circlearrowleft$}"] \ar[r, hook] \ar[d, hook] & ... \ar[d, hook]\\
\overset{N_{n,1}}{\bullet} \ar[ur, phantom, "\scalebox{1.5}{$\circlearrowleft$}"] \ar[r, hook] & \overset{N_{n,2}}{\bullet} \ar[ur, phantom, "\scalebox{1.5}{$\circlearrowleft$}"] \ar[r, hook] & ... \ar[ur, phantom, xshift=5, "\scalebox{1.5}{$\circlearrowleft$}"] \ar[r, hook] & \overset{N_{n,n-1}}{\bullet}
\end{tikzcd},
\end{equation*}
where $\dim(N_{i,j}) = \text{r}^w_{i,j}$ and $N_{i,j} \subset E_i$. Crucially, due to the inclusion structure on each row, it suffices to require only $N_{i,n-1} \subseteq E_i$. Therefore, we can add a column corresponding to the fixed flag $(0 \subset E_1 \subset E_2 \subset \cdots \subset E_n = \mathbb{C}^n)$ to the above diagram and obtain (the inclusions in the last column are given by $\iota$'s):
\begin{equation*}
\begin{tikzcd}[sep=large]
\overset{N_{1,1}}{\bullet} \ar[r, hook] \ar[d, hook]  & \overset{N_{1,2}}{\bullet} \ar[r, hook] \ar[d, hook] & ...\ar[r, hook] \ar[d, hook] & \overset{N_{1,n-1}}{\bullet} \ar[r,hook] \ar[d, hook]   & \overset{E_1}{\bullet}  \ar[d, hook]  \\
\overset{N_{2,1}}{\bullet} \ar[ur, phantom, "\scalebox{1.5}{$\circlearrowleft$}"] \ar[r, hook] \ar[d, hook] & \overset{N_{2,2}}{\bullet} \ar[ur, phantom, "\scalebox{1.5}{$\circlearrowleft$}"] \ar[r, hook] \ar[d, hook]  & ... \ar[ur, phantom,xshift=5, "\scalebox{1.5}{$\circlearrowleft$}"] \ar[r, hook] \ar[d, hook] & \overset{N_{2,n-1}}{\bullet} \ar[r,hook] \ar[d, hook] \ar[ur, phantom,xshift=5, "\scalebox{1.5}{$\circlearrowleft$}"] & \overset{E_2}{\bullet} \ar[d, hook] \\
... \ar[ur, phantom,xshift=5, "\scalebox{1.5}{$\circlearrowleft$}"] \ar[r, hook] \ar[d, hook] & ... \ar[ur, phantom,xshift=5, "\scalebox{1.5}{$\circlearrowleft$}"] \ar[r, hook] \ar[d, hook] & ... \ar[ur, phantom, xshift=5, "\scalebox{1.5}{$\circlearrowleft$}"] \ar[r, hook] \ar[d, hook] & ... \ar[r,hook] \ar[d, hook] \ar[ur, phantom,xshift=5, "\scalebox{1.5}{$\circlearrowleft$}"] & ... \ar[d, hook] \\
\overset{N_{n,1}}{\bullet} \ar[ur, phantom, "\scalebox{1.5}{$\circlearrowleft$}"] \ar[r, hook] & \overset{N_{n,2}}{\bullet} \ar[ur, phantom, "\scalebox{1.5}{$\circlearrowleft$}"] \ar[r, hook] & ... \ar[ur, phantom, xshift=5, "\scalebox{1.5}{$\circlearrowleft$}"] \ar[r, hook]  & \overset{N_{n,n-1}}{\bullet} \ar[r,hook] \ar[ur, phantom,xshift=5, "\scalebox{1.5}{$\circlearrowleft$}"] & \overset{E_n}{\bullet} 
\end{tikzcd}.
\end{equation*}
Now we can forget about the subrepresentation condition $N_{i,j} \subset E_i$, and we only have the condition $\dim(N_{i,j}) = \text{r}^w_{i,j}$. The commutativity of the newly added squares follows from the fact that all the arrows are inclusions.

The above diagram with the conditions $\dim(N_{i,j}) = \text{r}^w_{i,j}$ is the same as the configuration in figure \ref{fig:quiver geometry}, up to a rotation of 90 degrees and a relabeling of the permutation. This rotation is due to a difference in conventions: the quantity denoted $\text{r}_{i,j}^w$ in \cite{iezzi2025quivergrassmanniansbottsamelsonresolution} corresponds to what we denote by $\text{r}_{j,i}^w$ in equation \eqref{rw defn}. The relabeling is due to the fact that we use $\text{r}^{w_0 w}_{i,j}$ instead of $\text{r}^w_{i,j}$ in the definition of Schubert varieties.

\newpage
\section{Schubert defects in Fl\texorpdfstring{$(4)$}{(4)}}\label{app:Fl(4) defects}
\subsection*{Coupled systems defining the Schubert defects}
\begin{table}[h]
\centering
\begin{subtable}{0.45\textwidth}
\centering
 \begin{equation*}
        \begin{array}{|c||c|c|c|}
    \hline
        w&{\rm General\,\,proposal} & {\rm Reduced \,\,quiver}\\
        \hline
        \hline
        \raisebox{9ex}{$(1\,2\,3\,4)$}& \includegraphics[scale = 0.65]{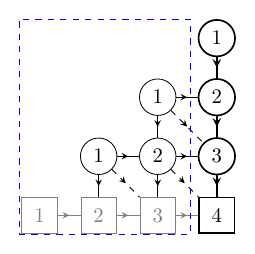}&\includegraphics[scale = 0.65]{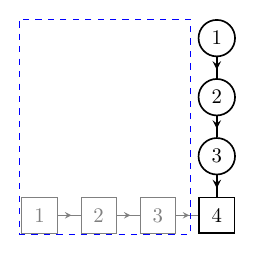}\\ 
\hline
\raisebox{9ex}{$(1\,2\,4\,3)$}& \includegraphics[scale = 0.65]{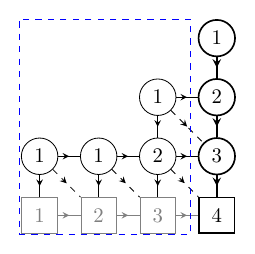}&\includegraphics[scale = 0.65]{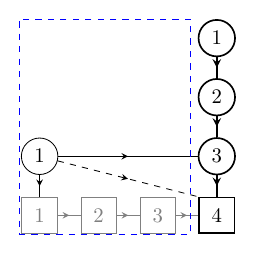}\\ 
\hline
\raisebox{9ex}{$(1\,3\,2\,4)$}& \includegraphics[scale = 0.65]{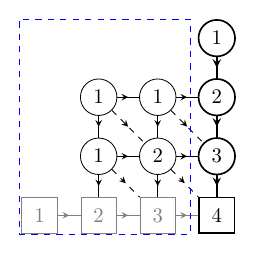}&\includegraphics[scale = 0.65]{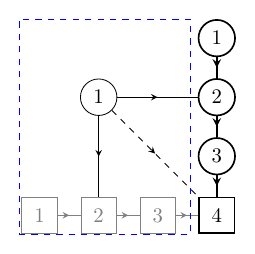}\\ 
\hline
\raisebox{9ex}{$(1\,3\,4\,2)$}& \includegraphics[scale = 0.65]{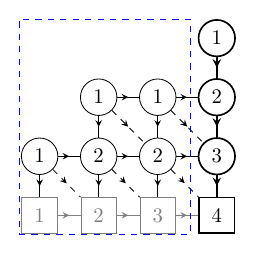}&\includegraphics[scale = 0.65]{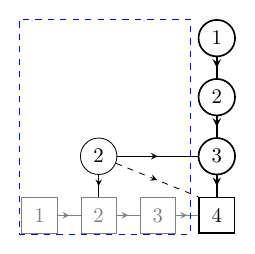}\\ 
\hline
\raisebox{9ex}{$(1\,4\,2\,3)$}& \includegraphics[scale = 0.65]{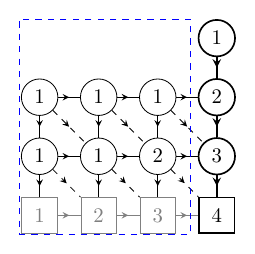}&\includegraphics[scale = 0.65]{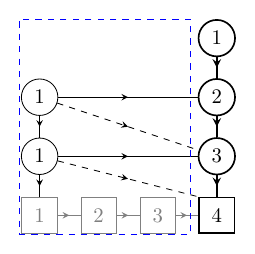}\\ 
\hline
\raisebox{9ex}{$(1\,4\,3\,2)$}& \includegraphics[scale = 0.65]{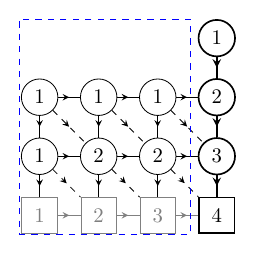}&\includegraphics[scale = 0.65]{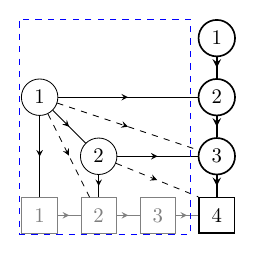}\\ 
           \hline
           
        \end{array}
    \end{equation*}
\end{subtable}
\hfill
\begin{subtable}{0.45\textwidth}
\centering
\begin{equation*}
        \begin{array}{|c||c|c|c|}
    \hline
        w&{\rm General\,\,proposal}&  {\rm Reduced \,\,quiver}\\
        \hline
        \hline
    \raisebox{9ex}{$(2\,1\,3\,4)$}& \includegraphics[scale = 0.65]{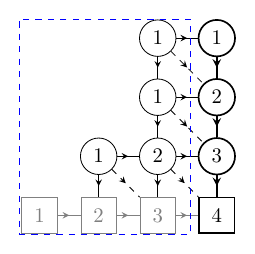}&\includegraphics[scale = 0.65]{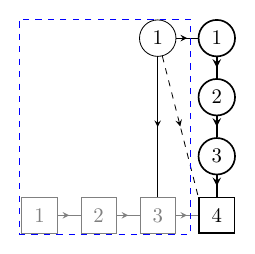}\\ 
\hline
\raisebox{9ex}{$(2\,1\,4\,3)$}& \includegraphics[scale = 0.65]{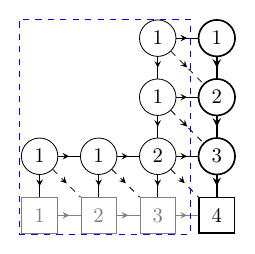}&\includegraphics[scale = 0.65]{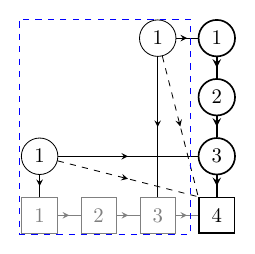}\\ 
\hline
\raisebox{9ex}{$(2\,3\,1\,4)$}& \includegraphics[scale = 0.65]{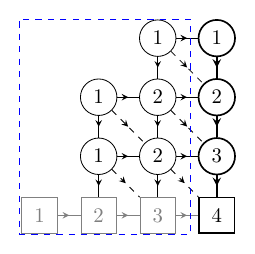}&\includegraphics[scale = 0.65]{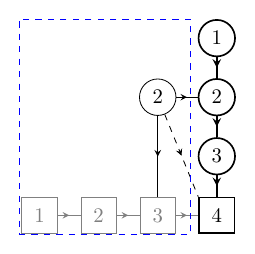}\\ 
\hline
\raisebox{9ex}{$(2\,3\,4\,1)$}& \includegraphics[scale = 0.65]{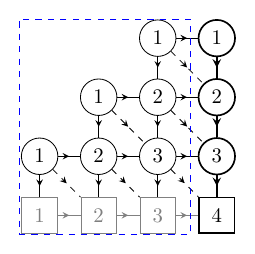}&\includegraphics[scale = 0.65]{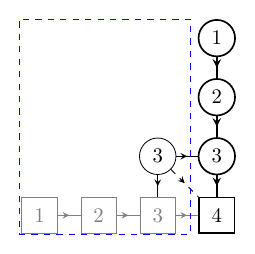}\\ 
\hline
\raisebox{9ex}{$(2\,4\,1\,3)$}& \includegraphics[scale = 0.65]{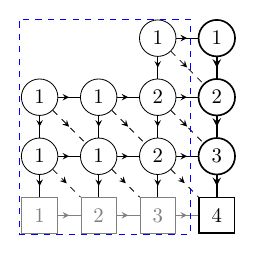}&\includegraphics[scale = 0.65]{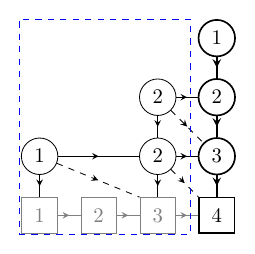}\\ 
\hline
\raisebox{9ex}{$(2\,4\,3\,1)$}& \includegraphics[scale = 0.65]{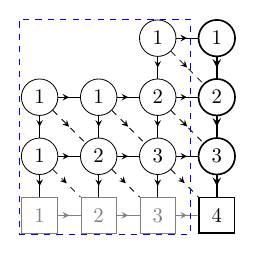}&\includegraphics[scale = 0.65]{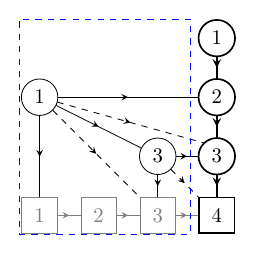}\\ 
           \hline
        \end{array}
    \end{equation*}
\end{subtable}
\end{table}

\newpage

\begin{table}[h]
\centering
\begin{subtable}{0.45\textwidth}
\centering
\begin{equation*}
        \begin{array}{|c||c|c|c|}
    \hline
        w&{\rm General\,\,proposal} & {\rm Reduced \,\,quiver}\\
        \hline
        \hline
\raisebox{9ex}{$(3\,1\,2\,4)$}& \includegraphics[scale = 0.65]{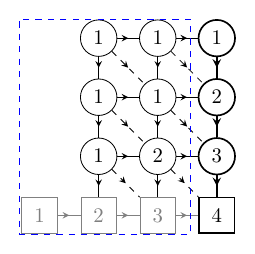}&\includegraphics[scale = 0.65]{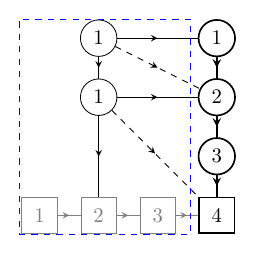}\\ 
\hline
\raisebox{9ex}{$(3\,1\,4\,2)$}& \includegraphics[scale = 0.65]{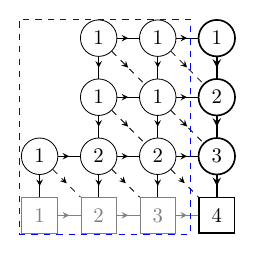}&\includegraphics[scale = 0.65]{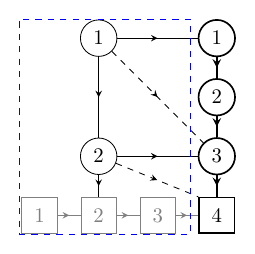}\\ 
\hline
\raisebox{9ex}{$(3\,2\,1\,4)$}& \includegraphics[scale = 0.65]{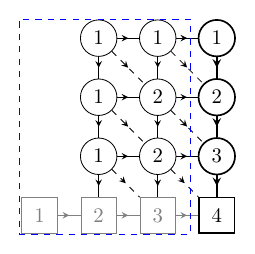}&\includegraphics[scale = 0.65]{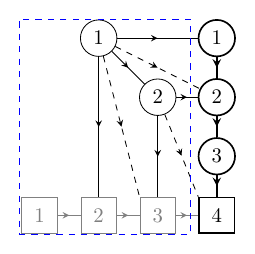}\\ 
\hline
\raisebox{9ex}{$(3\,2\,4\,1)$}& \includegraphics[scale = 0.65]{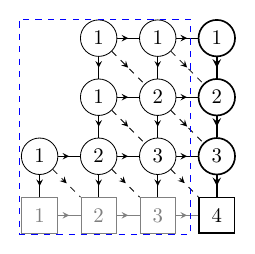}&\includegraphics[scale = 0.65]{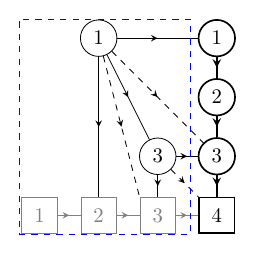}\\ 
\hline
\raisebox{9ex}{$(3\,4\,1\,2)$}& \includegraphics[scale = 0.65]{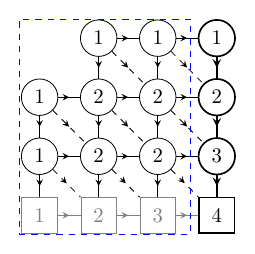}&\includegraphics[scale = 0.65]{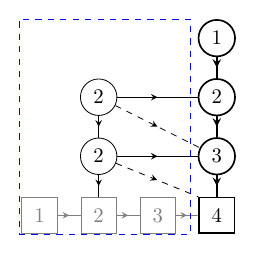}\\ 
\hline
\raisebox{9ex}{$(3\,4\,2\,1)$}& \includegraphics[scale = 0.65]{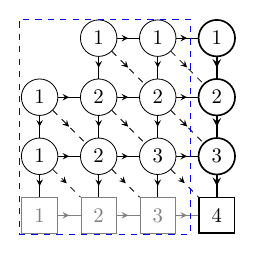}&\includegraphics[scale = 0.65]{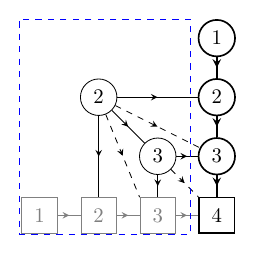}\\ 
           \hline
           
        \end{array}
    \end{equation*}
\end{subtable}
\hfill
\begin{subtable}{0.45\textwidth}
\centering
\begin{equation*}
        \begin{array}{|c||c|c|c|}
    \hline
        w&{\rm General\,\,proposal}&  {\rm Reduced \,\,quiver}\\
        \hline
        \hline
\raisebox{9ex}{$(4\,1\,2\,3)$}& \includegraphics[scale = 0.65]{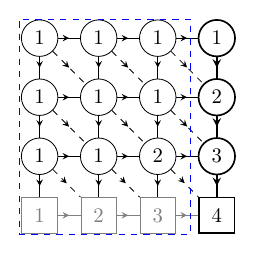}&\includegraphics[scale = 0.65]{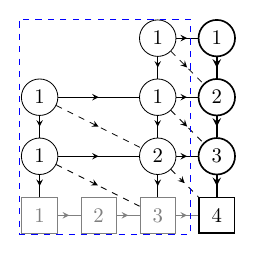}\\ 
\hline
\raisebox{9ex}{$(4\,1\,3\,2)$}& \includegraphics[scale = 0.65]{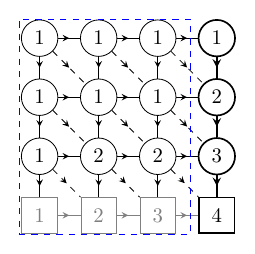}&\includegraphics[scale = 0.65]{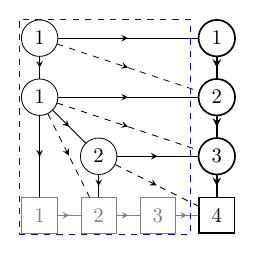}\\ 
\hline
\raisebox{9ex}{$(4\,2\,1\,3)$}& \includegraphics[scale = 0.65]{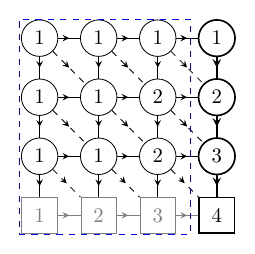}&\includegraphics[scale = 0.65]{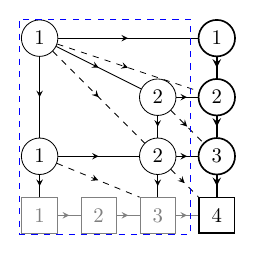}\\ 
\hline
\raisebox{9ex}{$(4\,2\,3\,1)$}& \includegraphics[scale = 0.65]{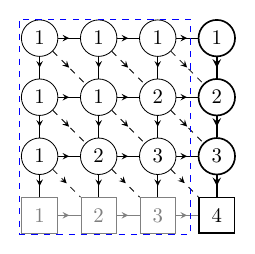}&\includegraphics[scale = 0.65]{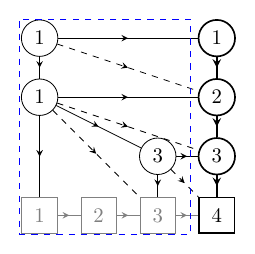}\\ 
\hline
\raisebox{9ex}{$(4\,3\,1\,2)$}& \includegraphics[scale = 0.65]{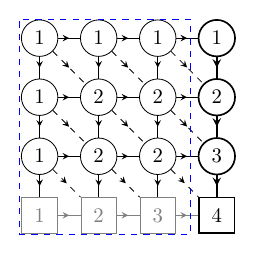}&\includegraphics[scale = 0.65]{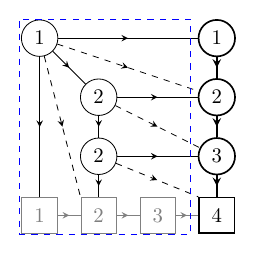}\\ 
\hline
\raisebox{9ex}{$(4\,3\,2\,1)$}& \includegraphics[scale = 0.65]{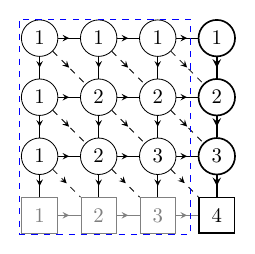}&\includegraphics[scale = 0.65]{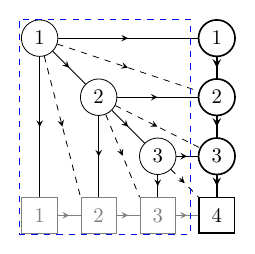}\\ 
           \hline
        \end{array}
    \end{equation*}
\end{subtable}
\end{table}

\newpage
\begin{landscape}
\subsection*{1d indices of Schubert line defects in Fl\texorpdfstring{$(4)$}{(4)}}

    \renewcommand{\arraystretch}{2}
    \begin{longtable}{|p{.08\textwidth} || p{.60\textwidth} | p{.65\textwidth}|}
    \hline
    $w$ & $\mathcal{I}_{w}^{({\rm 1d})}(x,y)$ & $\mathcal{I}_{w}^{({\rm 1d})}(x,y;\qk)$\\
    \hline
    \hline
    $(1\,2\,3\,4)$ & $1$ & $1$\\
    \hline
    $(2\,1\,3\,4)$ &  $1-\frac{e_1(x^{(1)})}{y_1}$ & $1-\frac{x_1}{y_1}$ \\
    \hline
    $(1\,3\,2\,4)$ &  $ 1 - \frac{e_2(x^{(2)})}{y_1 y_2} $ & $1-\frac{1}{1-q_1}\frac{x_1 x_2}{y_1 y_2}$\\
    \hline
    $(1\,2\,4\,3)$ &$1- \frac{e_3(x^{(3)})}{y_1 y_2 y_3}$ & $1- \frac{1}{(1-q_1)(1-q_2)} \frac{x_1 x_2 x_3}{y_1 y_2 y_3}$\\
    \hline
    $(2\,3\,1\,4)$ &  $1 - \frac{e_1(x^{(2)})}{y_1} + \frac{e_2 (x^{(2)})}{y_1^2} $ & $1 - \frac{x_1 + x_2}{y_1}+\frac{1}{1-q_1}\frac{x_1 x_2}{y_1^2}$\\
    \hline
    $(3\,1\,2\,4)$ & $1-e_1(x^{(1)})\left(\frac{1}{y_1} + \frac{1}{y_2}\right) + \frac{e_1(x^{(1)}) e_1(x^{(2)}) } {y_1 y_2} - \frac{e_2(x^{(2)})}{y_1 y_2}  $ & $1 - x_1 \left(\frac{1}{y_1}+\frac{1}{y_2} \right) + \frac{x_1(x_1+x_2)}{y_1 y_2} - \frac{1}{(1-q_1)}\frac{x_1 x_2}{y_1 y_2}$\\
    \hline
    $(2\,1\,4\,3)$ & $\left(1- \frac{e_1(x^{(1)})}{y_1}\right) \left(1 - \frac{e_3(x^{(3)})}{y_1 y_2 y_3}\right)$ & $\left(1- \frac{x_1}{y_1}\right) \left(1 - \frac{1}{(1-q_1)(1-q_2)}\frac{x_1 x_2 x_3}{y_1 y_2 y_3}\right)$ \\
    \hline
    $(1\,3\,4\,2)$ &  $1 - \frac{e_2(x^{(3)})}{y_1 y_2} + \frac{e_{3}(x^{(3)})}{y_1 y_2}\left(\frac{1}{y_1} + \frac{1}{y_2} \right)$ &  $1- \frac{1}{y_1 y_2}\left( \frac{x_1 x_2}{1-q_1} +\frac{x_2 x_3}{1-q_2} + x_1 x_3\right) + \frac{x_1 x_2 x_3}{(1-q_1)(1-q_2) y_1 y_2}\left(\frac{1}{y_1}+\frac{1}{y_2} \right)$\\
    \hline
    $(1\,4\,2\,3)$  & $1- e_2(x^{(2)})\left( \frac{1}{y_1 y_2} + \frac{1}{y_1 y_3} + \frac{1}{y_2 y_3}\right) + \frac{e_2(x^{(2}) e_1(x^{(3)})}{y_1 y_2 y_3} - \frac{e_3(x^{(3)})}{y_1 y_2 y_3}$ & $1 - \frac{x_1 x_2}{1-q_1}\left( \frac{1}{y_1 y_2} + \frac{1}{y_1 y_3} + \frac{1}{y_2 y_3}\right) + \frac{x_1 x_2 (x_1 + x_2)}{(1-q_1) y_1 y_2 y_3}  - \frac{q_2  x_1 x_2 x_3}{(1-q_1)(1-q_2)y_1 y_2 y_3}$\\
    \hline
    $(2\,3\,4\,1)$ & $1- \frac{e_1(x^{(3)})}{y_1} + \frac{e_2(x^{(3)})}{y_1^2} - \frac{e_3(x^{(3)})}{y_1^3}$ & $1- \frac{x_1 + x_2 + x_3}{y_1} + \frac{1}{y_1^2} \left( \frac{x_1 x_2}{(1-q_1)} + \frac{x_2 x_3}{(1-q_2)} + x_1 x_3\right) - \frac{x_1 x_2 x_3}{(1-q_1)(1-q_2) y_1^3}$ \\
    \hline
    $(3\,2\,1\,4)$ & $\left(1- \frac{e_1(x^{(1)})}{y_2}\right)\left(1- \frac{e_1(x^{(2)})}{y_1} + \frac{e_2(x^{(2)})}{y_1^2}\right)$ & $\left( 1- \frac{x_1}{y_2}\right)\left(1-\frac{x_1 + x_2}{y_1} +\frac{x_1 + x_2}{(1-q_1)y_1^2}\right)$\\
    \hline
    $(3\,2\,4\,1)$ & $\left(1-\frac{e_1(x^{(1)})}{y_2}\right)\left(1-\frac{e_1(x^{(3)})}{y_1} + \frac{e_2(x^{(3)})}{y_1^2} - \frac{e_3(x^{(3)})}{y_1^3}\right)$ & $\left(1-\frac{x_1}{y_2} \right)\vadjust{\vspace{6pt}}\newline \times \left( 1 - \frac{x_1 + x_2 + x_3}{y_1} + \frac{x_1 x_2 }{(1-q_1) y_1^2} + \frac{x_2 x_3}{(1-q_2)y_1^2}+ \frac{ x_1 x_3}{y_1^2} - \frac{x_1 x_2 x_3}{(1-q_1)(1-q_2)y_1^3}\right)$\\
    \hline
    $(2\,4\,3\,1)$ &$\left(1-\frac{e_2(x^{(2)})}{y_2 y_3}\right) \left(1-\frac{e_1(x^{(3)})}{y_1} + \frac{e_2(x^{(3)})}{y_1^2} - \frac{e_3(x^{(3)})}{y_1^3}\right)$ & $\left(1-\frac{x_1 x_2}{(1-q_1)y_2 y_3}\right) \vadjust{\vspace{6pt}}\newline \times\left( 1 - \frac{x_1 + x_2 + x_3}{y_1} + \frac{x_1 x_2 }{(1-q_1) y_1^2} + \frac{x_2 x_3}{(1-q_2) y_1^2}  + \frac{ x_1 x_3}{y_1^2} - \frac{x_1 x_2 x_3}{(1-q_1)(1-q_2)y_1^3}\right)$\\
    \hline
    $(3\,4\,2\,1)$ & $\left(1 - \frac{e_1(x^{(2)})}{y_2} + \frac{e_2(x^{(2)})}{y_2^2} \right)\left(1-\frac{e_1(x^{(3)})}{y_1} + \frac{e_2(x^{(3)})}{y_1^2} - \frac{e_3(x^{(3)})}{y_1^3}\right)$ &$\left( 1 - \frac{x_1+x_2}{y_2} + \frac{x_1 x_2}{(1-q_1 )y_2^2}\right)\vadjust{\vspace{6pt}} \newline \times\left( 1 - \frac{x_1 + x_2 + x_3}{y_1} + \frac{x_1 x_2}{(1-q_1) y_1^2} + \frac{x_2 x_3}{(1-q_2) y_1^2} + \frac{ x_1 x_3}{y_1^2} - \frac{x_1 x_2 x_3}{(1-q_1)(1-q_2)y_1^3}\right)$ \\
    \hline
    $(2\,4\,1\,3)$ & $1-\frac{e_1(x^{(2)})}{y_1} + \frac{e_2(x^{(2)})}{y_1^2} - \frac{e_2(x^{(2)})}{y_2 y_3} + \frac{e_2(x^{(2)}) e_1 (x^{(3)}) - e_3(x^{(3)})}{y_1 y_2 y_3} \vadjust{\vspace{6pt}} \newline - \frac{e_2(x^{(2)}) e_2(x^{(3)}) - e_1(x^{(2)})e_3(x^{(3)})}{y_1^2 y_2 y_3}$ & $ 1 - \frac{x_1 + x_2}{y_1}+ \frac{x_1 x_2}{(1-q_1)}\left(\frac{1}{y_1^2} - \frac{1}{y_2 y_3}\right) + \frac{x_1 x_2\left(x_1  + x_2 + x_3\right)}{(1-q_1)y_1 y_2 y_3} \vadjust{\vspace{6pt}} \newline - \frac{x_1 x_2 x_3}{(1-q_1)(1-q_2)y_1 y_2 y_3} + \frac{(x_1 + x_2)x_1 x_2 x_3}{(1-q_1)(1-q_2)y_1^2 y_2 y_3} \vadjust{\vspace{6pt}} \newline - \frac{x_1 x_2}{(1-q_1)y_1^2 y_2 y_3}\left( \frac{x_1 x_2}{1-q_1} + \frac{x_2 x_3}{1-q_2} + x_1 x_3 \right)$ \\
    \hline
    $(3\,1\,4\,2)$ & $1- \left(\frac{e_1(x^{(1)})}{y_1} + \frac{e_1(x^{(1)})}{y_2} \right) + \frac{e_1(x^{(1)})e_1(x^{(3)}) - e_2(x^{(3)})}{y_1 y_2} \vadjust{\vspace{6pt}} \newline + \left(\frac{e_3(x^{(3)})}{y_1^2 y_2} + \frac{e_3(x^{(3)})}{y_1 y_2^2} \right) - \frac{e_1(x^{(1)}) e_3(x^{(3)})}{y_1^2 y_2^2} $ & $1-\left(\frac{x_1}{y_1} + \frac{x_1}{y_1}\right) + \frac{1}{y_1 y_2}\left( x_1^2 - \frac{q_1 x_1 x_2}{1-q_1} - \frac{x_2 x_3}{1-q_2} \right) + \frac{x_1 x_2 x_3(y_1 + y_2)}{(1-q_1)(1-q_2) y_1^2 y_2^2}\vadjust{\vspace{6pt}} \newline - \frac{x_1^2 x_2 x_3}{(1-q_1)(1-q_2) y_1^2 y_2^2}$\\
    \hline
    $(4\,1\,2\,3)$ &$1-e_1(x^{(1)})\left(\frac{1}{y_1} + \frac{1}{y_2} + \frac{1}{y_3}\right) \vadjust{\vspace{6pt}}\newline + \left(e_1(x^{(1)}) e_1(x^{(2)}) - e_2(x^{(2)})\right)\left(\frac{1}{y_1 y_2}+\frac{1}{y_1 y_3}+\frac{1}{y_2 y_3}\right) \vadjust{\vspace{6pt}}\newline  -\frac{1}{y_1 y_2 y_3}\Big( e_1(x^{(1)}) e_1(x^{(2)})e_1(x^{(3)}) \vadjust{\vspace{6pt}}\newline {\quad \quad - e_2(x^{(2)})e_1(x^{(3)}) - e_1(x^{(1)})e_2(x^{(3)})+ e_3(x^{(3)}) \Big)} $ & $1- x_1 \left(\frac{1}{y_1} + \frac{1}{y_2} + \frac{1}{y_3}\right) + \frac{x_1(y_1 + y_2 + y_3)}{y_1 y_2 y_3}\left( x_1 - \frac{ q_1 x_2}{1-q_1}\right)  \vadjust{\vspace{6pt}}\newline  - \frac{x_1}{y_1 y_2 y_3}\left( x_1^2 - \frac{2 q_1 x_1 x_2+q_1 x_2^2}{1-q_1}  + \frac{q_1 q_2 x_2 x_3}{(1-q_1)(1-q_2)} \right)$\\
\hline
 $(4\,2\,3\,1)$ & $\left( 1- \frac{e_1(x^{(3)})}{y_1} + \frac{e_2(x^{(3)})}{y_1^2} - \frac{e_3(x^{(3)})}{y_1^3}\right)
 \vadjust{\vspace{6pt}}\newline \times \left( 1- e_1(x^{(1)}) \left(\frac{1}{y_2}+\frac{1}{y_3} \right) + \frac{e_1(x^{(1)})e_1(x^{(2)}) - e_2(x^{(2)})}{y_2 y_3} \right)$ & $\left( 1 - \frac{x_1+x_2+x_3}{y_1}+ \frac{1}{y_1^2}\left( \frac{x_1 x_2}{1-q_1} + \frac{x_2 x_3}{1-q_2} + x_1 x_3\right) - \frac{x_1 x_2 x_3}{y_1^3 (1-q_1)(1-q_2)}\right) \vadjust{\vspace{6pt}}\newline \times \left(1-\frac{x_1}{y_2} - \frac{x_1}{y_3}  + \frac{x_1}{y_2 y_3}\left( x_1 + x_2 - \frac{x_2}{1-q_1} \right)\right)$ \\
\hline
$(1\,4\,3\,2)$ & $ 1 - \frac{e_2(x^{(3)})}{y_1 y_2} - \frac{e_2(x^{(2)})}{y_3}\left(\frac{1}{y_1}+\frac{1}{y_2}\right) + \frac{e_3(x^{(3)})}{y_1 y_2}\left(\frac{1}{y_1}+\frac{1}{y_2}\right) \vadjust{\vspace{6pt}}\newline + \frac{e_2(x^{(2)})e_1(x^{(3)})}{y_1 y_2 y_3} - \frac{e_2(x^{(2)})e_3(x^{(3)})}{y_1^2 y_2^2 y_3}$ & $1 - \frac{1}{y_1 y_2}\left( \frac{x_1 x_2}{1-q_1} + \frac{x_2 x_3}{1-q_2} + x_1 x_3\right) - \frac{x_1 x_2(y_1 + y_2)}{(1-q_1)y_1 y_2 y_3} + \frac{x_1 x_2 (x_1 + x_2 + x_3)}{(1-q_1) y_1 y_2 y_3}
        \vadjust{\vspace{6pt}}\newline + \frac{x_1 x_2 x_3(y_1 + y_2)}{(1-q_1) (1-q_2) y_1^2 y_2^2 }- \frac{x_1^2 x_2^2 x_3}{(1-q_1)^2 (1-q_2)y_1^2 y_2^2 y_3}$\\
\hline
 $(3\,4\,1\,2)$ &  $1 - e_1(x^{(2)})\left(\frac{1}{y_1}+\frac{1}{y_2}\right) + e_2(x^{(2)})\left(\frac{1}{y_1^2}+\frac{1}{y_1 y_2}+\frac{1}{y_2^2}\right) \vadjust{\vspace{6pt}}\newline + \frac{e_1(x^{(2)})e_1(x^{(3)}) - e_2(x^{(3)})}{y_1 y_2} - \frac{e_2(x^{(2)})e_1(x^{(3)}) - e_3(x^{(3)})}{y_1 y_2}\left( \frac{1}{y_1} + \frac{1}{y_2} \right) \vadjust{\vspace{6pt}}\newline+ \frac{e_2(x^{(2)})e_2(x^{(3)}) - e_1(x^{(2)})e_3(x^{(3)})}{y_1^2 y_2^2}$ & $1 - \left(x_1 + x_2\right)\left(\frac{1}{y_1}+\frac{1}{y_2}\right) + \frac{x_1^2+x_2^2}{y_1 y_2} + \frac{x_1 x_2}{1-q_1}\left(\frac{1}{y_1^2}+\frac{1}{y_2^2}\right) + \frac{2 x_1 x_2}{y_1 y_2} \vadjust{\vspace{6pt}}\newline- \frac{q_2 x_2 x_3}{(1-q_2)y_1 y_2} - \frac{x_1 x_2 (y_1 + y_2)}{(1-q_1)y_1^2 y_2^2}\left( x_1 + x_2 -\frac{q_2 x_3}{1-q_2}\right) \vadjust{\vspace{6pt}}\newline+ \frac{x_1^2 x_2}{(1-q_1)y_1^2 y_2^2} \left(\frac{x_2}{1-q_1} - \frac{q_2 x_3}{1-q_2}\right)$\\
\hline
$(4\,2\,1\,3)$ & $1-\frac{e_1(x^{(2)})}{y_1} -e_1(x^{(1)})\left(\frac{1}{y_2}+ \frac{1}{y_3}\right) + e_2(x^{(2)}) \left(\frac{1}{y_1^2} - \frac{1}{y_2 y_3}\right) 
\vadjust{\vspace{6pt}}\newline +  \frac{e_1(x^{(1)})e_1(x^{(2)})}{y_1 y_2 y_3}(y_1 + y_2 +y_3) - \frac{e_1(x^{(1)})e_2(x^{(2)})}{y_1^2 y_2 y_3}\left(y_1 + y_2 +y_3\right) 
\vadjust{\vspace{6pt}}\newline + \frac{1}{y_1 y_2 y_3} \left(e_1(x^{(3)})e_2(x^{(2)})- e_1(x^{(1)})e_1(x^{(2)})e_1(x^{(3)}) \right) \vadjust{\vspace{6pt}}\newline + \frac{1}{y_1 y_2 y_3}\left( e_1(x^{(1)})e_2(x^{(3)})-e_3(x^{(3)})\right) 
\vadjust{\vspace{6pt}}\newline + \frac{1}{y_1^2 y_2 y_3}\left(e_1(x^{(1)})e_2(x^{(2)})e_1(x^{(3)}) - e_2(x^{(2)})e_2(x^{(3)}) \right) \vadjust{\vspace{6pt}}\newline - \frac{1}{y_1^2 y_2 y_3}\left(e_1(x^{(1)})e_3(x^{(3)})- e_1(x^{(2)})e_3(x^{(3)}) \right)$ 
& $1- \left(\frac{x_1+x_2}{y_1}+\frac{x_1}{y_2}+\frac{x_1}{y_3}\right) + \left(x_1^2 + x_1 x_2 \right) \left(\frac{1}{y_1 y_2} + \frac{1}{y_1 y_3} + \frac{1}{y_2 y_3}\right) 
\vadjust{\vspace{6pt}}\newline  +\frac{x_1 x_2}{1-q_1}\left( \frac{1}{y_1^2} - \frac{1}{y_2 y_3}\right) -\frac{(1-q_1)x_1^3+(1-2q_1)x_1^2 x_2 - q_1 x_1 x_2^2}{(1-q_1 )y_1 y_2 y_3} 
\vadjust{\vspace{6pt}}\newline - \frac{q_1 q_2 x_1 x_2 x_3}{(1-q_1)(1-q_2)y_1 y_2 y_3} - \frac{(y_2 + y_3)x_1^2 x_2}{(1-q_1)y_1^2 y_2 y_3} + \frac{x_1^3 x_2}{(1-q_1)y_1^2 y_2 y_3} - \frac{q_1 x_1^2 x_2^2}{(1-q_1)^2 y_1^2 y_2 y_3}$ \\
\hline
$(4\,1\,3\,2)$ & $1 - e_{1}(x^{(1)})\left( \frac{1}{y_1} + \frac{1}{y_2} + \frac{1}{y_3}\right) + \frac{e_1(x^{(1)})e_1(x^{(3)}) - e_2(x^{(3)})}{y_1 y_2} 
\vadjust{\vspace{6pt}}\newline + \left(e_1(x^{(1)})e_1(x^{(2)}) - e_2(x^{(2)})\right)\left(\frac{1}{y_1 y_3}+\frac{1}{y_2 y_3}\right) 
\vadjust{\vspace{6pt}}\newline + \frac{e_3(x^{(3)})}{y_1 y_2}\left(\frac{1}{y_1}+\frac{1}{y_2} \right)  - \frac{e_1(x^{(1)}) e_3(x^{(3)})  }{y_1 y_2}\left( \frac{1}{y_1 y_2}  + \frac{1}{y_1 y_3} + \frac{1}{y_2 y_3}\right)  
\vadjust{\vspace{6pt}}\newline + \frac{\left( e_1(x^{(1)}) e_2(x^{(3)}) - e_1(x^{(1)}) e_1(x^{(2)})e_1(x^{(3)}) + e_2(x^{(2)}) e_1(x^{(3)}) \right)}{y_1 y_2 y_3}
\vadjust{\vspace{6pt}}\newline + \frac{e_1(x^{(1)})e_1(x^{(2)})e_3(x^{(3)}) - e_2(x^{(2)}) e_3(x^{(3)})    }{y_1^2 y_2^2 y_3}$ & $1 - \left(\frac{x_1}{y_1} +\frac{x_1}{y_2} + \frac{x_1}{y_3}\right) + \frac{x_1 (y_1 + y_2 + y_3)}{y_1 y_2 y_3} \left( x_1 - \frac{q_1 x_2}{1-q_1}\right) - \frac{x_2 x_3}{(1-q_2)y_1 y_2} 
\vadjust{\vspace{6pt}}\newline - \frac{x_1^3}{y_1 y_2 y_3} + \frac{q_1 \left( 2x_1^2 x_2 + x_1 x_2^2\right)}{(1-q_1)y_1 y_2 y_3}+ \frac{(1-q_1 q_2) x_1 x_2 x_3}{(1-q_1)(1-q_2)y_1 y_2 y_3} 
\vadjust{\vspace{6pt}}\newline - \frac{x_1 x_2 x_3(x_1 - y_1 - y_2)}{(1-q_1)(1-q_2)y_1^2 y_2^2} + \frac{x_1^2 x_2 x_3(x_1 - y_1 - y_2)}{(1-q_1)(1-q_2) y_1^2 y_2^2 y_3} - \frac{q_1 x_1^2 x_2^2 x_3}{(1-q_1)^2 (1-q_2) y_1^2 y_2^2 y_3} $\\
\hline
$(4\,3\,1\,2)$ & $\left(1-\frac{e_1(x^{(1)})}{y_3}\right) \vadjust{\vspace{6pt}}\newline \times \left( 1 - e_1(x^{(2)})\left(\frac{1}{y_1}+\frac{1}{y_2} \right) + e_2(x^{(2)}) \left(\frac{1}{y_1^2}+\frac{1}{y_2^2}  \right) \right.     
\vadjust{\vspace{6pt}}\newline + \left. \frac{e_2(x^{(2)}) + e_1(x^{(2)}) e_1(x^{(3)}) - e_2(x^{(3)})}{y_1 y_2} - \frac{e_2(x^{(2)})e_1(x^{(3)})}{y_2 y_2}\left( \frac{1}{y_1} + \frac{1}{y_2}\right) \right. 
\vadjust{\vspace{6pt}}\newline \left. + \frac{e_3(x^{(3)})}{y_2 y_2}\left( \frac{1}{y_1} + \frac{1}{y_2}\right) + \frac{e_2(x^{(2)})e_2(x^{(3)}) - e_1(x^{(2)})e_3(x^{(3)})}{y_1^2 y_2^2}\right)$ & $\left(1-\frac{x_1}{y_3}\right) 
\vadjust{\vspace{6pt}}\newline \times \left(1 - (x_1 + x_2)\left(\frac{1}{y_1} + \frac{1}{y_2} \right)  + \frac{x_1 x_2}{(1-q_1)}\left(\frac{1}{y_1^2} +\frac{1}{y_2^2} \right)+ \frac{(x_1+x_2)^2}{y_1 y_2}\right.
\vadjust{\vspace{6pt}}\newline \left. - \frac{q_2x_2 x_3}{(1-q_2)y_1 y_2} - \frac{x_1 x_2 (y_1 +y_2)\left((1-q_2)x_1 + (1-q_2)x_2 - q_2 x_3\right)}{(1-q_1)(1-q_2)y_1^2 y_2^2} \right.    \vadjust{\vspace{6pt}}\newline \left. + \frac{x_1^2 x_2 \left( (1-q_2)x_2 - (1-q_1) q_2 x_3 \right)}{(1-q_1)^2(1-q_2)y_1^2 y_2^2}\right)
$\\ \hline
$(4\,3\,2\,1)$ & $\left(1-\frac{e_1(x^{(1)})}{y_3}\right)\left(1-\frac{e_1(x^{(2)})}{y_2}+\frac{e_2(x^{(2)})}{y_2^2}\right)
\vadjust{\vspace{6pt}}\newline \times \left(1-\frac{e_1(x^{(3)})}{y_1}+\frac{e_2(x^{(3)})}{y_1^2} - \frac{e_3(x^{(3)})}{y_1^3}\right)$ & $\left(1-\frac{x_1}{y_3}\right)\left(1-\frac{x_1 + x_2}{y_2}+\frac{x_1 x_2}{(1-q_1)y_2^2}\right)
\vadjust{\vspace{6pt}}\newline \times \left(1-\frac{x_1 + x_2 + x_3}{y_1}+\frac{1}{y_1^2}\left( \frac{x_1 x_2}{1-q_1} + \frac{x_2 x_3}{1-q_2} + x_1 x_3 \right) - \frac{x_1 x_2 x_3}{(1-q_1)(1-q_2)y_1^3}\right)$\\\hline
\end{longtable}
\end{landscape}

\newpage
\begin{landscape}
    \subsection*{0d partition functions of Schubert point defects in Fl\texorpdfstring{$(4)$}{(4)}}
    \renewcommand{\arraystretch}{2}
        \begin{longtable}{|p{.08\textwidth} || p{.65\textwidth} | p{.7\textwidth}|}
    \hline
    $w$ & $\mathcal{I}_{w}^{({\rm 0d})}(x,y)$ & $\mathcal{I}_{w}^{({\rm 0d})}(x,y;\qcoh)$\\
    \hline
    \hline
    $(1\,2\,3\,4)$ & $1$ & $1$\\
    \hline
    $(2\,1\,3\,4)$ & $e_1(\widetilde{\sigma}^{(1)})-m_1$ & $\sigma_1 - m_1$ \\
    \hline
    $(1\,3\,2\,4)$ & $e_1(\widetilde{\sigma}^{(2)}) - m_1 -m_2$ & $\sigma_1 + \sigma_2 - m_1 - m_2$ \\
    \hline
    $(1\,2\,4\,3)$ & $e_1(\widetilde{\sigma}^{(3)}) - m_1 - m_2 - m_3$ &$\sigma_1 + \sigma_2 + \sigma_3 - m_1 - m_2 - m_3$ \\
    \hline
    $(2\,3\,1\,4)$ & $e_2(\widetilde{\sigma}^{(2)}) -  m_1 e_1(\widetilde{\sigma}^{(2)}) + m_1^2$ & $\sigma_1 \sigma_2 - m_1 (\sigma_1 + \sigma_2) + m_1^2 + \qcoh_1$ \\
    \hline
    $(3\,1\,2\,4)$ & $(e_1(\widetilde{\sigma}^{(1)})-m_1)(e_1(\widetilde{\sigma}^{(1)})-m_2) -  e_1(\widetilde{\sigma}^{(1)})^2 
    \vadjust{\vspace{6pt}}\newline + e_1(\widetilde{\sigma}^{(1)}) e_1(\widetilde{\sigma}^{(2)})  - e_2(\widetilde{\sigma}^{(2)}))$ & $\sigma_1^2 - (m_1 + m_2)\sigma_1 + m_1 m_2 - \qcoh_1$  \\
    \hline
    $(2\,1\,4\,3)$ & $(e_1(\widetilde{\sigma}^{(1)})-m_1)(e_1(\widetilde{\sigma}^{(3)}) - m_1 - m_2 - m_3)$ & $(\sigma_1-m_1)(\sigma_1+\sigma_2 +\sigma_3 - m_1 - m_2 - m_3)$\\
    \hline
    $(1\,3\,4\,2)$ & $e_2(\widetilde{\sigma}^{(3)}) - e_1(\widetilde{\sigma}^{(3)}) (m_1 + m_2) + m_1^2 + m_1 m_2 + m_2^2$ & $\sigma_1 \sigma_2 + \sigma_1 \sigma_3 + \sigma_2 \sigma_3 - (\sigma_1 + \sigma_2 + \sigma_3)(m_1 + m_2) + m_1^2 +m_1 m_2 \vadjust{\vspace{6pt}}\newline + m_2^2 +\qcoh_1 + \qcoh_2$\\
    \hline
    $(1\,4\,2\,3)$ & $e_1(\widetilde{\sigma}^{(2)}) \left( e_1(\widetilde{\sigma}^{(3)}) - m_1 -m_2 -m_3 \right) - e_2(\widetilde{\sigma}^{(3)}) + m_1 m_2 
    \vadjust{\vspace{6pt}}\newline + m_1 m_3 + m_2 m_3$ & $\sigma_1^2 + \sigma_1 \sigma_2 +\sigma_2^2 - (\sigma_1 + \sigma_2)(m_1 + m_2 +m_3) + m_1 m_2+ m_1 m_3 
    \vadjust{\vspace{6pt}}\newline + m_2 m_3 - \qcoh_1 -\qcoh_2 $\\
    \hline
    $(2\,3\,4\,1)$ & $e_3(\widetilde{\sigma}^{(3)}) - e_2(\widetilde{\sigma}^{(3)}) m_1 + e_1(\widetilde{\sigma}^{(3)})m_1^2 - m_1^3$ & $\sigma_1 \sigma_2 \sigma_3 - (\sigma_1 \sigma_2 + \sigma_1 \sigma_3 + \sigma_2 \sigma_3)m_1 + (\sigma_1 + \sigma_2+\sigma_3)m_1^2 - m_1^3
    \vadjust{\vspace{6pt}}\newline + \qcoh_1 \sigma_3 + \qcoh_2 \sigma_1 - \qcoh_1 m_1 - \qcoh_2 m_1 $\\
    \hline
    $(3\,2\,1\,4)$ & $( e_1(\widetilde{\sigma}^{(1)}) - m_2 ) (e_2(\widetilde{\sigma}^{(2)}) - e_1(\widetilde{\sigma}^{(2)})m_1 +m_1^2 )$ & $(\sigma_1 - m_2)\left(\sigma_1 \sigma_2 - (\sigma_1 + \sigma_2)m_1 + m_1^2 + \qcoh_1\right)$  \\
    \hline
    $(3\,2\,4\,1)$ & $(e_1(\widetilde{\sigma}^{(1)}) - m_2) \left( e_3(\widetilde{\sigma}^{(3)}) - e_2(\widetilde{\sigma}^{(3)}) m_1 + e_1(\widetilde{\sigma}^{(3)}) m_1^2 - m_1^3  \right)$ & $(\sigma_1 -m_2) \left( \sigma_1 \sigma_2 \sigma_3 - (\sigma_1 \sigma_2 + \sigma_1 \sigma_3 + \sigma_2 \sigma_3 + \qcoh_1 + \qcoh_2 ) m_1 \right. 
    \vadjust{\vspace{6pt}}\newline  \left. + (\sigma_1 + \sigma_2 + \sigma_3) m_1^3 + \qcoh_1 \sigma_3 + \qcoh_2 \sigma_1 - m_1^3 \right)$ \\
    \hline
    $(2\,4\,3\,1)$ & $\left(e_1(\widetilde{\sigma}^{(2)}) - m_2 -m_3 \right) \left( e_3(\widetilde{\sigma}^{(3)}) - e_2(\widetilde{\sigma}^{(3)})m_1  \right. 
    \vadjust{\vspace{6pt}}\newline  \left. + e_1(\widetilde{\sigma}^{(3)}) m_1^2 - m_1^3 \right)$ &  $(\sigma_1 + \sigma_2 -m_2 - m_3) \left( \sigma_1 \sigma_2 \sigma_3 - (\sigma_1 \sigma_2 + \sigma_1 \sigma_3 + \sigma_2 \sigma_3 + \qcoh_1 + \qcoh_2 ) m_1 \right. 
    \vadjust{\vspace{6pt}}\newline  \left. + (\sigma_1 + \sigma_2 + \sigma_3) m_1^3 + \qcoh_1 \sigma_3 + \qcoh_2 \sigma_1 - m_1^3 \right)$\\
    \hline
    $(3\,4\,2\,1)$ & $\left(e_2(\widetilde{\sigma}^{(2)}) - e_1(\widetilde{\sigma}^{(2)}) m_2 + m_2^3 \right) 
    \vadjust{\vspace{6pt}}\newline \times \left( e_3(\widetilde{\sigma}^{(3)}) - e_2(\widetilde{\sigma}^{(3)})m_1  + e_1(\widetilde{\sigma}^{(3)}) m_1^2 - m_1^3 \right)$ & $(\sigma_1\sigma_2 - (\sigma_1 + \sigma_2)m_2 + m_2^2+\qcoh_1) 
    \vadjust{\vspace{6pt}}\newline  \times \left( \sigma_1 \sigma_2 \sigma_3 - (\sigma_1 \sigma_2 + \sigma_1 \sigma_3 + \sigma_2 \sigma_3 + \qcoh_1 + \qcoh_2 ) m_1 \right.
    \vadjust{\vspace{6pt}}\newline \left. \quad + (\sigma_1 + \sigma_2 + \sigma_3) m_1^3 + \qcoh_1 \sigma_3 + \qcoh_2 \sigma_1 - m_1^3 \right)$ \\
    \hline
     $(2\,4\,1\,3)$ & $e_2(\widetilde{\sigma}^{(2)}) e_1(\widetilde{\sigma}^{(3)}) - e_3(\widetilde{\sigma}^{(3)}) - e_2(\widetilde{\sigma}^{(2)})(m_1 + m_2 + m_3) 
     \vadjust{\vspace{6pt}}\newline - e_1(\widetilde{\sigma}^{(2)})e_1(\widetilde{\sigma}^{(3)}) m_1 + e_2(\widetilde{\sigma}^{(3)}) m_1 - m_1^2 m_2 - m_1^2 m_3
     \vadjust{\vspace{6pt}}\newline + e_1(\widetilde{\sigma}^{(2)}) (m_1^2 + m_1 m_2 + m_1 m_3 )  $ & $(\sigma_1 + \sigma_2 - m_2 -m_3) ( \sigma_1 - m_1) ( \sigma_2 -m_1 )  - \qcoh_2  (\sigma_1-m_1 ) 
     \vadjust{\vspace{6pt}}\newline + \qcoh_1 (\sigma_1 + \sigma_2 - m_2 -m_3)$\\
    \hline
    $(3\,1\,4\,2)$ & $e_1(\widetilde{\sigma}^{(1)}) e_2(\widetilde{\sigma}^{(3)}) - e_3(\widetilde{\sigma}^{(3)}) - e_1(\widetilde{\sigma}^{(1)}) e_1(\widetilde{\sigma}^{(3)}) (m_1 + m_2) 
    \vadjust{\vspace{6pt}}\newline + e_1(\widetilde{\sigma}^{(1)}) \left(m_1^2 + m_1 m_2+m_2^2 \right) + e_1(\widetilde{\sigma}^{(3)}) m_1 m_2 
    \vadjust{\vspace{6pt}}\newline - m_1^2 m_2 - m_1 m_2^2$ & $(\sigma_1 - m_1)(\sigma_1 -m_2)(\sigma_2 + \sigma_3 - m_1 -m_2) + \qcoh_1 (\sigma_1 - \sigma_3)$\\
    \hline
    $(4\,1\,2\,3)$ & $e_1(\widetilde{\sigma}^{(1)}) e_1(\widetilde{\sigma}^{(2)})e_1(\widetilde{\sigma}^{(3)}) - e_2(\widetilde{\sigma}^{(2)})e_1(\widetilde{\sigma}^{(3)}) + e_3(\widetilde{\sigma}^{(3)})
    \vadjust{\vspace{6pt}}\newline  - e_1(\widetilde{\sigma}^{(1)})e_2(\widetilde{\sigma}^{(3)})  - e_1(\widetilde{\sigma}^{(1)}) e_1(\widetilde{\sigma}^{(2)}) (m_1 + m_2+m_3) 
    \vadjust{\vspace{6pt}}\newline + e_2(\widetilde{\sigma}^{(2)})(m_1 + m_2+m_3) - m_1 m_2 m_3
    \vadjust{\vspace{6pt}}\newline + e_1(\widetilde{\sigma}^{(1)}) (m_1 m_2 + m_1 m_3 + m_2 m_3)  $ & $\sigma_1^3 - \sigma_1^2 (m_1 + m_2 +m_3)+ \sigma_1 (m_1 m_2 + m_1 m_3 + m_2 m_3) 
    \vadjust{\vspace{6pt}}\newline  + \qcoh_1 (m_1 + m_2 + m_3 -2\sigma_1 -\sigma_2) - m_1 m_2 m_3$\\
    \hline
    $(4\,2\,3\,1)$ & $\left( e_3(\widetilde{\sigma}^{(3)}) - e_2(\widetilde{\sigma}^{(3)}) m_1 + e_1(\widetilde{\sigma}^{(3)}) m_1^2 - m_1^3\right) \times
 \vadjust{\vspace{6pt}}\newline  \left(e_1(\widetilde{\sigma}^{(1)}) e_1(\widetilde{\sigma}^{(2)}) - e_2(\widetilde{\sigma}^{(2)}) - e_1(\widetilde{\sigma}^{(1)}) (m_2 + m_3) + m_2 m_3 \right)$ & $\left( (\sigma_1 - m_2 )(\sigma_1 - m_3 ) - \qcoh_1\right)\left( \sigma_1 \sigma_2 \sigma_3 - m_1 (\sigma_1 \sigma_2 + \sigma_1 \sigma_3 + \sigma_2 \sigma_3) \right. 
 \vadjust{\vspace{6pt}}\newline  \left.+ m_1^2 (\sigma_1 + \sigma_2 + \sigma_3)+ \qcoh_1 \sigma_3 + \qcoh_2 \sigma_1 - m_1(\qcoh_1 + \qcoh_2 ) - m_1^3 \right)$\\
    \hline
$(1\,4\,3\,2)$ & $e_1(\widetilde{\sigma}^{(2)}) e_2(\widetilde{\sigma}^{(3)}) - e_3(\widetilde{\sigma}^{(3)}) - e_1(\widetilde{\sigma}^{(2)})e_1(\widetilde{\sigma}^{(3)})(m_1 +m_2) 
\vadjust{\vspace{6pt}}\newline - e_2(\widetilde{\sigma}^{(3)})m_3 + e_1(\widetilde{\sigma}^{(3)}) (m_1 m_2 + m_1 m_3 + m_2 m_3) 
\vadjust{\vspace{6pt}}\newline + e_1(\widetilde{\sigma}^{(2)})(m_1^2 + m_1 m_2 +m_2^2)  - m_1^2 m_2 - m_1 m_2^2 - m_1^2 m_3
\vadjust{\vspace{6pt}}\newline - m_2^2 m_3 - m_1 m_2 m_3 $ & $(\sigma_1 + \sigma_2 - m_3)(\sigma_1 \sigma_2 + \sigma_1 \sigma_3 + \sigma_2 \sigma_3 + \qcoh_1 + \qcoh_2) - \sigma_1 \sigma_2 \sigma_3 
\vadjust{\vspace{6pt}}\newline - (\sigma_1 + \sigma_2 + \sigma_3)(\sigma_1 + \sigma_2)(m_1 + m_2) +(\sigma_1 + \sigma_2 + \sigma_3)(m_1 m_2 
\vadjust{\vspace{6pt}}\newline + m_1 m_3 + m_2 m_3) + (\sigma_1 + \sigma_2) (m_1^2 +m_1m_2 + m_2^2)- \qcoh_1 \sigma_3 
\vadjust{\vspace{6pt}}\newline - \qcoh_2 \sigma_2 - m_1^2 m_2 - m_1 m_2^2 - m_1^2 m_3 - m_2^2 m_3 - m_1 m_2 m_3 $\\
\hline
$(3\,4\,1\,2)$ & $e_2(\widetilde{\sigma}^{(2)}) e_2(\widetilde{\sigma}^{(3)}) - e_1(\widetilde{\sigma}^{(2)})e_3(\widetilde{\sigma}^{(3)}) + e_3(\widetilde{\sigma}^{(3)})(m_1 + m_2) 
\vadjust{\vspace{6pt}}\newline - e_2(\widetilde{\sigma}^{(2)}) e_1(\widetilde{\sigma}^{(3)}) (m_1 + m_2) + e_2(\widetilde{\sigma}^{(2)})(m_1^2 + m_1 m_2 + m_2^2) 
\vadjust{\vspace{6pt}}\newline + \left(e_1(\widetilde{\sigma}^{(2)}) e_1(\widetilde{\sigma}^{(3)})  - e_2(\widetilde{\sigma}^{(3)}) \right) m_1 m_2 
\vadjust{\vspace{6pt}}\newline - e_1(\widetilde{\sigma}^{(2)}) (m_1^2 m_2 + m_1 m_2^2) + m_1^2 m_2^2      $ & $(\sigma_1 \sigma_2 + \sigma_1 \sigma_3 + \sigma_2 \sigma_3 + \qcoh_1 + \qcoh_2) (\sigma_1 \sigma_2 + \qcoh_1) - (\sigma_1 \sigma_2 \sigma_3 + \qcoh_1 \sigma_3 
\vadjust{\vspace{6pt}}\newline + \qcoh_2 \sigma_1)(\sigma_1 + \sigma_2 -m_1 -m_2) - (\sigma_1 + \sigma_2 + \sigma_3)(\sigma_1 \sigma_2 + \qcoh_1)(m_1 
\vadjust{\vspace{6pt}}\newline + m_2) + (\sigma_1+\sigma_2 +\sigma_3)(\sigma_1 + \sigma_2)m_1 m_2 + (\sigma_1 \sigma_2 + \qcoh_1) (m_1^2 + m_2^2) 
\vadjust{\vspace{6pt}}\newline - (\sigma_1 \sigma_3 + \sigma_2 \sigma_3 + \qcoh_2) m_1 m_2 - (\sigma_1 + \sigma_2)(m_1^2 m_2 + m_1 m_2^2) + m_1^2 m_2^2$\\
\hline
$(4\,3\,2\,1)$ & $\left( e_3(\widetilde{\sigma}^{(3)}) - e_2(\widetilde{\sigma}^{(3)}) m_1 + e_1(\widetilde{\sigma}^{(3)}) m_1^2 - m_1^3 \right)
\vadjust{\vspace{6pt}}\newline \times \left( e_2(\widetilde{\sigma}^{(2)}) - e_1(\widetilde{\sigma}^{(2)}) m_2 + m_2^2 \right) (e_1(\widetilde{\sigma}^{(1)}) - m_3)$ & $(\sigma_1 - m_3)\left(\sigma_1 \sigma_2 - (\sigma_1 + \sigma_2) m_2 + m_2^2 + \qcoh_1\right)
\vadjust{\vspace{6pt}}\newline \times \left( \sigma_1 \sigma_2 \sigma_3 - m_1 (\sigma_1 \sigma_2 + \sigma_1 \sigma_3 + \sigma_2 \sigma_3 + \qcoh_1 + \qcoh_2) \right. 
\vadjust{\vspace{6pt}}\newline \left. \quad + m_1^2 (\sigma_1 + \sigma_2 +\sigma_3) - m_1^3 + \qcoh_1 \sigma_3 + \qcoh_2 \sigma_1\right)$ \\ \hline
$(4\,2\,1\,3)$ & $e_1(\widetilde{\sigma}^{(2)}) e_3(\widetilde{\sigma}^{(3)})- e_2(\widetilde{\sigma}^{(2)}) e_2(\widetilde{\sigma}^{(3)}) - e_1(\widetilde{\sigma}^{(1)}) e_3(\widetilde{\sigma}^{(3)}) 
    \vadjust{\vspace{6pt}}\newline + e_1(\widetilde{\sigma}^{(1)}) e_1(\widetilde{\sigma}^{(3)}) e_2(\widetilde{\sigma}^{(2)})  -  e_1(\widetilde{\sigma}^{(1)})  e_1(\widetilde{\sigma}^{(2)})  e_1(\widetilde{\sigma}^{(3)}) m_1  
    \vadjust{\vspace{6pt}}\newline -  e_1(\widetilde{\sigma}^{(1)})  e_2(\widetilde{\sigma}^{(2)})(m_1 + m_2)  +  e_2(\widetilde{\sigma}^{(2)})  e_1(\widetilde{\sigma}^{(3)}) m_1 
    \vadjust{\vspace{6pt}}\newline +  e_1(\widetilde{\sigma}^{(1)})  e_2(\widetilde{\sigma}^{(3)}) m_1 -  e_3(\widetilde{\sigma}^{(3)}) m_1 +  e_1(\widetilde{\sigma}^{(1)})  e_1(\widetilde{\sigma}^{(2)}) m_1^2 
    \vadjust{\vspace{6pt}}\newline -  e_2(\widetilde{\sigma}^{(2)}) m_1^2 +  e_1(\widetilde{\sigma}^{(1)})  e_1(\widetilde{\sigma}^{(2)}) m_1 m_2 -  e_1(\widetilde{\sigma}^{(1)}) m_1^2 m_2 
    \vadjust{\vspace{6pt}}\newline -  e_1(\widetilde{\sigma}^{(1)})  e_2(\widetilde{\sigma}^{(2)}) m_3 +  e_1(\widetilde{\sigma}^{(1)})  e_1(\widetilde{\sigma}^{(2)}) m_1 m_3 + m_1^2 m_2 m_3
    \vadjust{\vspace{6pt}}\newline  -  e_1(\widetilde{\sigma}^{(1)}) m_1^2 m_3 +  e_2(\widetilde{\sigma}^{(2)}) m_2 m_3 -  e_1(\widetilde{\sigma}^{(2)}) m_1 m_2 m_3 $ & 
    $\sigma_1 \sigma_2 \left( (\sigma_1 - m_2)(\sigma_1 - m_3) -\qcoh_1 \right) -\qcoh_1^2 - \qcoh_1 \qcoh_2 - \qcoh_1 m_1^2  
    \vadjust{\vspace{6pt}}\newline - (\sigma_1 + \sigma_2)\left((\sigma_1 - m_2)(\sigma_1 - m_3) - \qcoh_1 \right) m_1 \vadjust{\vspace{6pt}}\newline + (\sigma_1 - m_2)(\sigma_1 - m_3) (\qcoh_1 + m_1^2 ) $\\
\hline
$(4\,3\,1\,2)$ & $(e_1(\widetilde{\sigma}^{(1)}) - m_3) \left( e_2(\widetilde{\sigma}^{(2)}) e_2(\widetilde{\sigma}^{(3)}) - e_1(\widetilde{\sigma}^{(2)}) e_3(\widetilde{\sigma}^{(3)}) \right. 
\vadjust{\vspace{6pt}}\newline  \left.- e_2(\widetilde{\sigma}^{(2)}) e_1(\widetilde{\sigma}^{(3)}) (m_1+m_2) + e_3(\widetilde{\sigma}^{(3)}) (m_1 +m_2) \right.
\vadjust{\vspace{6pt}}\newline  \left. + e_2(\widetilde{\sigma}^{(2)}) (m_1^2 + m_1 m_2 + m_2^2) + e_1(\widetilde{\sigma}^{(2)})e_1(\widetilde{\sigma}^{(3)})m_1 m_2 \right.
\vadjust{\vspace{6pt}}\newline  \left. - e_2(\widetilde{\sigma}^{(3)}) m_1 m_2 - e_1(\widetilde{\sigma}^{(2)}) (m_1^2 m_2 + m_1 m_2^2 ) + m_1^2 m_2^2\right)$ &
$ (\sigma_1 - m_3) \left( (\sigma_1 -m_1) (\sigma_1 - m_2)(\sigma_2 - m_1)(\sigma_2 - m_2) + \qcoh_1 m_1^2 \right.
\vadjust{\vspace{6pt}}\newline \left. + \qcoh_1^2 + \qcoh_1 \qcoh_2 - \qcoh_1 (\sigma_1+\sigma_2)m_1 + \qcoh_2 (\sigma_1 - m_2) m_1 \right.
\vadjust{\vspace{6pt}}\newline \left. + \qcoh_1 ( 2\sigma_1 - m_2) \sigma_2 - (\sigma_1 - m_2)(\sigma_1 \qcoh_2 + m_2 \qcoh_1)\right)
$\\ \hline
$(4\,1\,3\,2)$ & $e_1(\widetilde{\sigma}^{(1)}) e_1(\widetilde{\sigma}^{(2)}) e_2(\widetilde{\sigma}^{(3)}) -e_2(\widetilde{\sigma}^{(2)}) e_2(\widetilde{\sigma}^{(3)}) 
\vadjust{\vspace{6pt}}\newline - e_1(\widetilde{\sigma}^{(1)}) e_3(\widetilde{\sigma}^{(3)}) + e_2(\widetilde{\sigma}^{(2)})e_1(\widetilde{\sigma}^{(3)}) (m_1 + m_2) 
\vadjust{\vspace{6pt}}\newline - e_1(\widetilde{\sigma}^{(1)}) e_1(\widetilde{\sigma}^{(2)}) e_1(\widetilde{\sigma}^{(3)}) (m_1 + m_2) 
\vadjust{\vspace{6pt}}\newline - e_1(\widetilde{\sigma}^{(1)}) e_2(\widetilde{\sigma}^{(3)}) m_3 + e_3(\widetilde{\sigma}^{(3)}) m_3
\vadjust{\vspace{6pt}}\newline + \left(e_1(\widetilde{\sigma}^{(1)})e_1(\widetilde{\sigma}^{(2)}) - e_2(\widetilde{\sigma}^{(2)}) \right) (m_1^2 + m_1 m_2 + m_2^2) 
\vadjust{\vspace{6pt}}\newline  + e_1(\widetilde{\sigma}^{(1)}) e_1(\widetilde{\sigma}^{(3)}) (m_1 m_2 + m_1 m_3 + m_2 m_3) 
\vadjust{\vspace{6pt}}\newline   - e_1(\widetilde{\sigma}^{(1)}) (m_1^2 m_2 + m_1^2 m_3 + m_1 m_2^2 + m_2^2 m_3 + m_1 m_2 m_3) 
\vadjust{\vspace{6pt}}\newline   - e_1(\widetilde{\sigma}^{(3)}) m_1 m_2 m_3 + m_1^2 m_2 m_3 + m_1 m_2^2 m_3    $ &
$ (\sigma_1 - m_1)(\sigma_1 - m_2)(\sigma_1 - m_3)(\sigma_2 + \sigma_3 - m_1 - m_2) - \qcoh_1^2 - \qcoh_1 \qcoh_2 
\vadjust{\vspace{6pt}}\newline  - \qcoh_1 (m_1^2 + m_1 m_2 + m_2^2) + \qcoh_1 (\sigma_1 + \sigma_2 + \sigma_3)(m_1+m_2) 
\vadjust{\vspace{6pt}}\newline  - \qcoh_1 (\sigma_1 \sigma_2 + 2 \sigma_1 \sigma_3 + \sigma_2\sigma_3) + \qcoh_1 \sigma_1^2 - \qcoh_1 (\sigma_1 - \sigma_3)m_3
$\\
\hline
    \end{longtable}
\end{landscape}

\newpage
\bibliography{flagGLSM} 
\bibliographystyle{JHEP} 
\end{document}